\documentclass[twocolumn]{aastex631}

\received{\today}
\revised{\today}
\accepted{\today}
\submitjournal{ApJ}

\shorttitle{Pulse independence of Oscillatory Reconnection periodicity}

\shortauthors{Karampelas et al.}
\graphicspath{{./}{figures/}}

\begin{document}

\title{The independence of oscillatory reconnection periodicity from the initial pulse}

\correspondingauthor{Konstantinos Karampelas}
\email{konstantinos.karampelas@northumbria.ac.uk}

\author[0000-0001-5507-1891]{Konstantinos Karampelas}
\affiliation{Department of Mathematics, Physics and Electrical Engineering, Northumbria University,\\ Newcastle upon Tyne, NE1 8ST, UK}
\affiliation{Centre for mathematical Plasma Astrophysics, Department of Mathematics, KU Leuven,\\ Celestijnenlaan 200B bus 2400, B-3001 Leuven, Belgium }
\author[0000-0002-7863-624X]{James A. McLaughlin}
\affiliation{Department of Mathematics, Physics and Electrical Engineering, Northumbria University,\\ Newcastle upon Tyne, NE1 8ST, UK}
\author[0000-0002-5915-697X]{Gert J. J. Botha}
\affiliation{Department of Mathematics, Physics and Electrical Engineering, Northumbria University,\\ Newcastle upon Tyne, NE1 8ST, UK}
\author[0000-0001-8954-4183]{St\'{e}phane R\'{e}gnier}
\affiliation{Department of Mathematics, Physics and Electrical Engineering, Northumbria University,\\ Newcastle upon Tyne, NE1 8ST, UK}

\begin{abstract}
Oscillatory reconnection can manifest through the interaction between the ubiquitous MHD waves and omnipresent null points in the solar atmosphere and is characterized by an inherent periodicity. In the current study, we focus on the relationship between the period of oscillatory reconnection and the strength of the wave pulse initially perturbing the null point, in a hot coronal plasma. We use the PLUTO code to solve the fully compressive, resistive  MHD equations for a 2D magnetic X-point. Using wave pulses with a wide range of amplitudes, we perform a parameter study to obtain values for the period, considering the presence and absence of anisotropic thermal conduction separately. In both cases, we find that the resulting period is independent of the strength of the initial perturbation. The addition of anisotropic thermal conduction only leads to an increase in the mean value for the period, in agreement with our previous study. We also  consider a different type of initial driver and we obtain an oscillation period matching the independent trend previously mentioned. Thus, we report for the first time on the independence between the type and strength of the initializing wave pulse and the resulting period of oscillatory reconnection in a hot coronal plasma. This makes oscillatory reconnection a promising mechanism to be used within the context of coronal seismology.
\end{abstract}

\keywords{Magnetohydrodynamics (1964); Solar magnetic reconnection (1504);
Solar coronal seismology (1994); Solar coronal waves (1995); Magnetohydrodynamical simulations (1966);}


\section{Introduction} \label{sec:introduction}
Magnetic reconnection is a well-known physical process in which the topology of the magnetic field in a magnetized plasma is rearranged. During  reconnection,  strong currents allow neighboring magnetic field lines to diffuse, due to electrical resistivity, leading to a change in connectivity (e.g. \citealt{Parker1957JGR,Sweet1958IAUS,Petschek1964NASSP}). In 2D, magnetic reconnection is generally expected to occur at null points \citep{PriestForbes2000book}, where it can be triggered via variation of the current density, leading to plasma heating \citep{2006A&A...452..343N} and wave generation \citep{2008ApJ...683L..83N, 2009ApJ...705L.217H}. The accumulation and rapid dissipation of electric current, via magnetic reconnection, causes shock heating, mass ejection, and particle acceleration, making reconnection the central engine behind highly energetic phenomena such as solar flares(e.g. \citealt{ShibataMagara2011LRSP,2015ApJ...812..105J}), while making null points preferential locations for flares to occur (e.g. \citealt{2011A&A...533A..18M}). Note that 3D reconnection is different from 2D, and readers are directed to reviews by, e.g. \citet{2003JGRA..108.1285P}, \citet{2009PhPl...16l2101P}, and \citet{2011AdSpR..47.1508P}.

Null points are magnetic field singularities, where the amplitude of the magnetic field, thus the Alfv\'{e}n speed, is zero. Potential field extrapolations from photospheric magnetograms indicate the omnipresence of null points in the solar atmosphere \citep{Galsgaard1997,BrownPriest2001AnA,Longcope2005LRSP,Regnier2008AnA}. Numerous studies have been conducted in the past, revealing the effect of null points on the ubiquitous waves in the solar atmosphere \citep{Gruszecki2011null,McLaughlin2011SSRv,Santamaria2015,Sabri2018MNRAS,Sabri2019AnA}, such as wave refraction and mode conversion \citep{McLaughlin2004,McLaughlin2005,McLaughlin2006a,McLaughlin2006b,Thurgood2012,  2013A&A...558A.127T}. Null points have also been shown to behave as resonant cavities \citep{Santamaria2018}, generating waves at frequencies depending on the plasma conditions, such as high-frequency wave trains ($\sim 80$\,mHz) \citep{Santamaria2016,Santamaria2017}. In \citet{Sabri2020ApJ}, the interaction of fast magnetoacoustic waves with a $2.5$D null point produced plasmoids due to the tearing mode instability, while a plethora of numerical studies have shown null points can act as sources of coronal jets, slow, fast and Alfv\'en waves (e.g. \citealt{2014SoPh..289.3043L, 2017ApJ...834...62K, Thurgood2017ApJ, 2018ApJ...862....6C}) when subjected to various wave-based driving motions. All the above results reveal, among other things, the importance of these wave$-$null point interactions for coronal seismology \citep{Uchida1970, RobertsEdwinBenz1984}.

Coming back to reconnection at null points, while considering the relaxation of a perturbed 2D X-point, \citet{CraigMcClymont1991ApJ} identified the process of oscillatory reconnection. The term describes a series of reconnection events associated with periodic changes in the magnetic connectivity and topology of the field, the periodicity of which is derived from the relaxation of a perturbed field from a finite, aperiodic driver. As a mechanism, oscillatory reconnection has been proposed as a possible driving force behind quasi-periodic pulsations (QPPs) of solar flares (e.g. \citealt{Kupriyanova2016SoPh,VanDoorsselaere2016SoPh,Pugh2017AnA,2019ApJ...886L..25Y,2020ApJ...895...50H, 2020A&A...639L...5L,2020ApJ...893....7L,2021ApJ...921..179L, 2021ApJ...910..123C}) and stellar flares (e.g. \citealt{2019A&A...629A.147B,2019A&A...622A.210G,2019MNRAS.482.5553J,2019ApJ...876...58N,2019ApJ...884..160V,2020A&A...636A..96M,2021SoPh..296..162R}). Detailed reviews of a wide variety of suggested mechanism(s) behind QPPs can be found in \citet{McLaughlin2018SSRv}, \citet{Kupriyanova2020STP}, and \citet{Zimovets2021SSRv}. 

Alongside QPPs, oscillatory reconnection can also be associated with quasi-periodic flows associated with spicules (e.g. \citealt{DePontieu2010ApJ,DePontieu2011Sci,2019Sci...366..890S, 2020ApJ...891L..21Y}), as well as with observed periodicities in jets (\citealt{2019ApJ...874..146H}) and in the formation, disappearance, and eruption of magnetic flux ropes (\citealt{2018ApJ...853....1S,2019ApJ...874L..27X}).  \citet{McLaughlin2012ApJ} were able to reproduce such observed periodicities through oscillatory reconnection in a 2D flux emergence model. Recent observations by the Parker Solar Probe can also be potentially explained through this mechanism (e.g. \citealt{2016SSRv..204...49B, 2019Natur.576..237B,2019Natur.576..228K}), including Alfv\'enic spikes/kinks  \citep{2021ApJ...913L..14H} and periodicities correlated with Type III radio bursts \citep{2021A&A...650A...6C}.

Numerically, the mechanism of magnetic oscillatory reconnection was first studied in a 2D X-point for a cold plasma by \citet{McLaughlin2009}, where the fully compressible resistive MHD equations were solved, allowing the mechanism itself to generate localized heating. The results of this study were expanded for a 3D null point in \citet{Thurgood2017ApJ}, where the generation of MHD waves was reported, as was mentioned earlier. Additional studies have focused on the effects of resistivity, initial perturbation amplitude, and the length of the initial current sheet on the frequency of the derived oscillating signal for the current density at the null \citep{McLaughlin2012A&A, 2018ApJ...855...50T, 2018PhPl...25g2105T, Thurgood2019A&A}. Other studies have considered the mechanism within the context of a realistic stratified solar atmosphere (\citealt{2009A&A...494..329M,McLaughlin2012ApJ}), while \citet{Stewart2022} reported on the merging of two current-carrying cylindrical flux ropes into a single flux rope through a process of oscillatory reconnection and on the generation of waves originating from the reconnection site. In \citet{Karampelas2022a}, oscillatory reconnection of a 2D magnetic X-point was considered for the first time in a hot, coronal plasma, in the absence of nonlinear tearing, while also introducing the effects of anisotropic thermal conduction. This study also opened the door for utilizing oscillatory reconnection as a seismological tool in the solar atmosphere, by deriving a relationship between the equilibrium background magnetic field and the resulting period of oscillatory reconnection.

In this paper, we will take the next step in investigating the potential for oscillatory reconnection to be utilized for coronal seismology. We investigate the dependence of the reconnection period upon the amplitude of the external wave pulse that perturbs the initial 2D magnetic X-point. We describe our physical domain, numerical setup, and code used in Section \ref{sec:setup}, and present the results of our parameter study (\S\ref{sec:results}) for simulations in the absence (\S\ref{sec:NoTC}) and presence (\S\ref{sec:TC}) of anisotropic thermal conduction. An overview of these results with some additional analysis can be found in \S\ref{sec:periods}, while a different type of driver will be used in \S\ref{sec:pinch} and have its results compared with the drivers used in \S\ref{sec:periods}. Finally, our conclusions and general discussion take place in \S\ref{sec:discussions}.


\section{Numerical setup} \label{sec:setup}

\begin{figure}[t]
    \centering
    \includegraphics[trim={0.cm 0.cm 0.cm 0.cm},clip,scale=0.45]{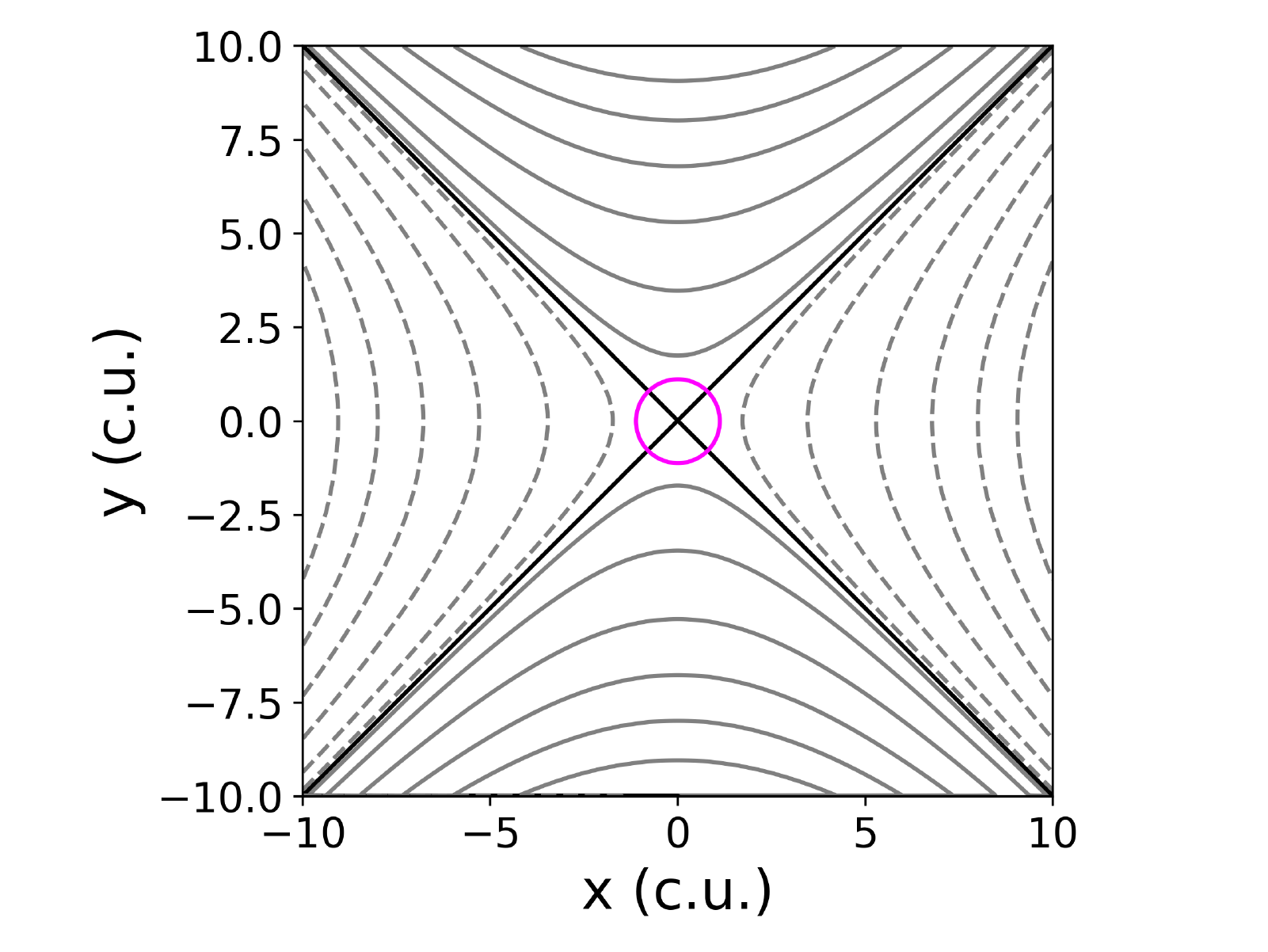}
    \caption{The initial magnetic field configuration for the basic X-point setup. The separatrices (solid black lines) indicate the regions with opposing polarities, as shown with the gray solid and dashed contours. The magenta circle corresponds to the equipartition layer, where the Alfv\'{e}n speed $V_A$ equals the sound speed $V_S$ for our setup with a background temperature of $1$\,MK.}
    \label{fig:profileB}
\end{figure}

\begin{figure*}[t]
    \centering
    \resizebox{\hsize}{!}{
    \includegraphics[trim={0.cm  1.08cm 0.cm 0.cm},clip,scale=0.45]{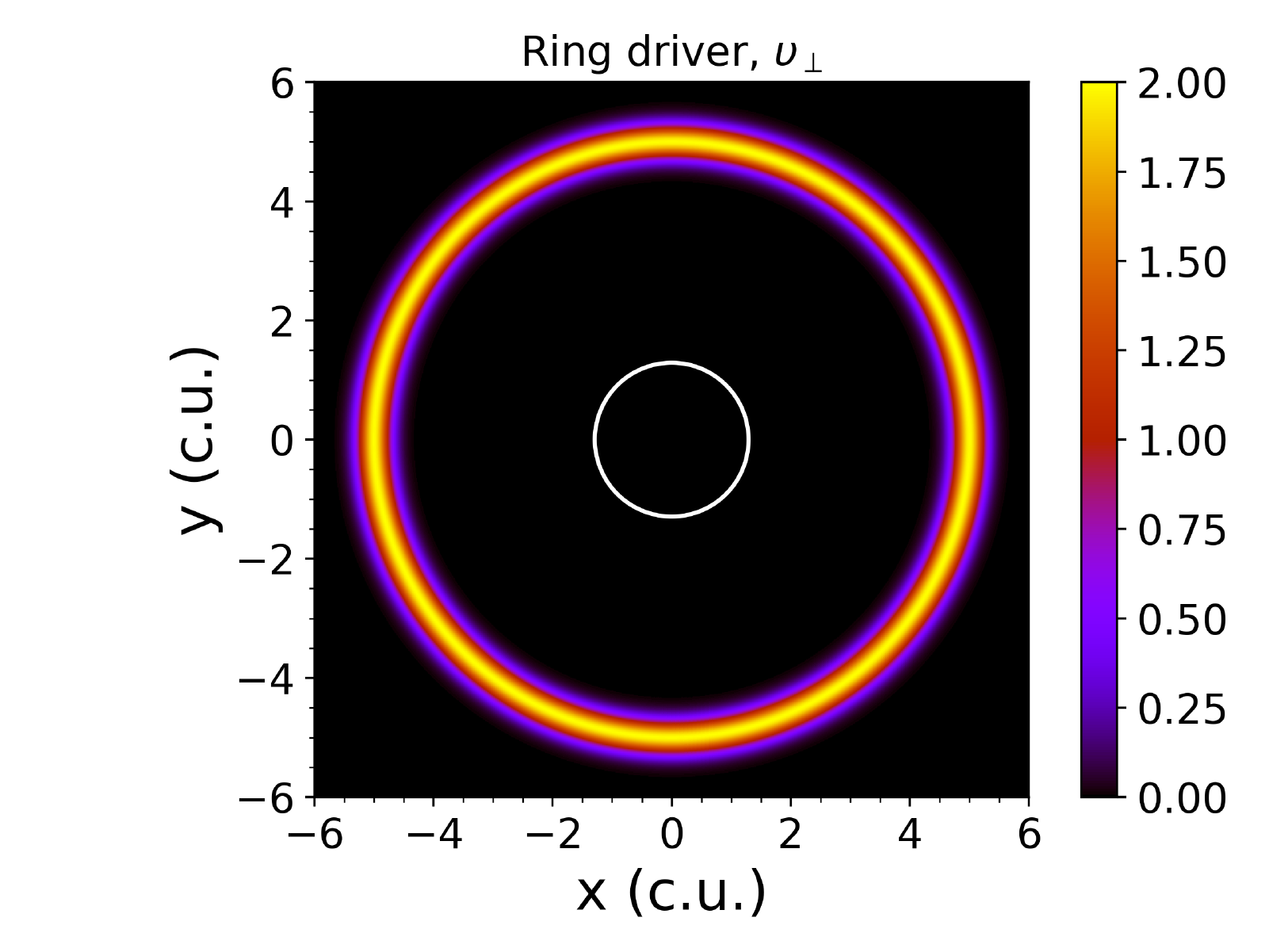}
    \includegraphics[trim={3.cm  1.08cm 0.cm 0.cm},clip,scale=0.45]{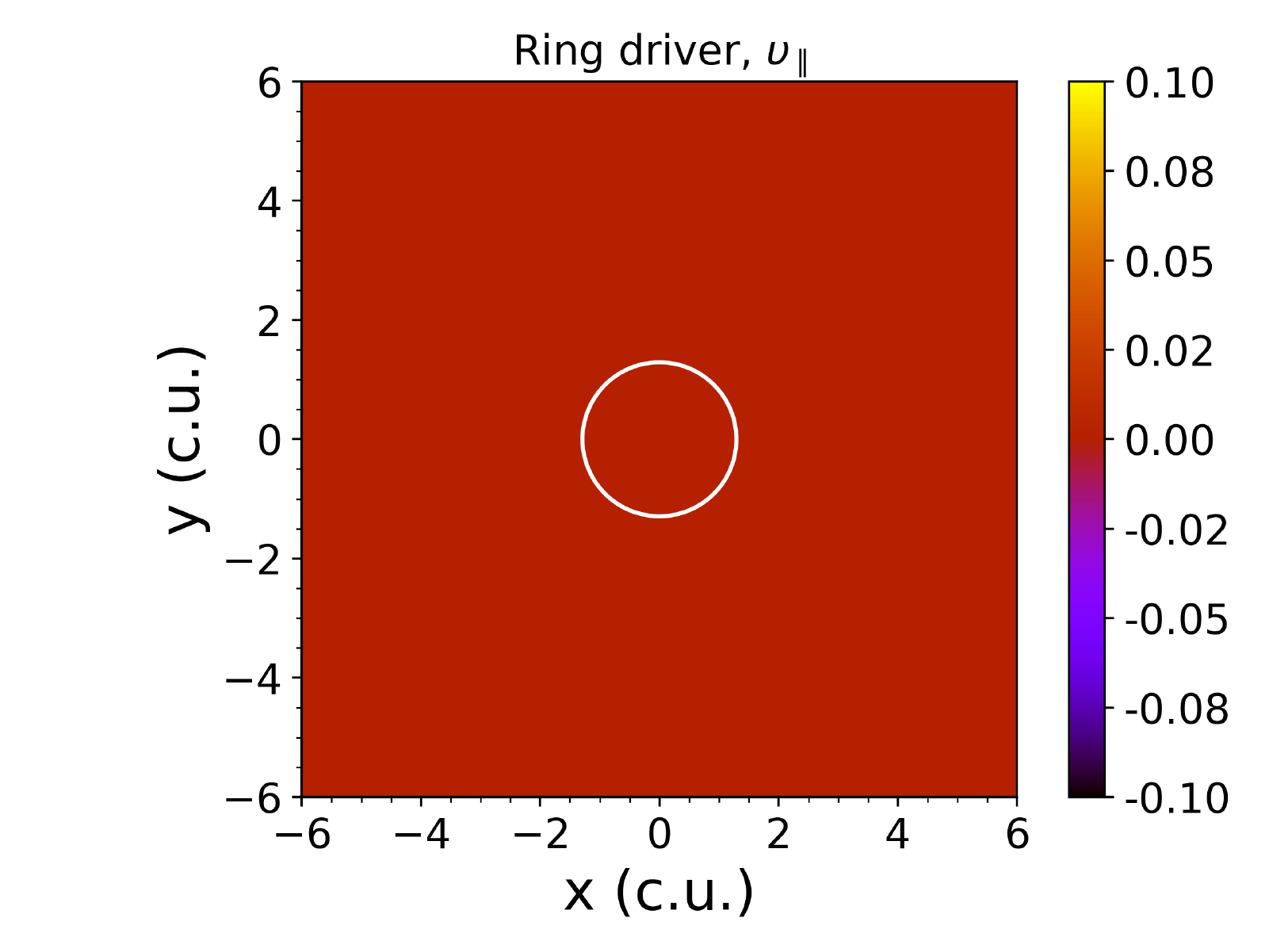}
    \includegraphics[trim={3.cm  1.08cm 0.cm 0.cm},clip,scale=0.45]{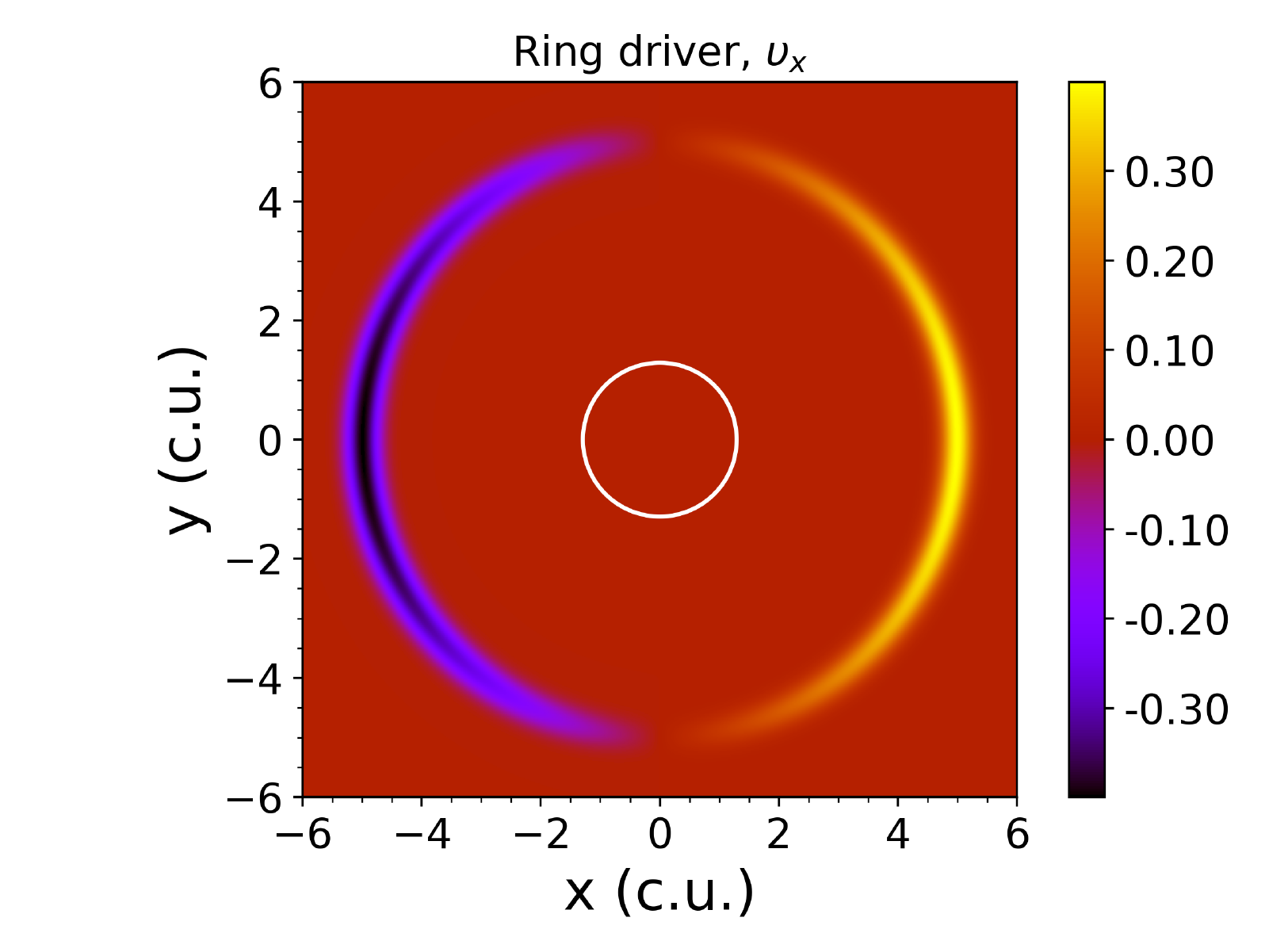}
    \includegraphics[trim={3.cm  1.08cm 0.cm 0.cm},clip,scale=0.45]{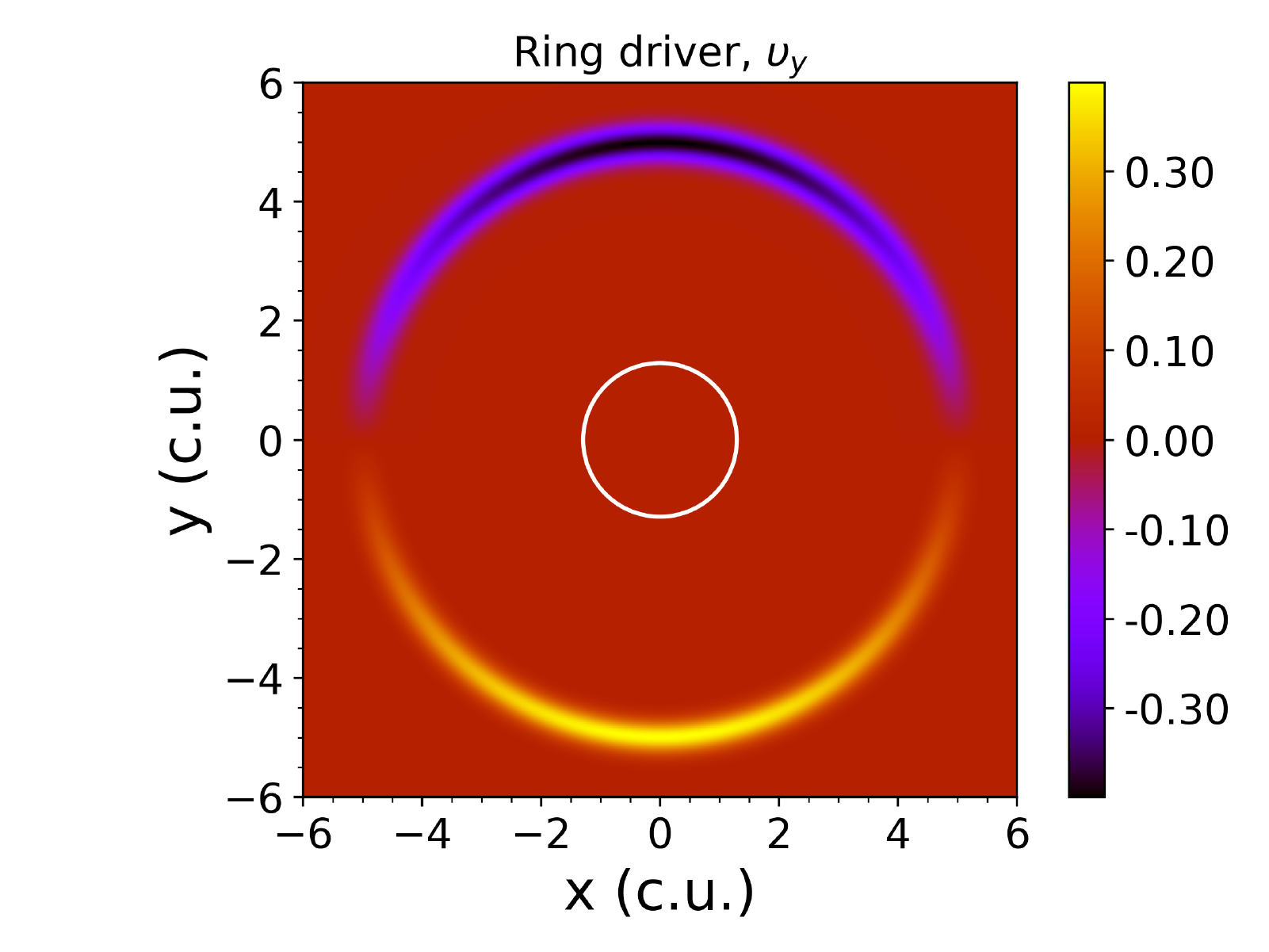}}
    \resizebox{\hsize}{!}{
    \includegraphics[trim={0.cm  0.cm 0.cm 0.cm},clip,scale=0.45]{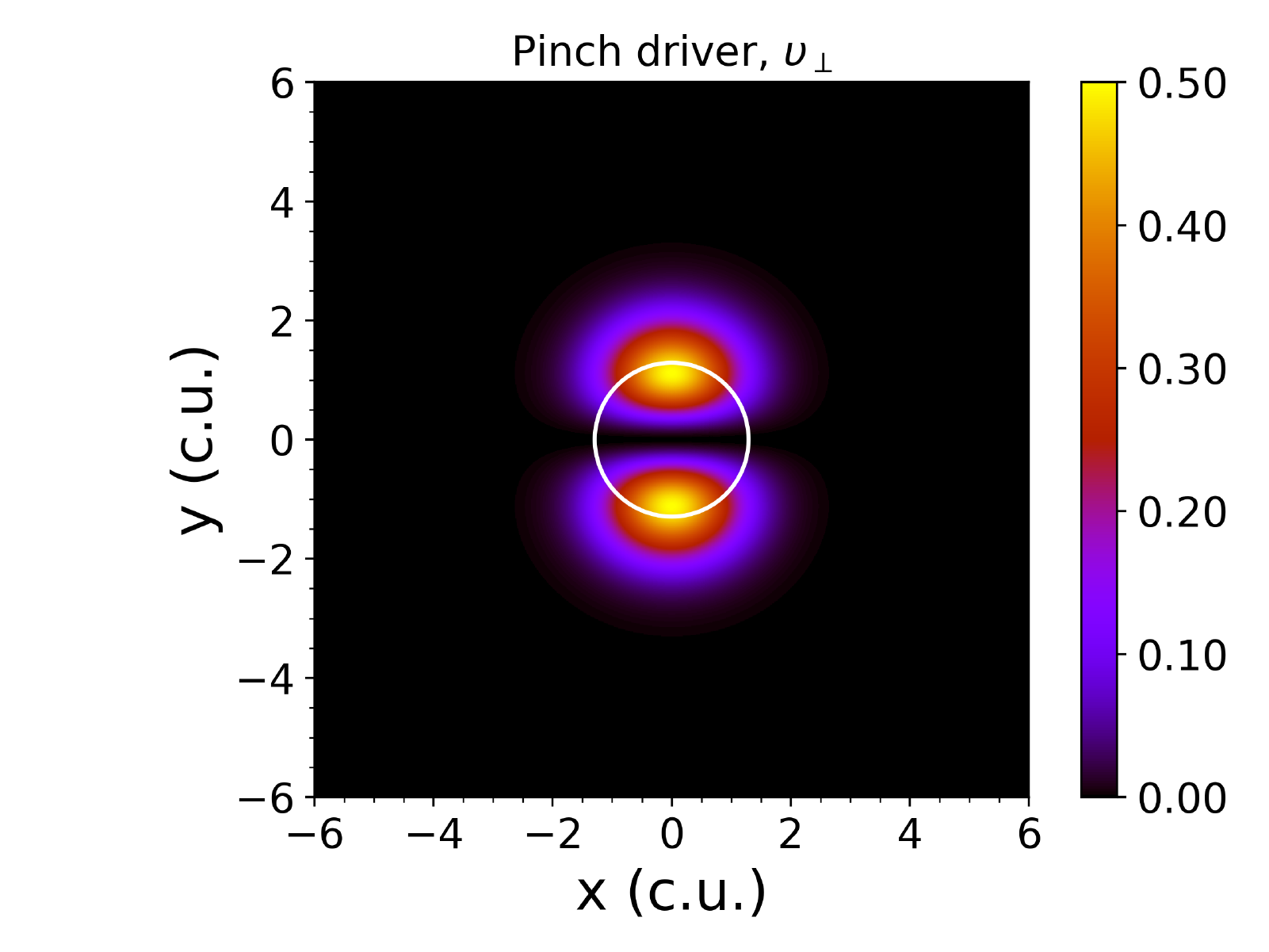}
    \includegraphics[trim={3.cm  0.cm 0.cm 0.cm},clip,scale=0.45]{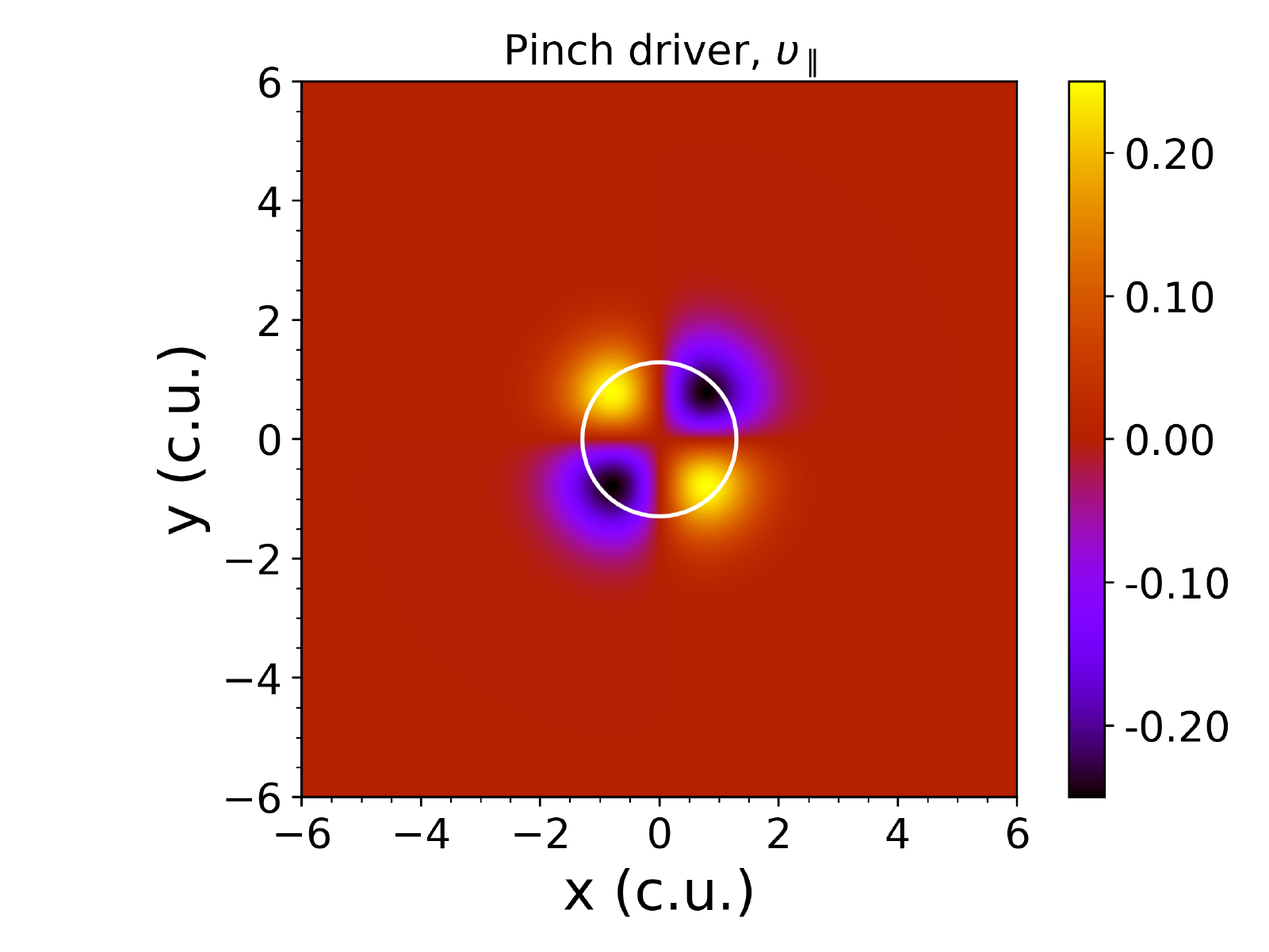}
    \includegraphics[trim={3.cm  0.cm 0.cm 0.cm},clip,scale=0.45]{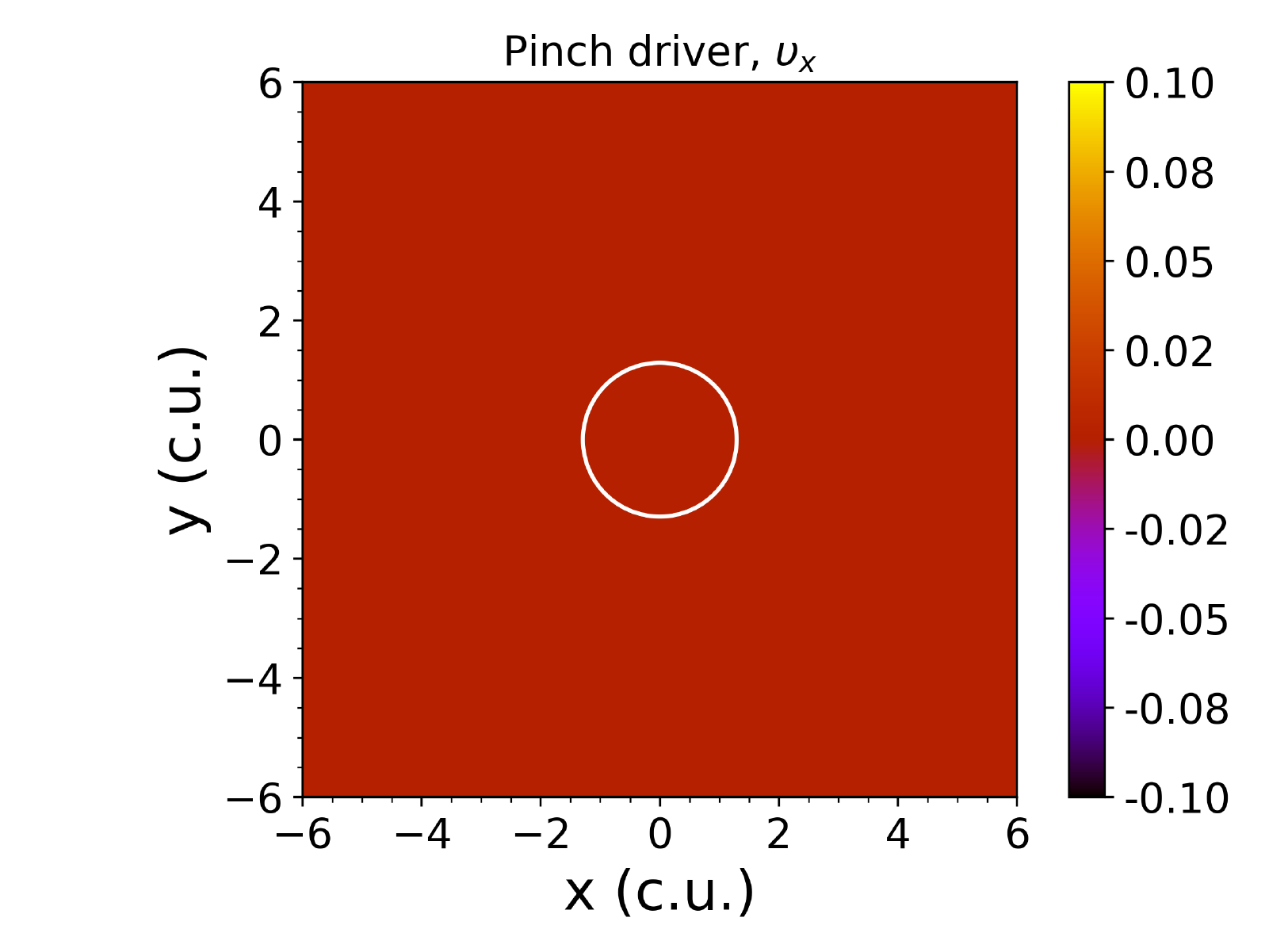}
    \includegraphics[trim={3.cm  0.cm 0.cm 0.cm},clip,scale=0.45]{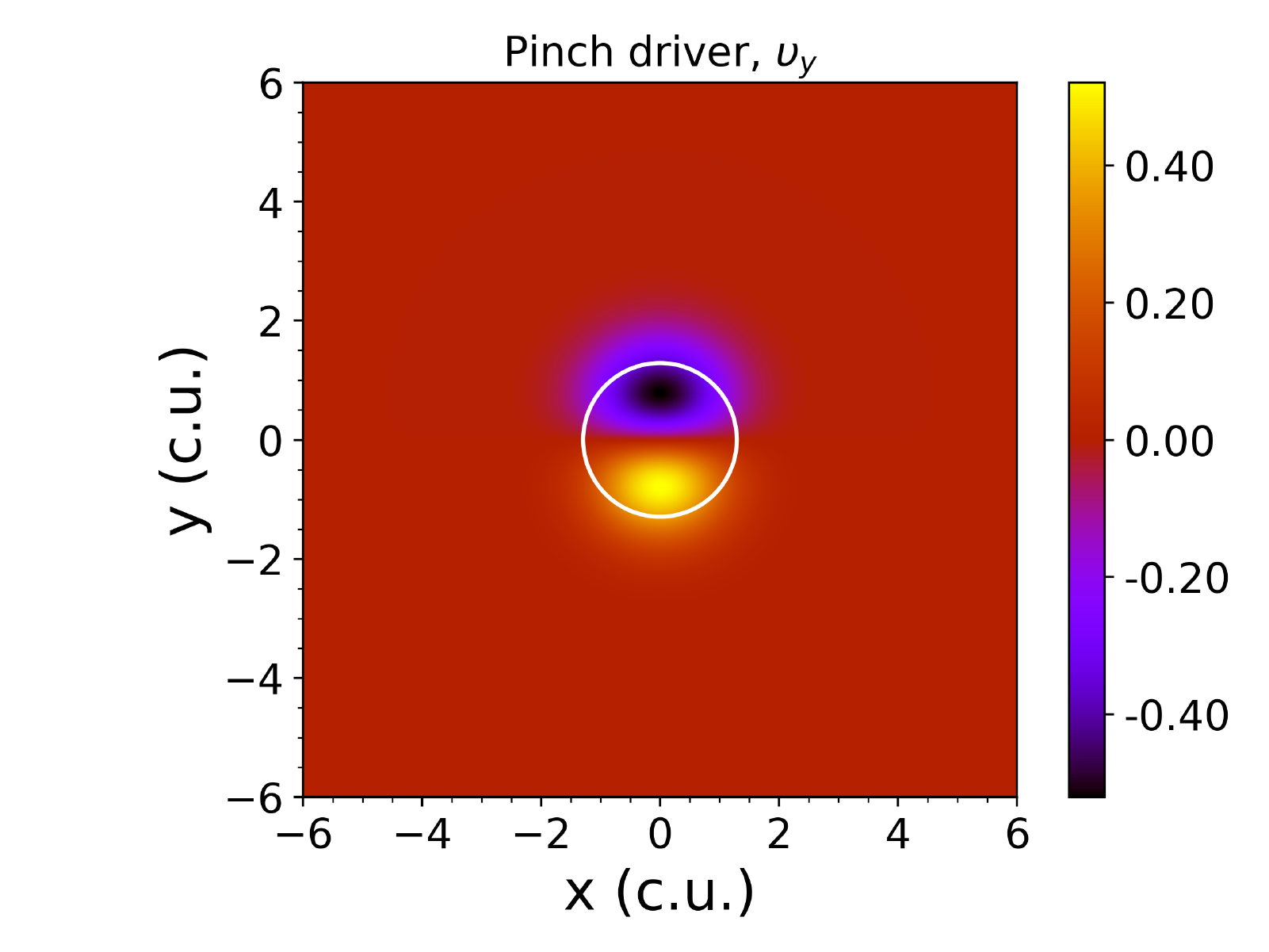}}
    \caption{2D profiles of the ring-style initial pulse for $C=1$ (top panel) and the pinch-style initial pulse (bottom panel). From left to right we show the $v_{\perp}$ and $v_{\parallel}$ pulse components, and the corresponding initial $v_x$ and $v_y$ velocities.}
    \label{fig:profiledriver}
\end{figure*}

\begin{figure}[t]
    \centering
    \includegraphics[trim={0.cm 0.cm 0.cm 0.cm},clip,scale=0.45]{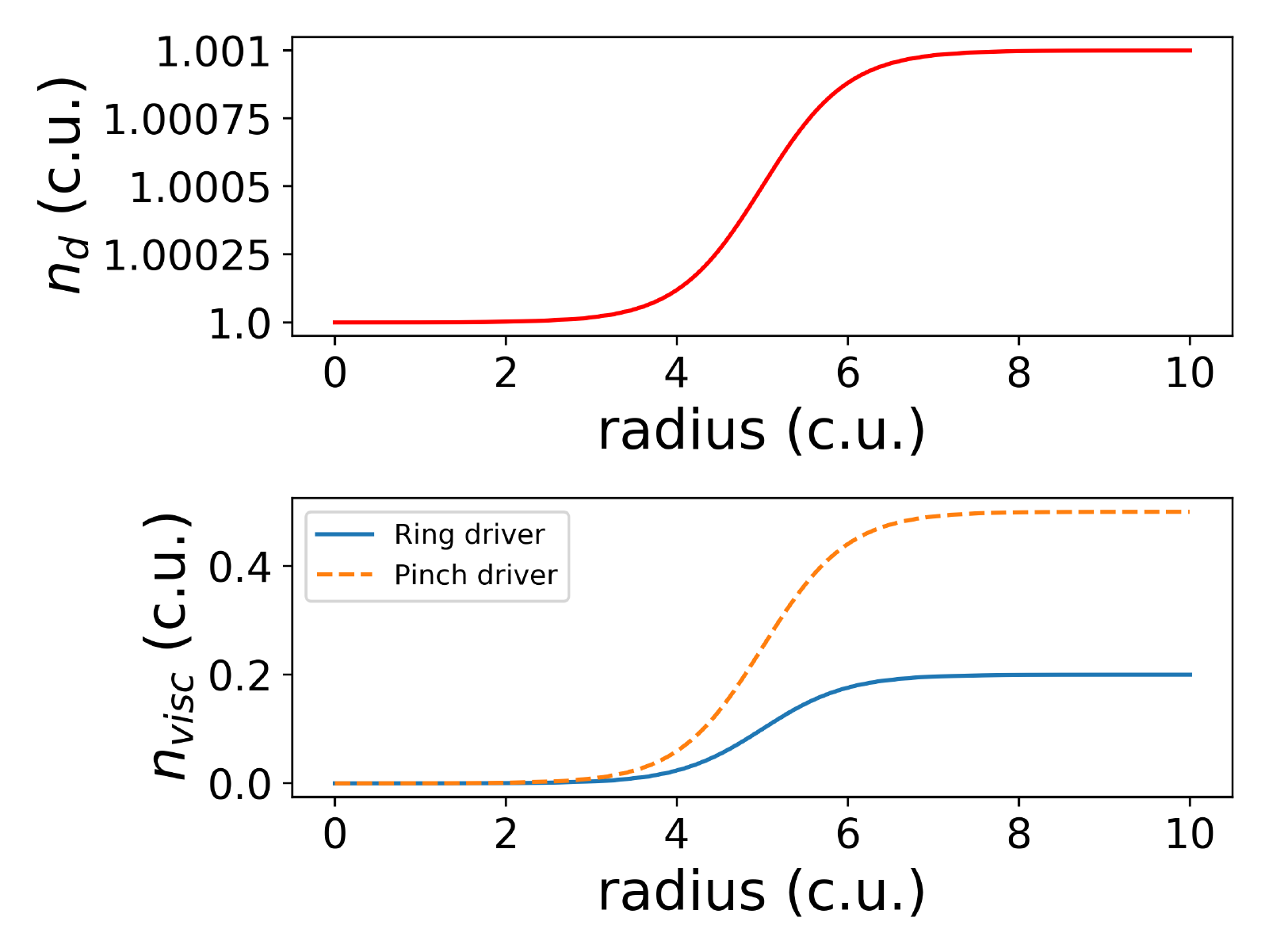}
    \caption{Profiles of the viscous ($n_{visc}$, top panel) and numerical ($n_d$, bottom panel) dissipation coefficients as a function of the radius.}
    \label{fig:profilecoeff}
\end{figure}

\begin{figure*}[t]
    \centering
    \resizebox{\hsize}{!}{
    \includegraphics[trim={2.cm 1.07cm 0.cm 0.cm},clip,scale=0.45]{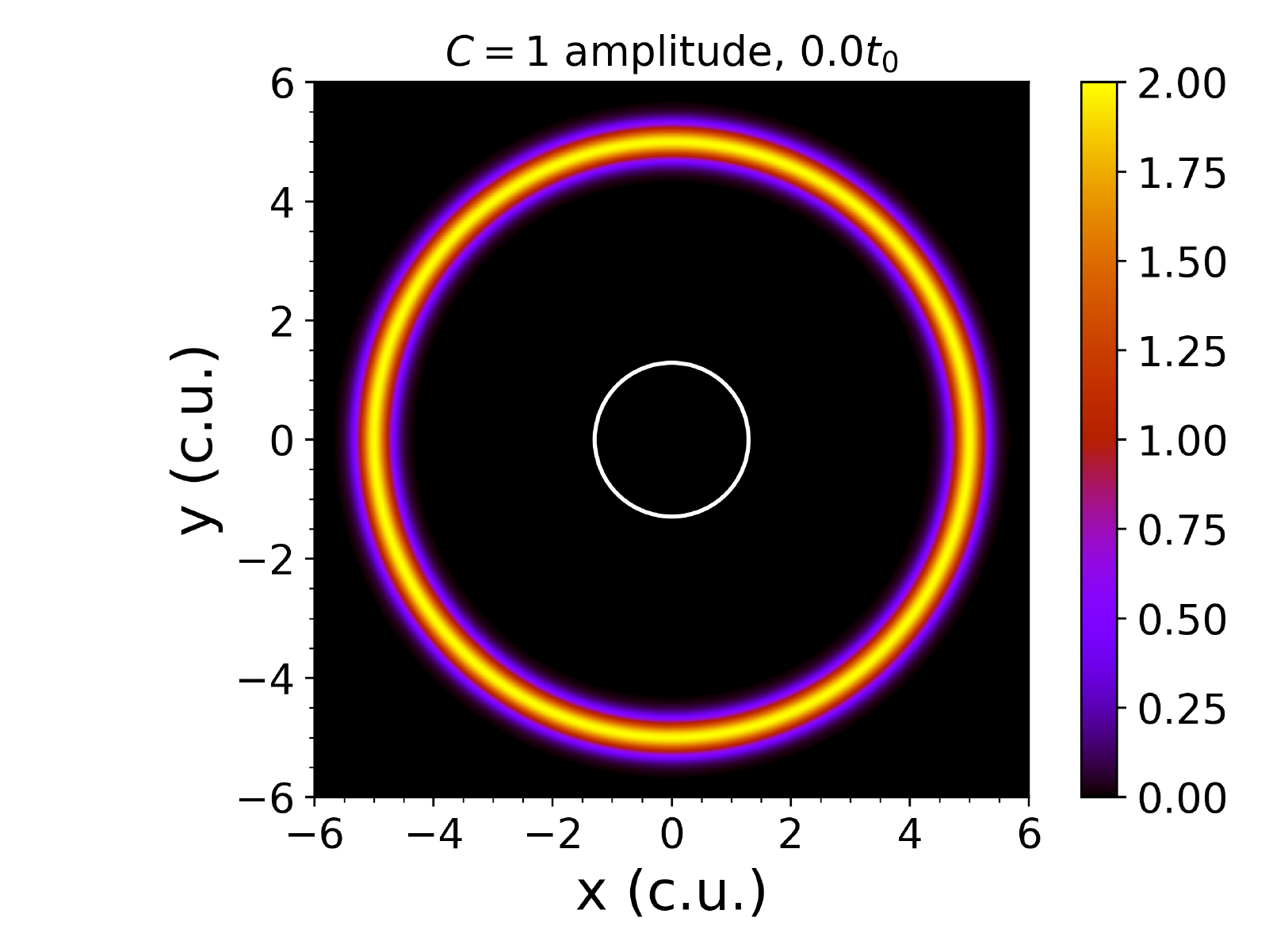}
    \includegraphics[trim={3.cm 1.07cm 0.cm 0.cm},clip,scale=0.45]{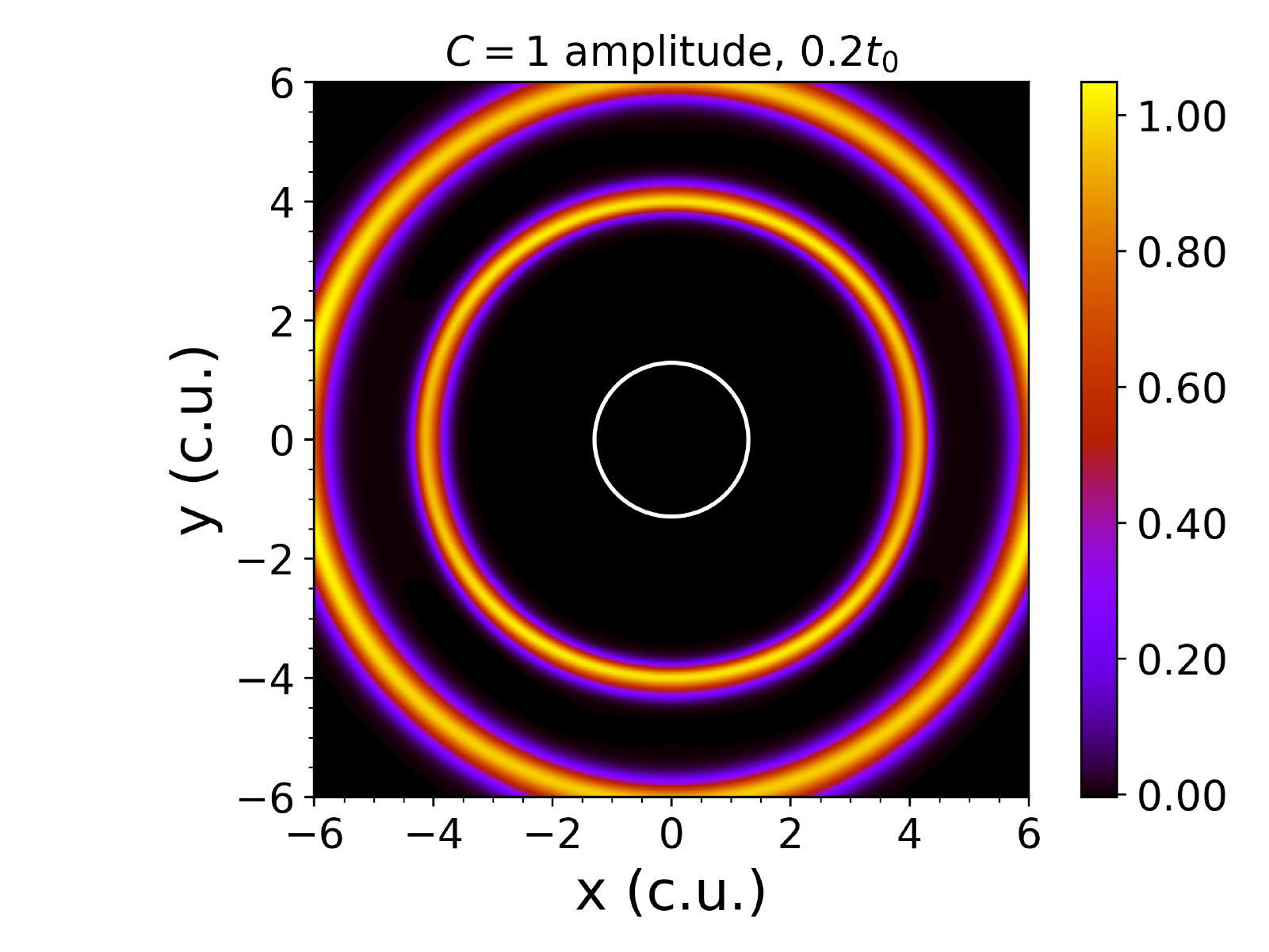}
    \includegraphics[trim={3.cm 1.07cm 0.cm 0.cm},clip,scale=0.45]{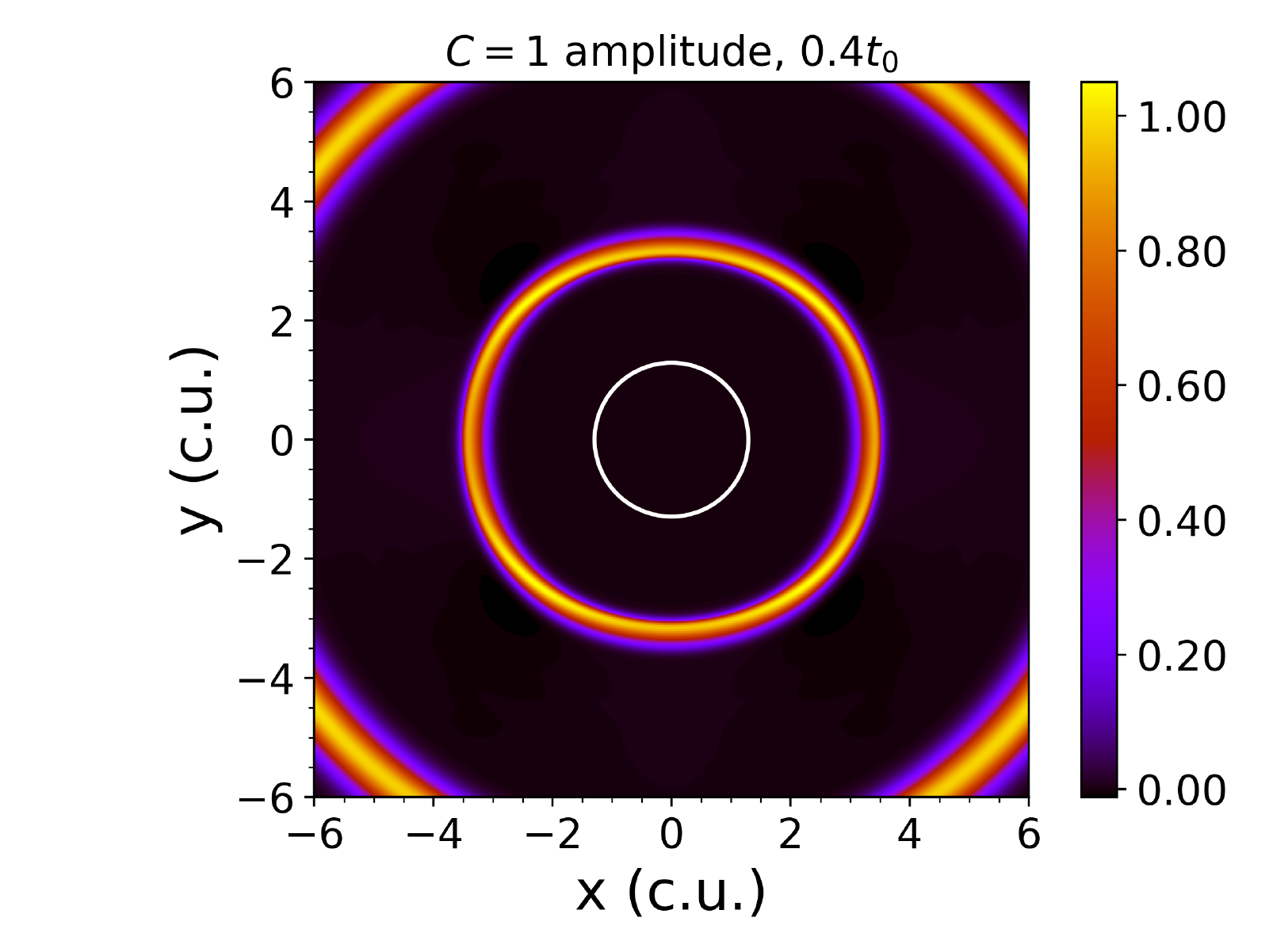}
    \includegraphics[trim={3.cm 1.07cm 0.cm 0.cm},clip,scale=0.45]{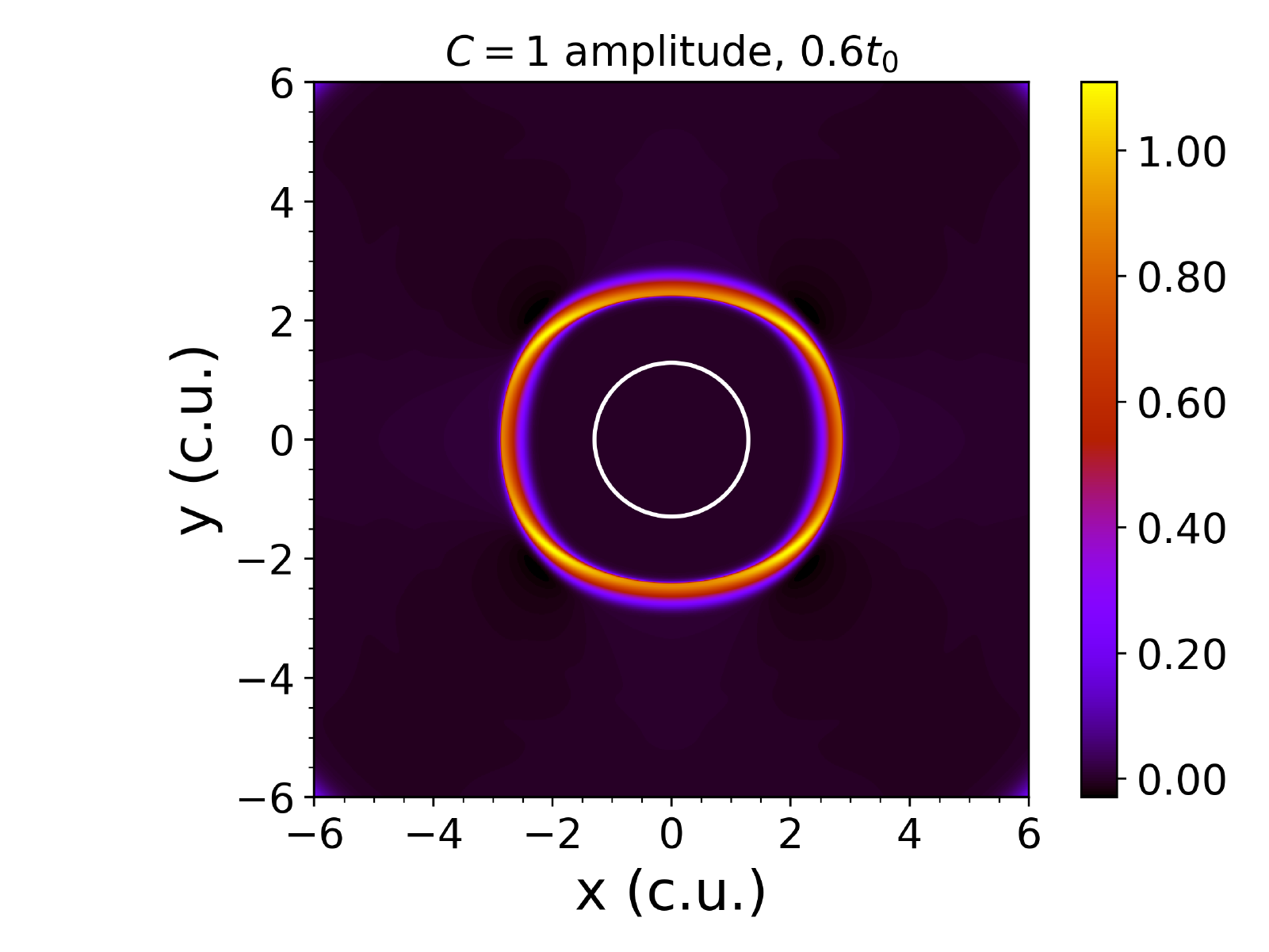}
    }
    \resizebox{\hsize}{!}{
    \includegraphics[trim={2.cm 1.07cm 0.cm 0.cm},clip,scale=0.45]{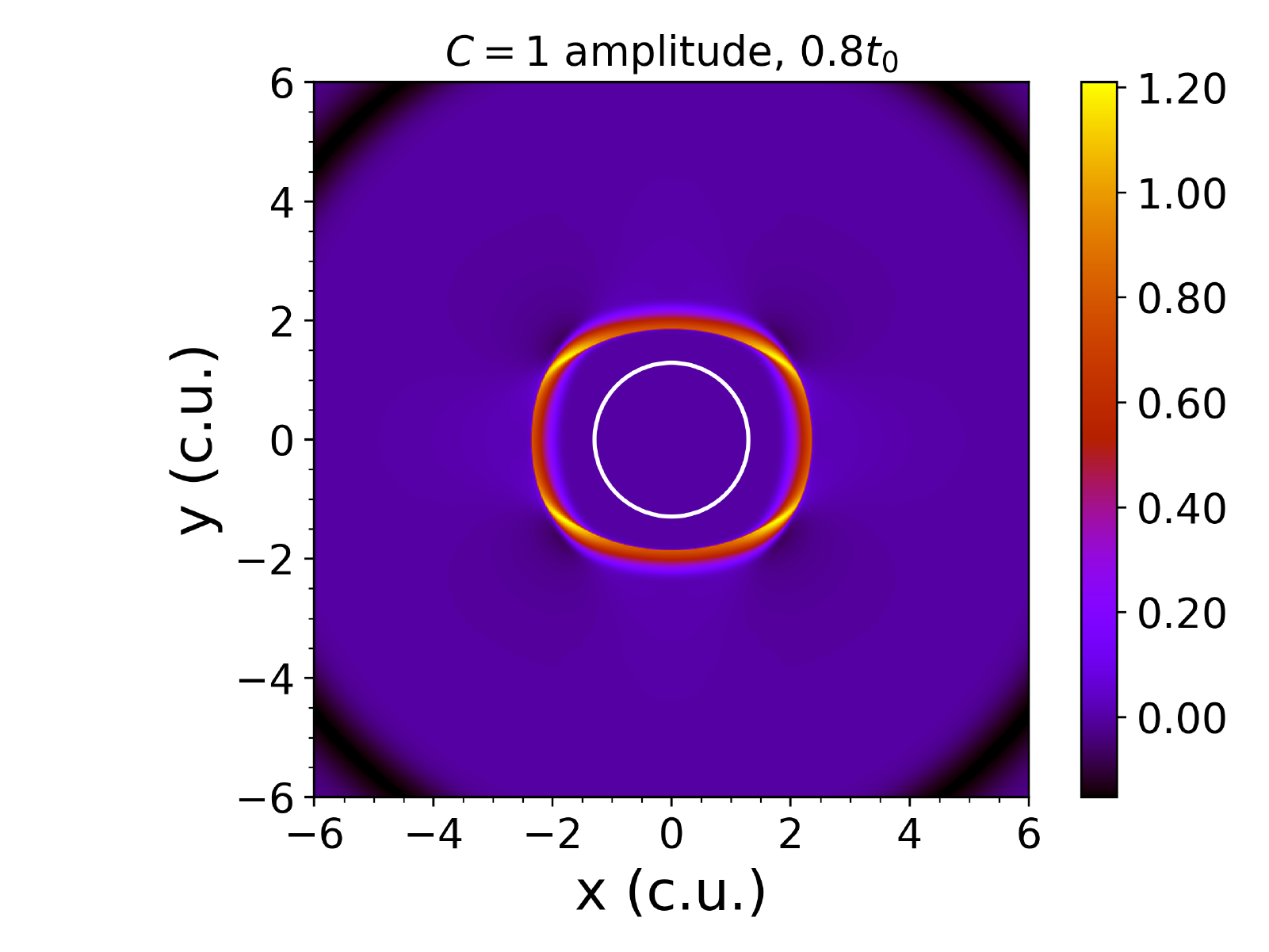}
    \includegraphics[trim={3.cm 1.07cm 0.cm 0.cm},clip,scale=0.45]{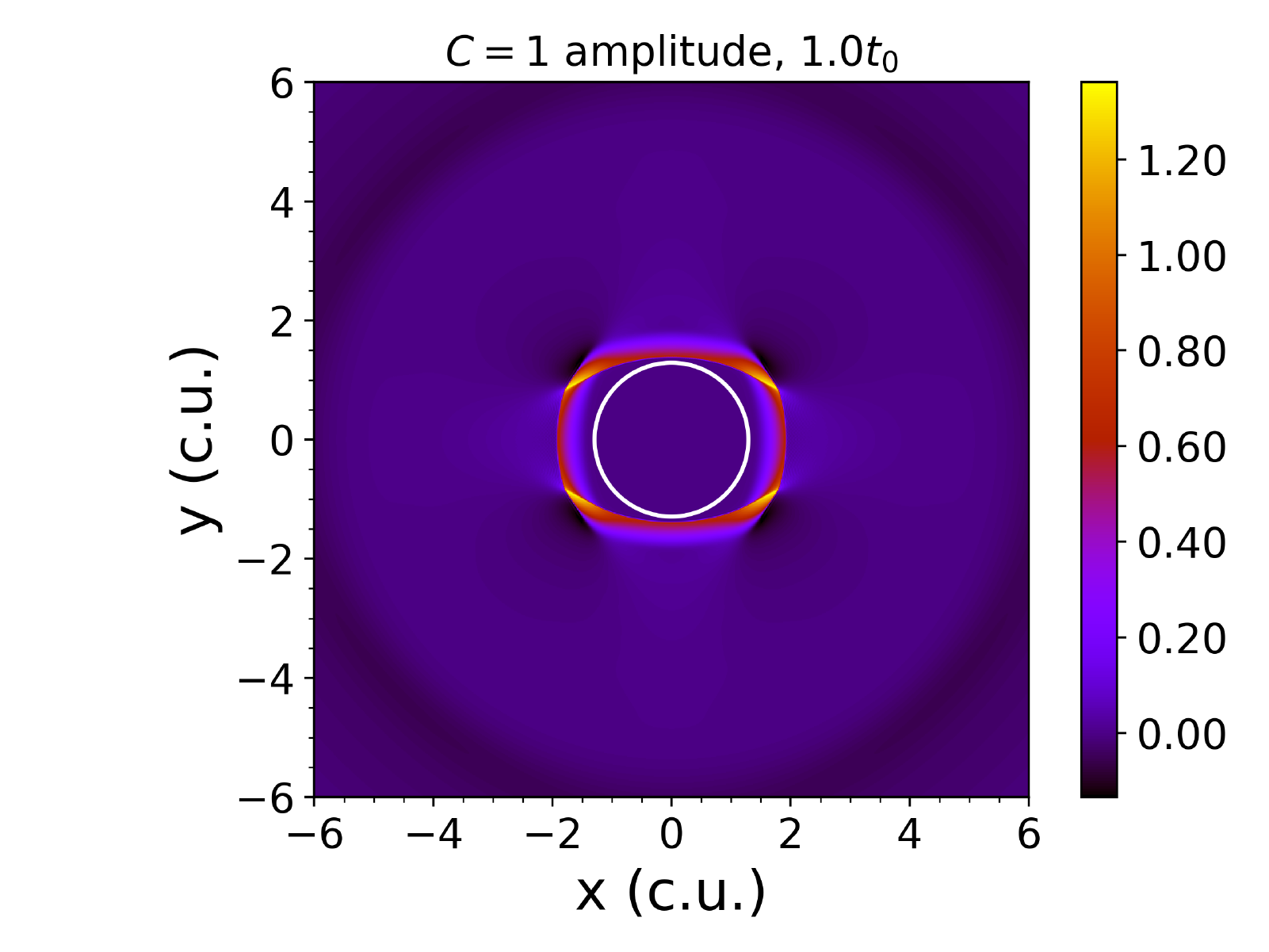}
    \includegraphics[trim={3.cm 1.07cm 0.cm 0.cm},clip,scale=0.45]{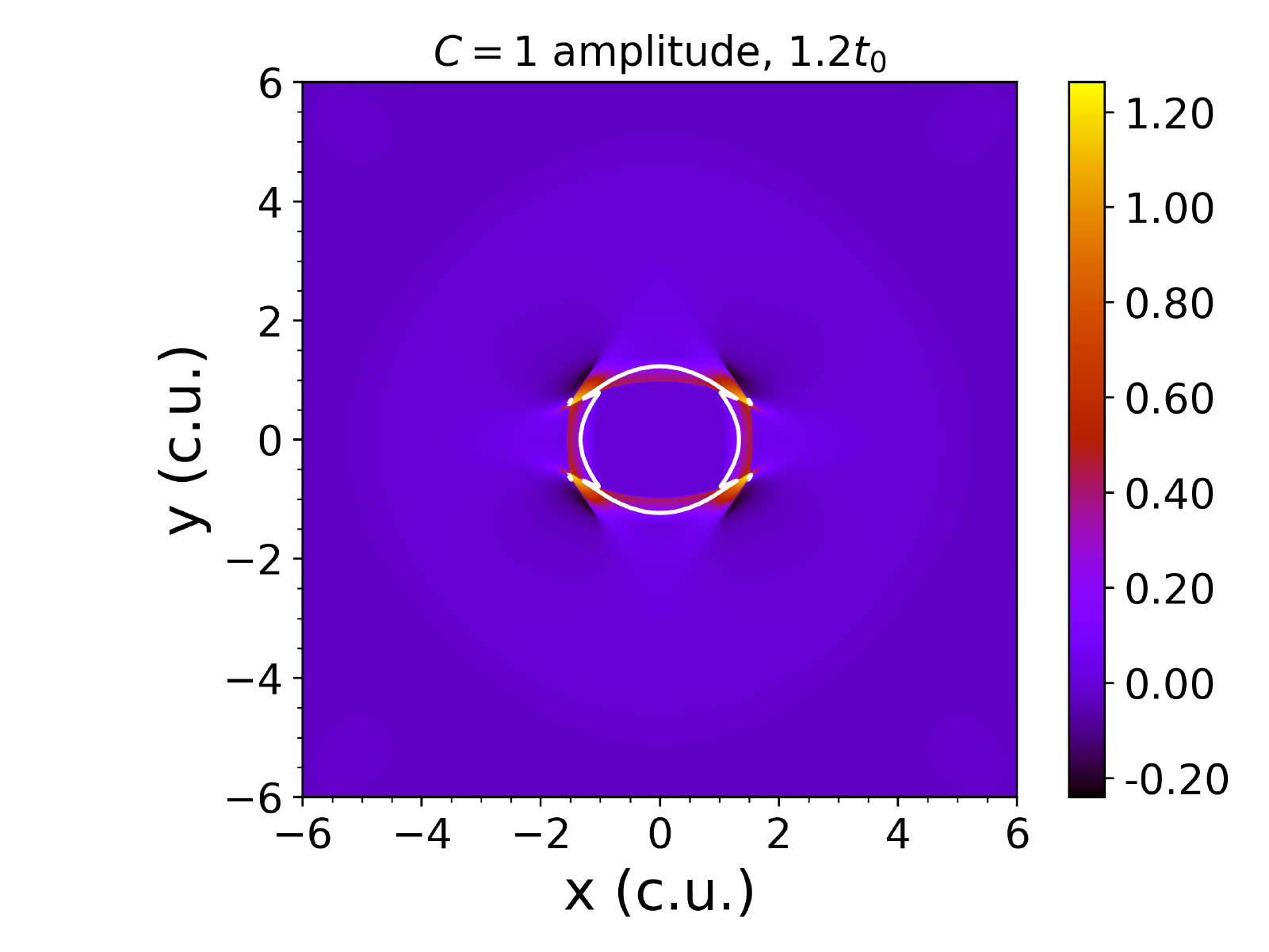}
    \includegraphics[trim={3.cm 1.07cm 0.cm 0.cm},clip,scale=0.45]{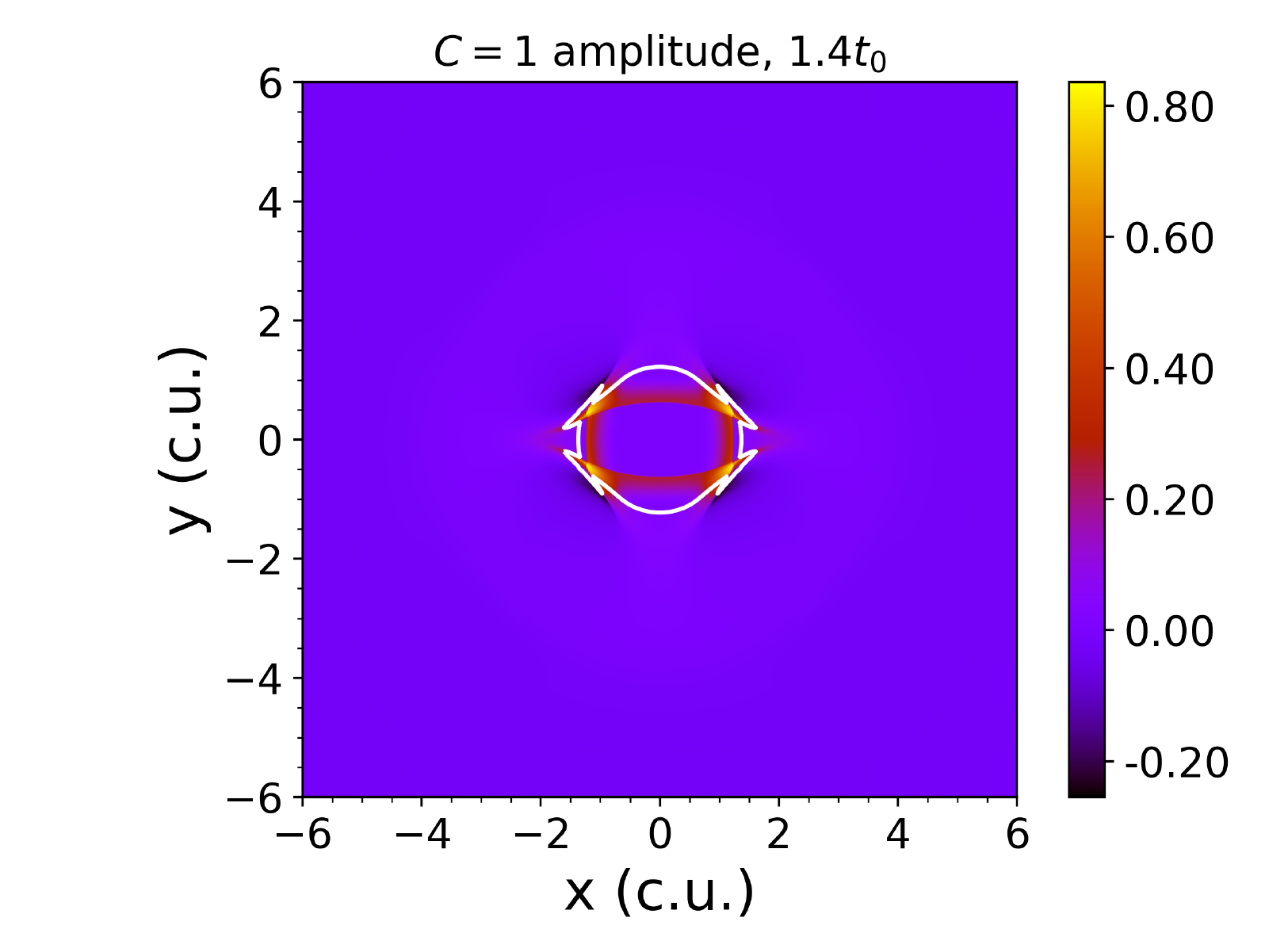}
    }
    \resizebox{\hsize}{!}{
    \includegraphics[trim={2.cm 0.cm 0.cm 0.cm},clip,scale=0.45]{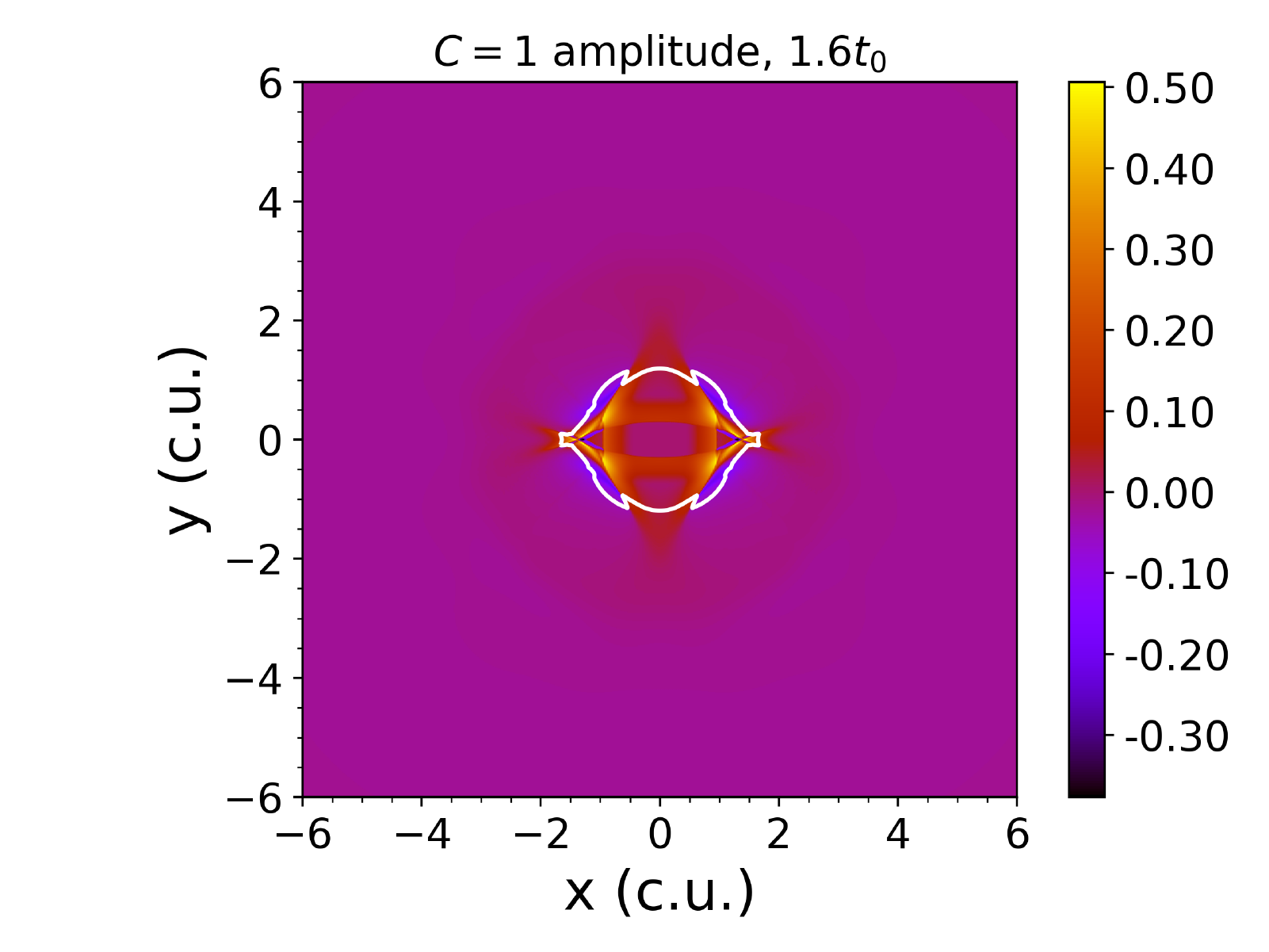}
    \includegraphics[trim={2.8cm 0.cm 0.cm 0.cm},clip,scale=0.45]{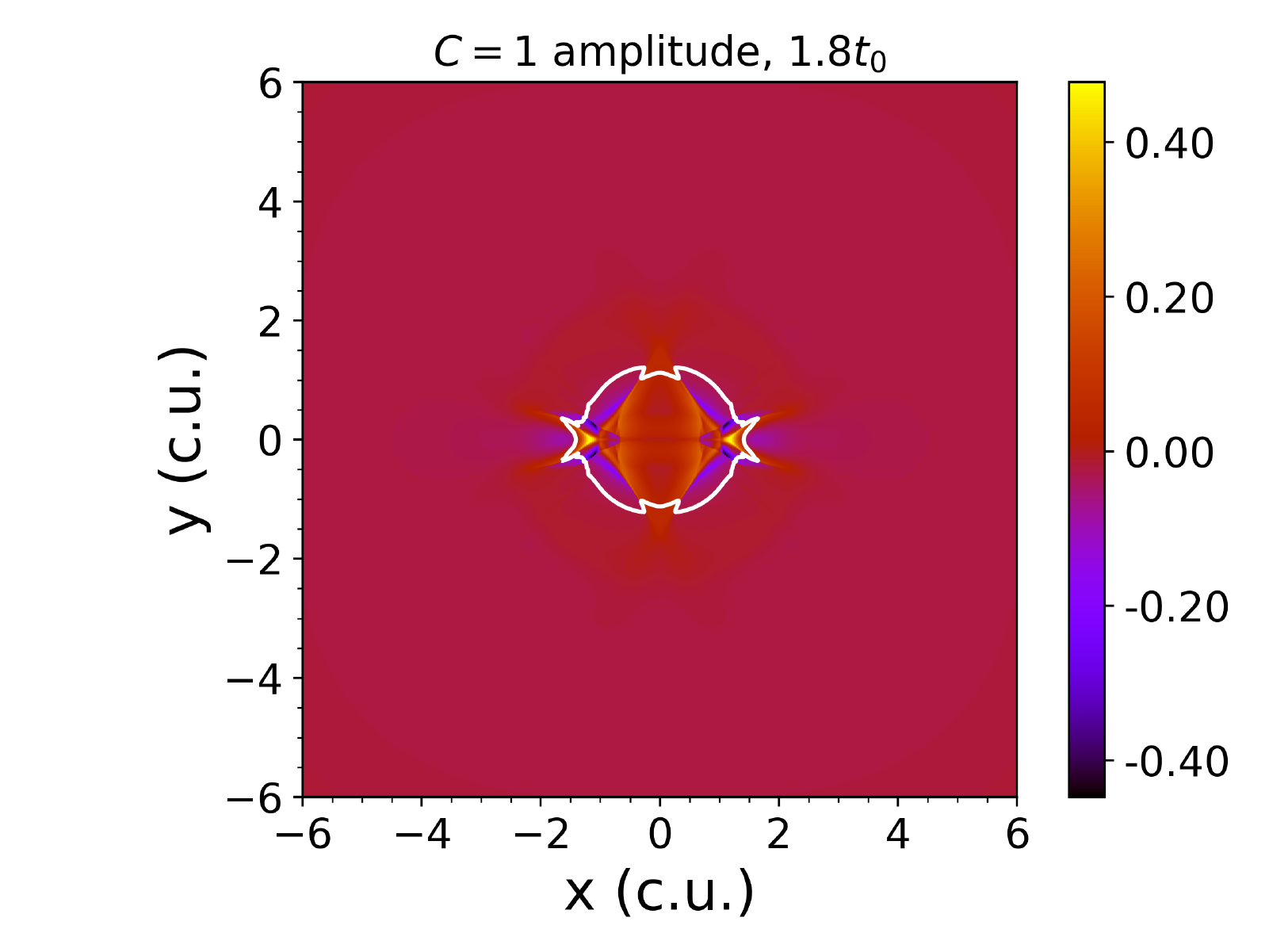}
    \includegraphics[trim={3.cm 0.cm 0.cm 0.cm},clip,scale=0.45]{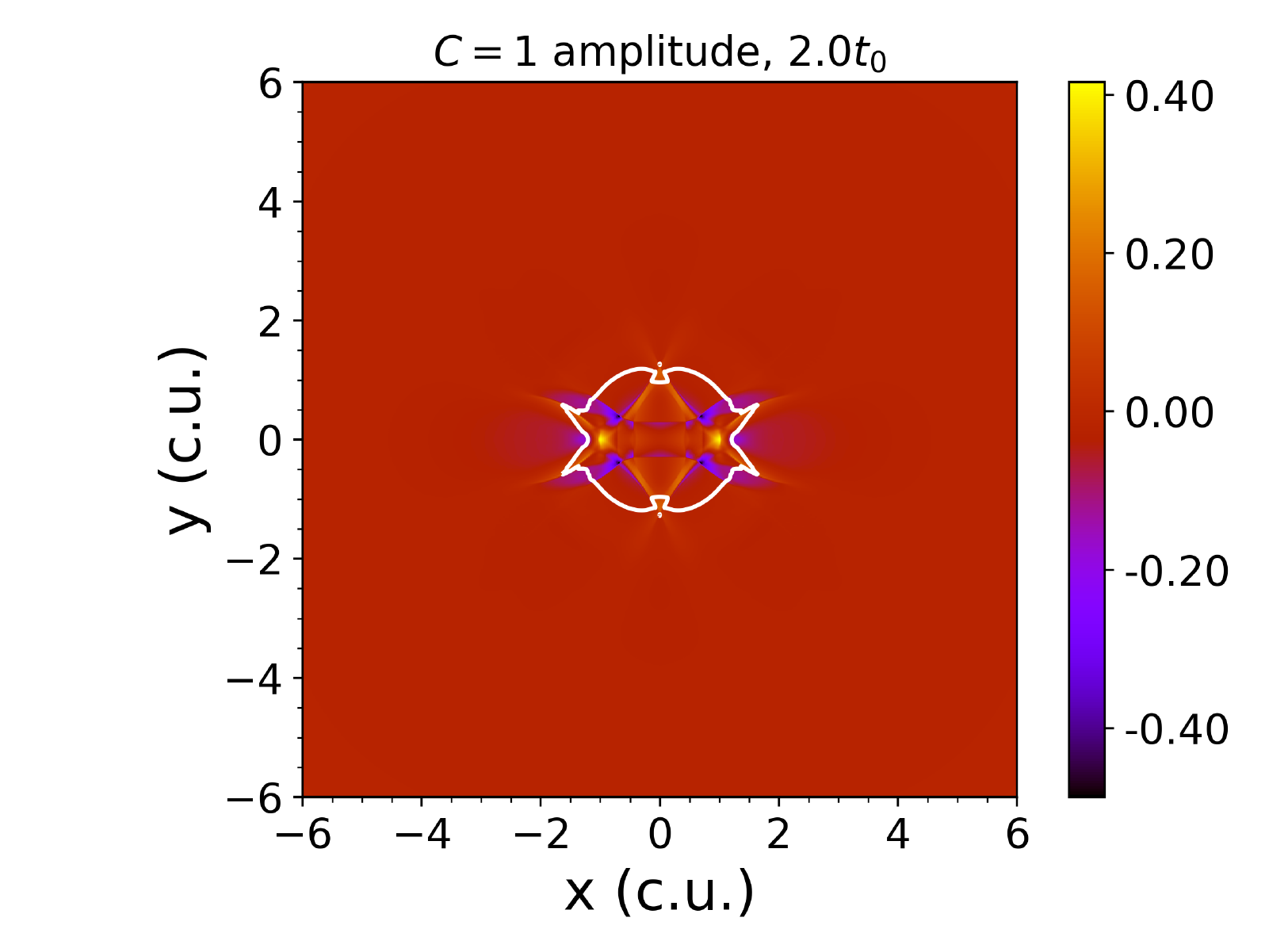}
    \includegraphics[trim={3.cm 0.cm 0.cm 0.cm},clip,scale=0.45]{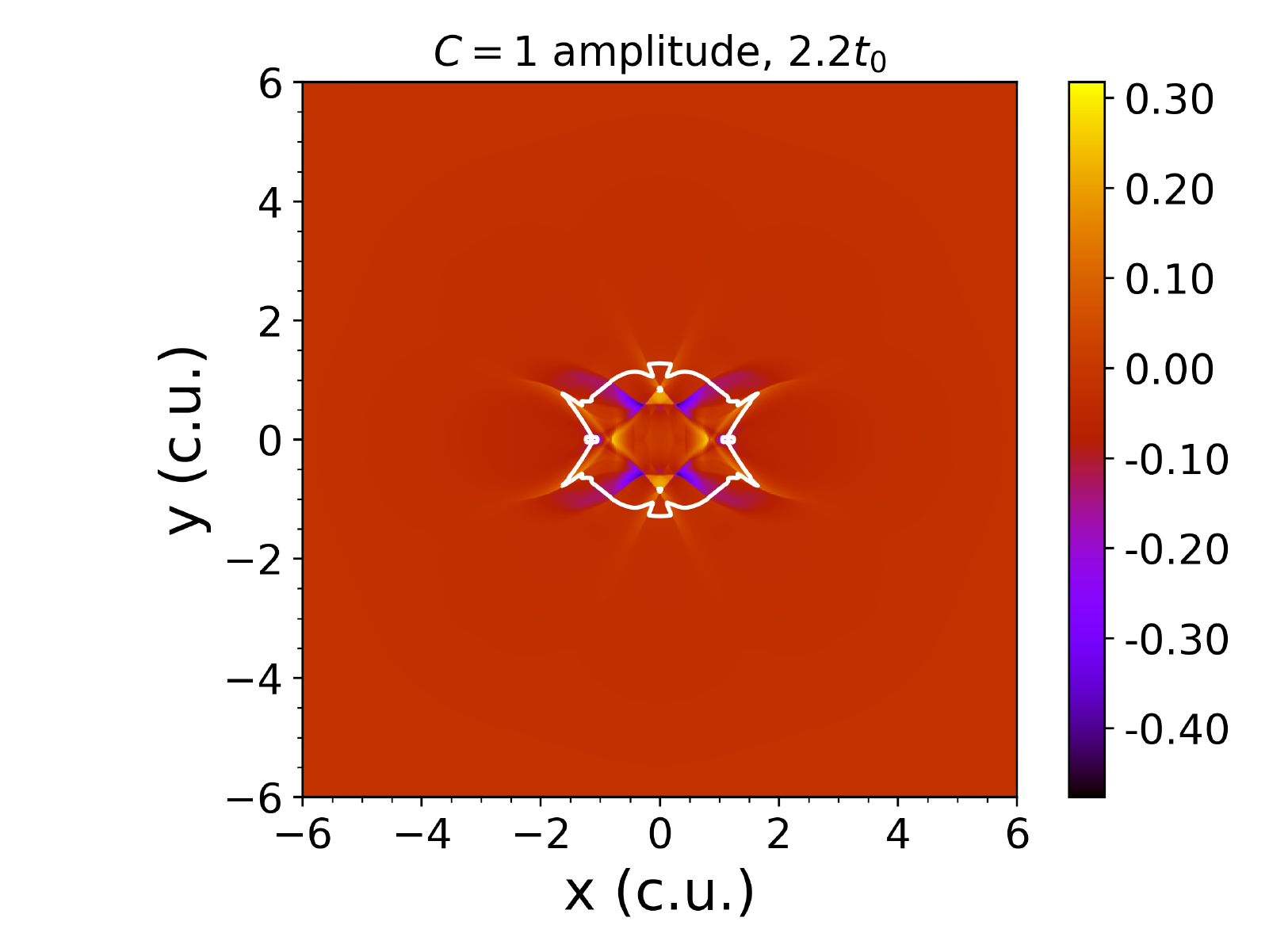}
    }
    \caption{Time evolution of the Ring driver for $C=1$, for the initial steps of the simulation. The propagation of the $v_{\perp}$ pulse (in code units) toward the null point is shown here, as well as the deformation of the equipartition layer (white contour) from the passing of the shock fronts. The time is shown (in $t_0$) for each panel.}
    \label{fig:ringdriver}
\end{figure*}

\begin{figure*}[t]
    \centering
    \resizebox{\hsize}{!}{
    \includegraphics[trim={0.4cm 0.cm 1.4cm 0.8cm},clip,scale=0.45]{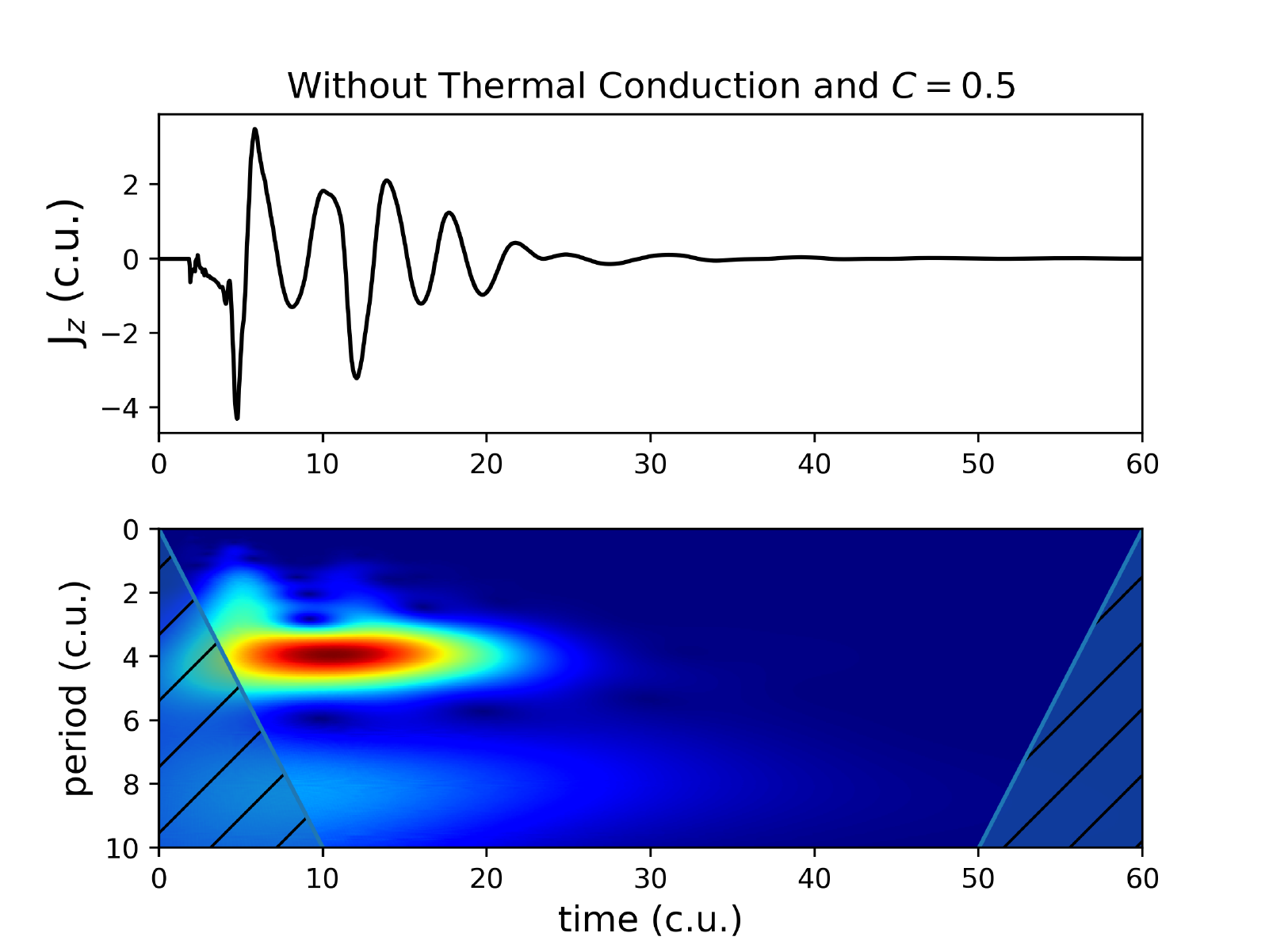}
    \includegraphics[trim={0.4cm 0.cm 1.4cm 0.8cm},clip,scale=0.45]{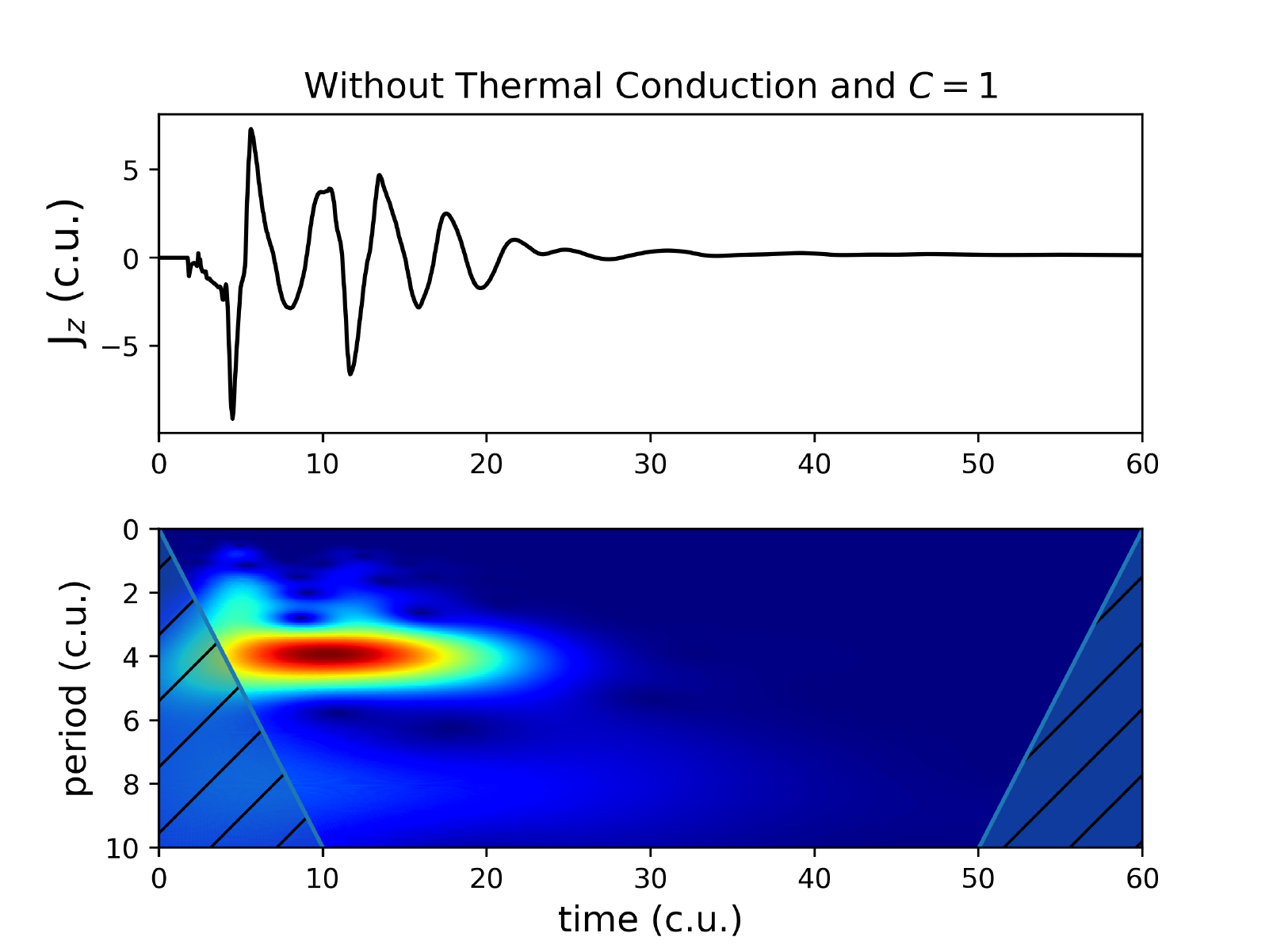}
    \includegraphics[trim={0.4cm 0.cm 1.4cm 0.8cm},clip,scale=0.45]{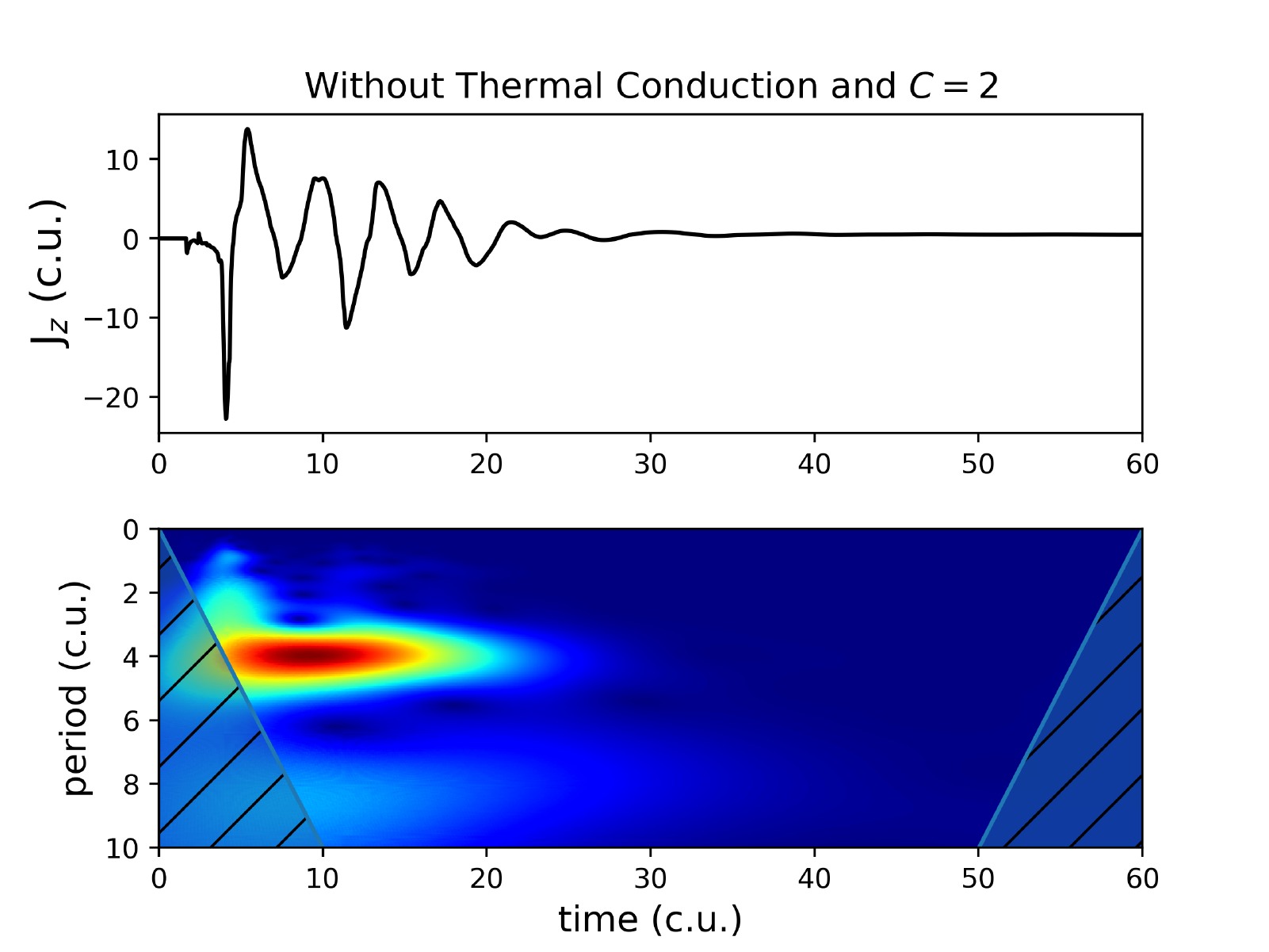}
    }
    \resizebox{\hsize}{!}{
    \includegraphics[trim={0.4cm 0.cm 1.4cm 0.8cm},clip,scale=0.45]{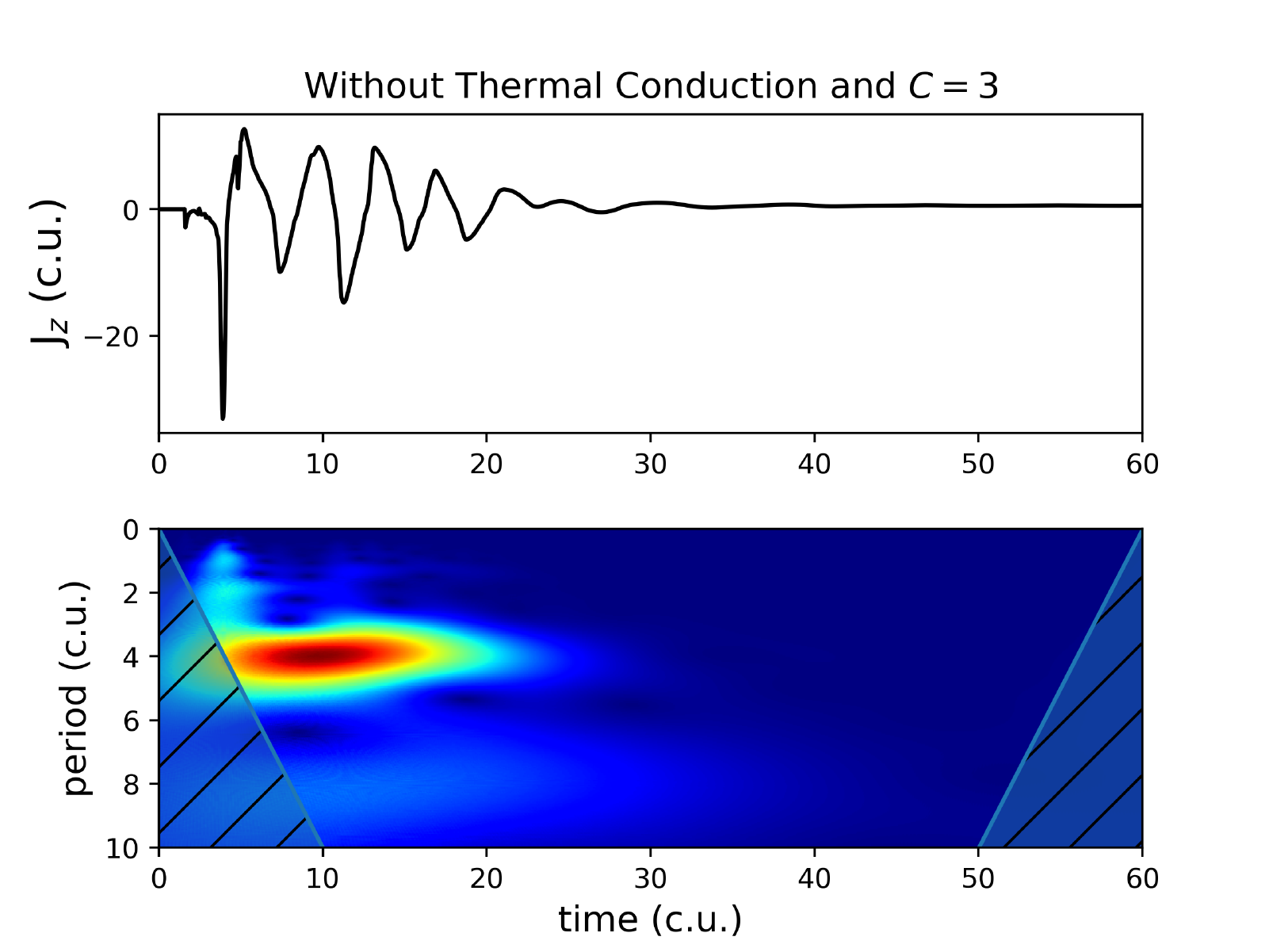}
    \includegraphics[trim={0.4cm 0.cm 1.4cm 0.8cm},clip,scale=0.45]{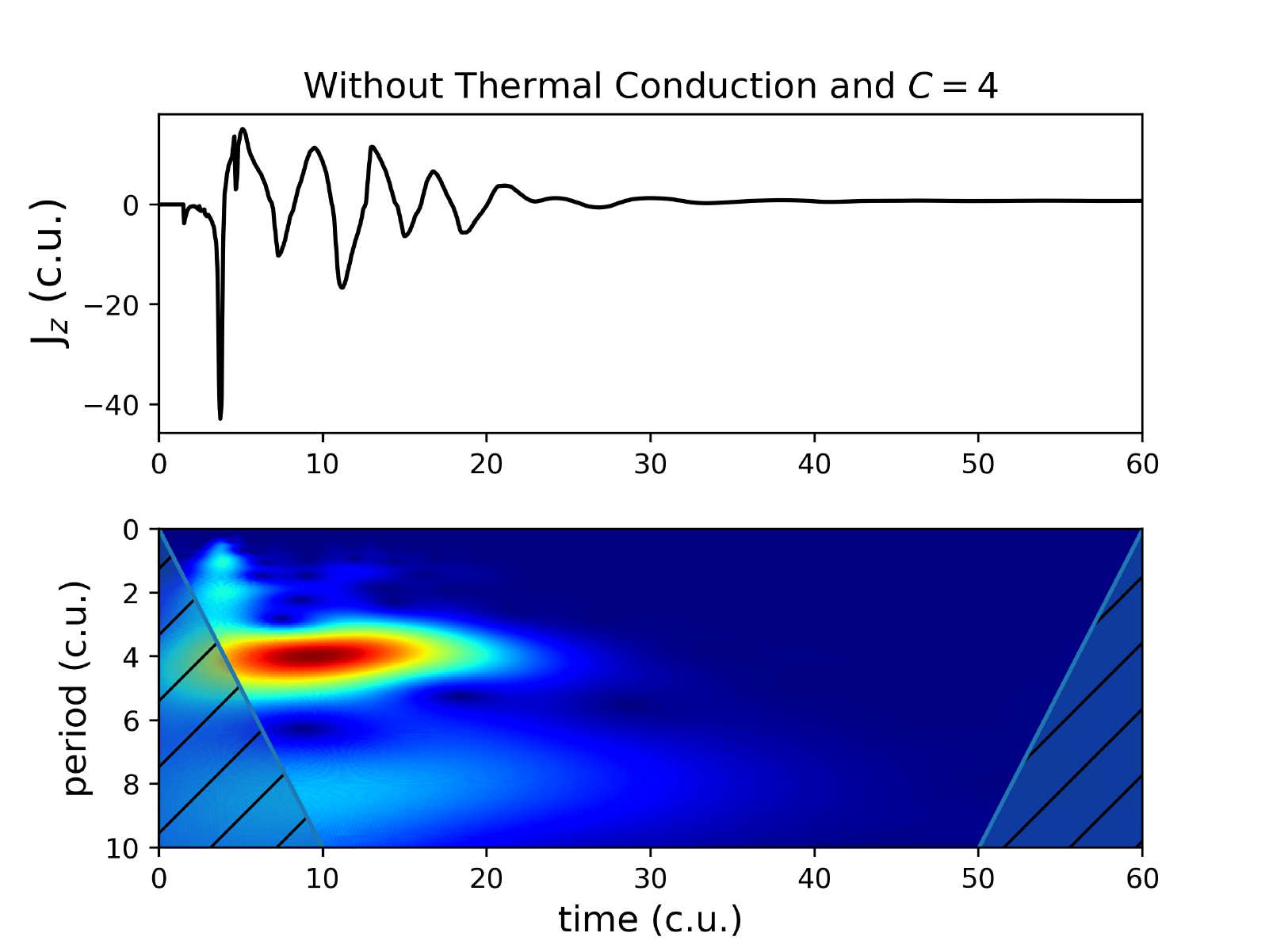}
    \includegraphics[trim={0.4cm 0.cm 1.4cm 0.8cm},clip,scale=0.45]{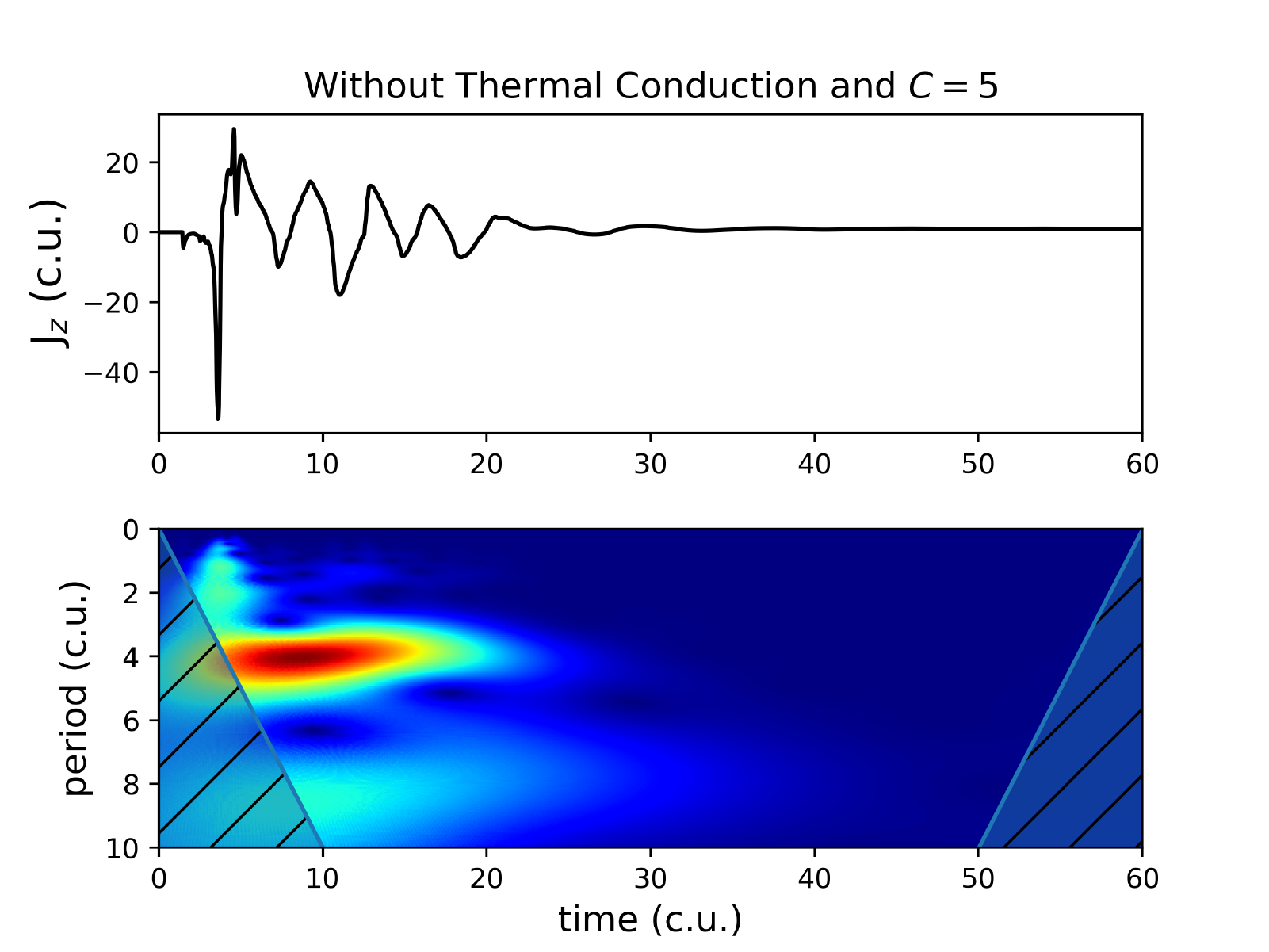}
    }
    \includegraphics[trim={0.4cm 0.cm 1.4cm 0.8cm},clip,scale=0.41]{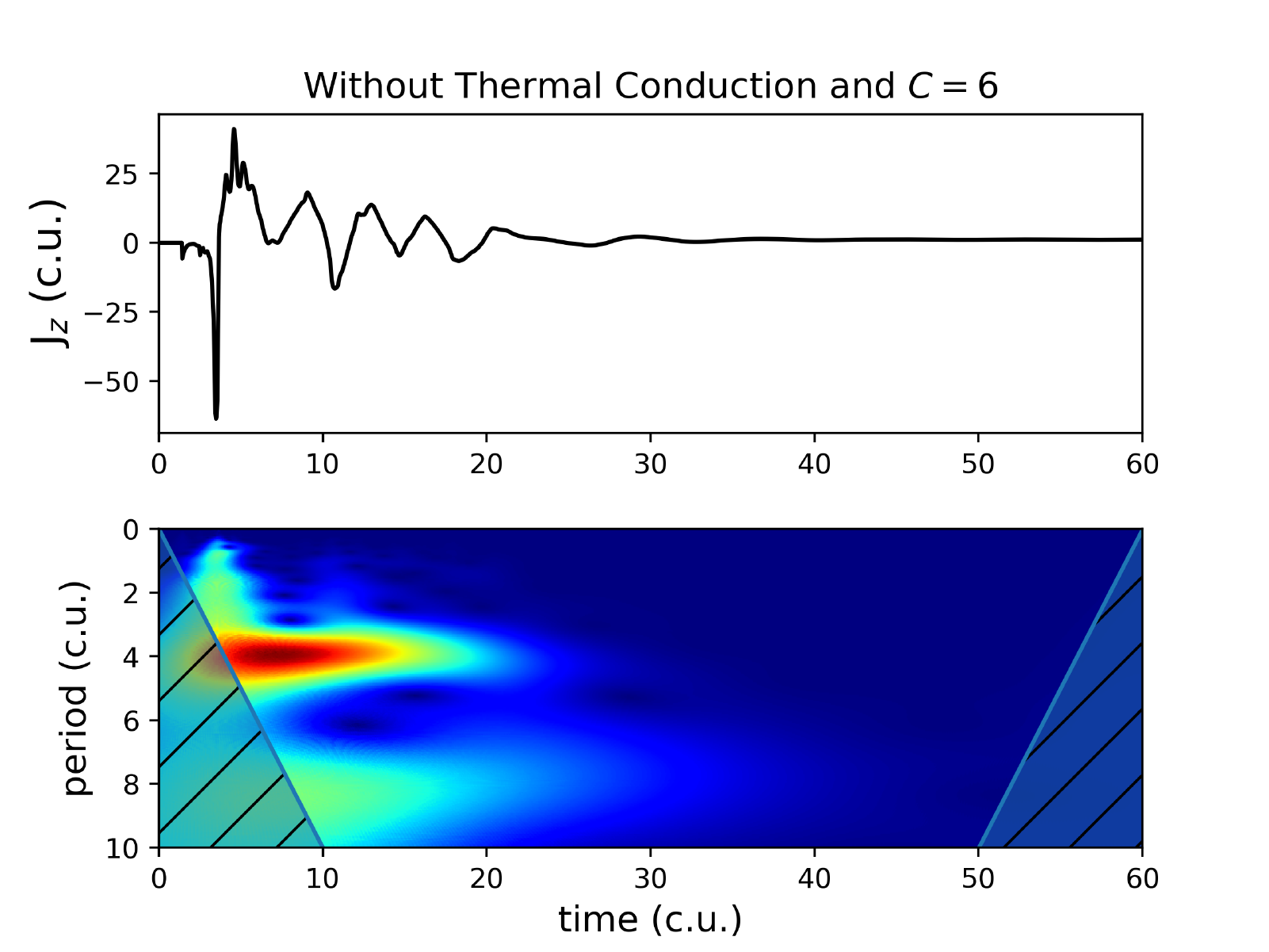}
    \caption{Oscillating profiles of the $J_z$ current density at the X-point as a function of time, for Ring drivers of different initial amplitudes ($C=0.5,\,1,\,2,\,3,\,4,\,5$ and $6$). Alongside the $J_z$ profiles, the density plots of the respective wavelet time series are shown. The setups considered here are all treated in resistive MHD, without thermal conduction.}
    \label{fig:ringwavelet}
\end{figure*}

\begin{figure*}[t]
    \centering
    \resizebox{\hsize}{!}{
    \includegraphics[trim={0.cm 0.cm 0.cm 0.cm},clip,scale=0.45]{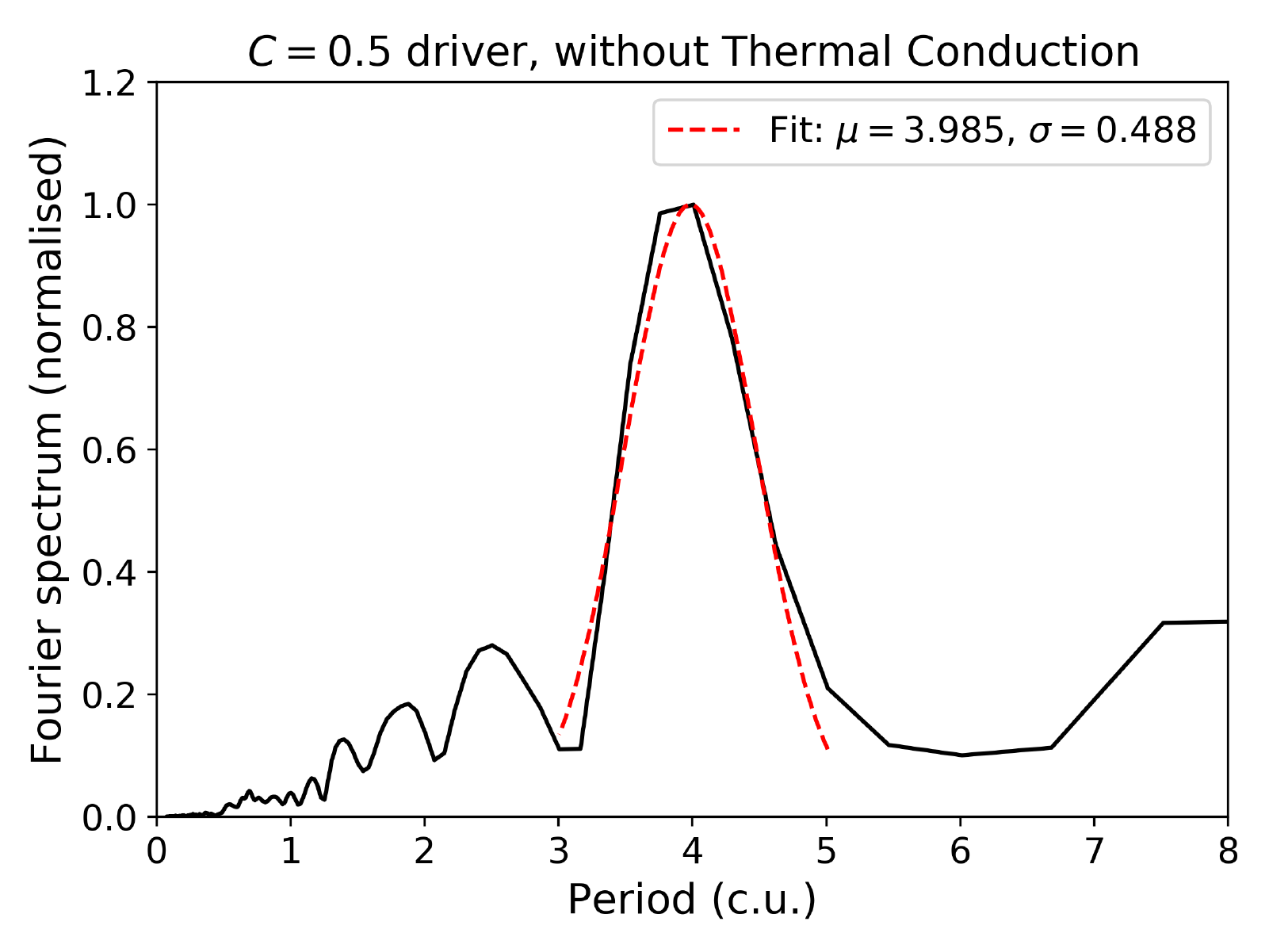}
    \includegraphics[trim={1.cm 0.cm 0.cm 0.cm},clip,scale=0.45]{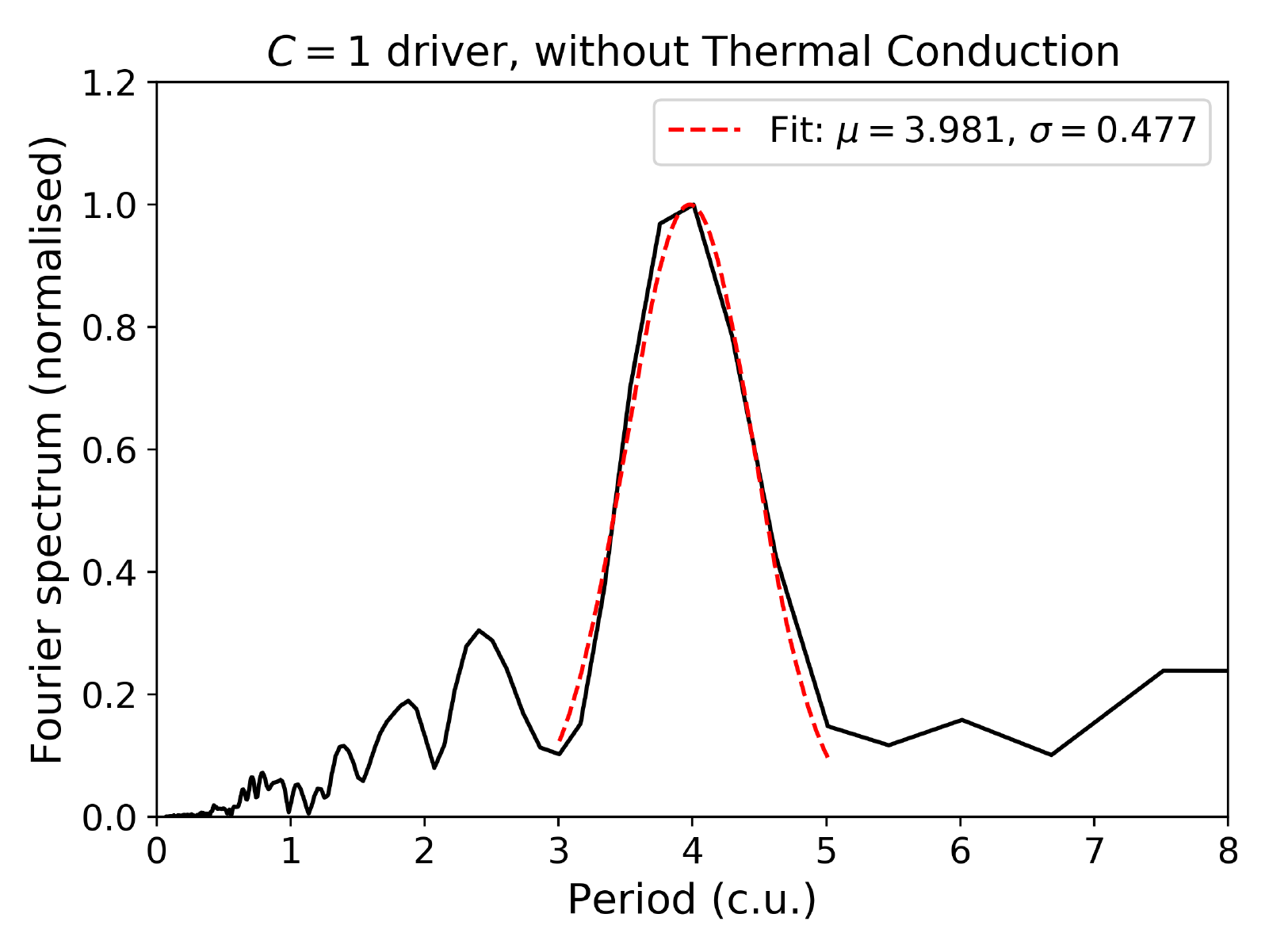}
    \includegraphics[trim={1.cm 0.cm 0.cm 0.cm},clip,scale=0.45]{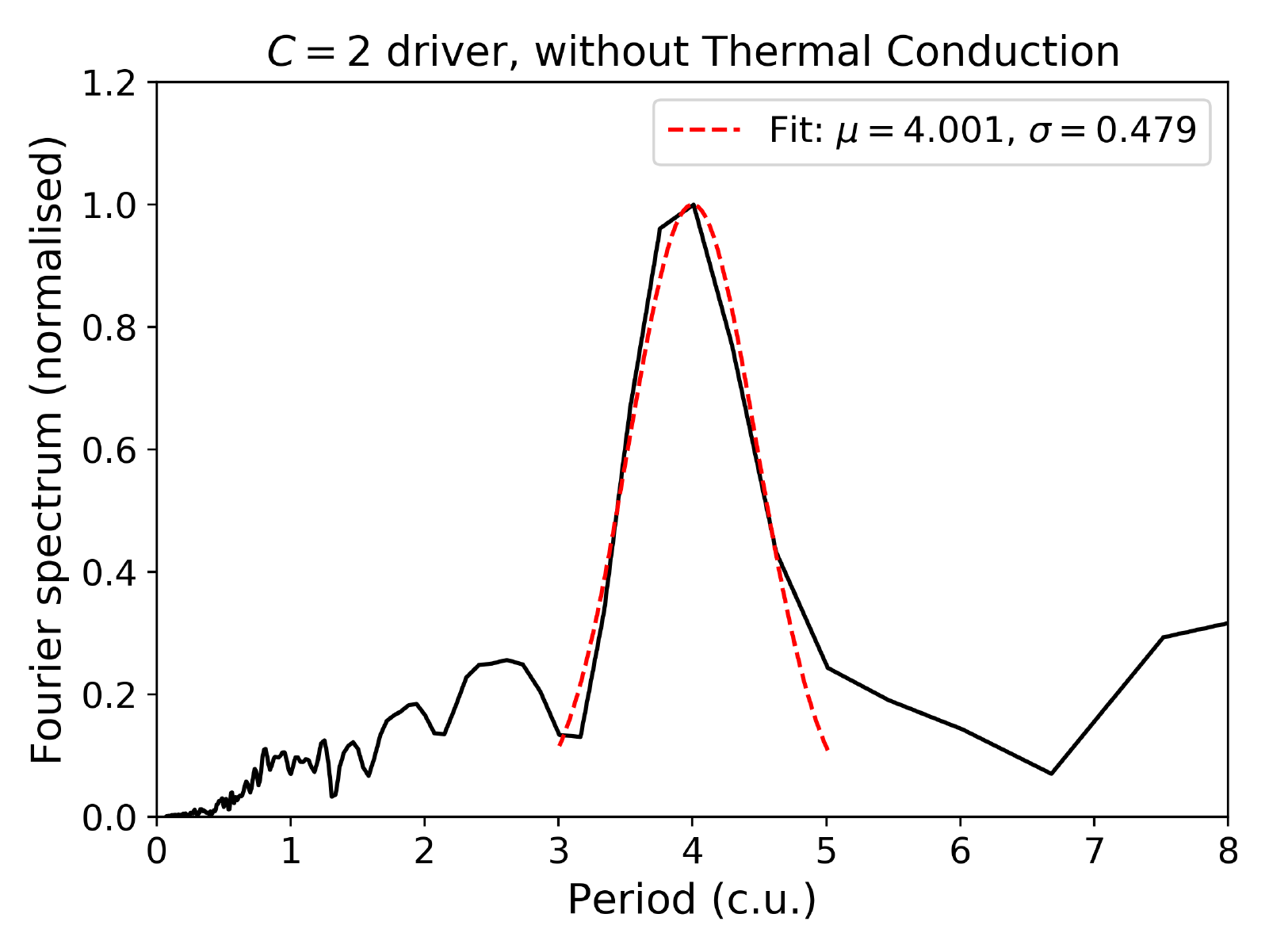}
    }
    \resizebox{\hsize}{!}{
    \includegraphics[trim={0.cm 0.cm 0cm 0.cm},clip,scale=0.45]{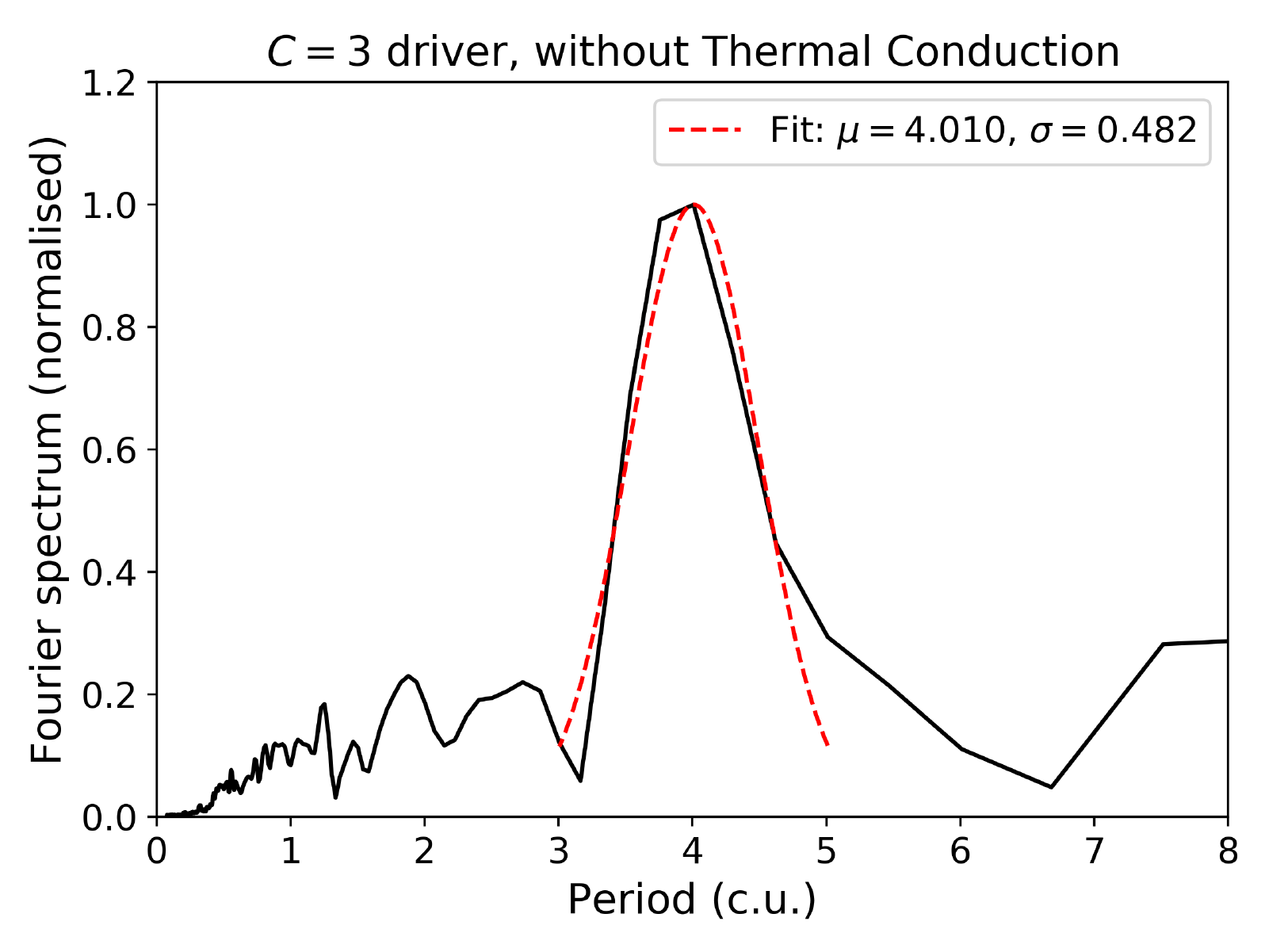}
    \includegraphics[trim={1.cm 0.cm 0cm 0.cm},clip,scale=0.45]{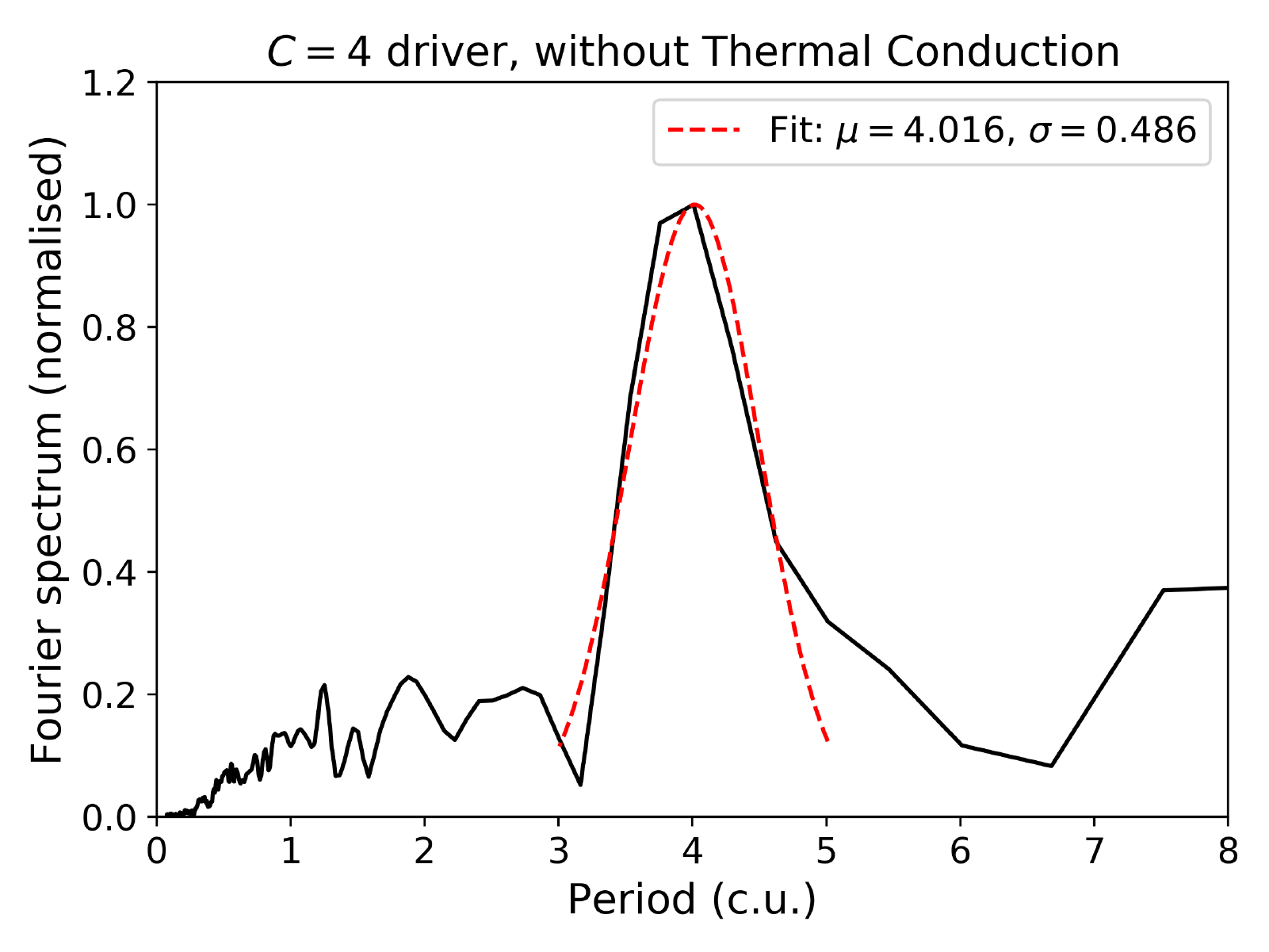}
    \includegraphics[trim={1.cm 0.cm 0cm 0.cm},clip,scale=0.45]{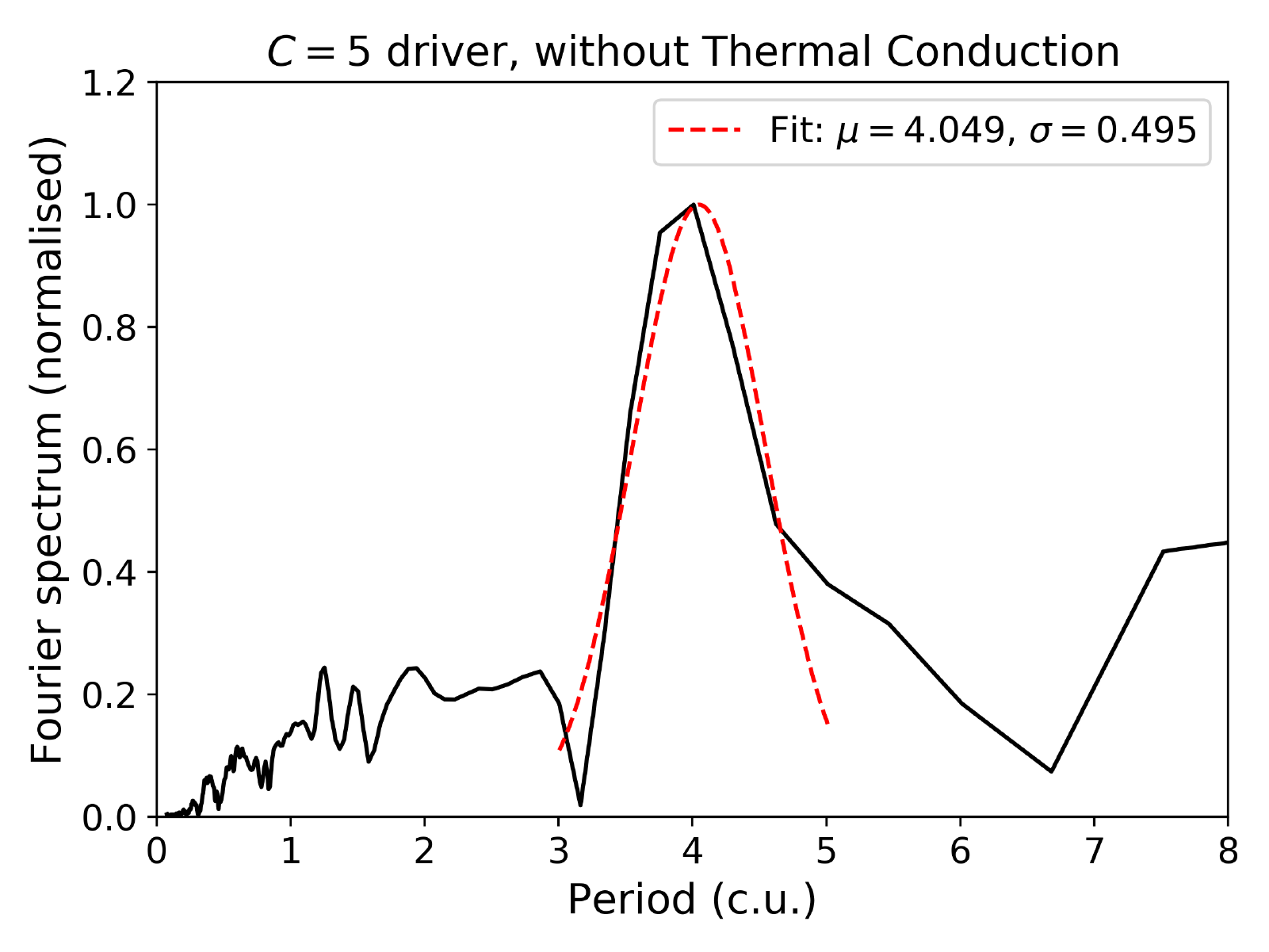}
    }
    \includegraphics[trim={0.cm 0.cm 0cm 0.cm},clip,scale=0.38]{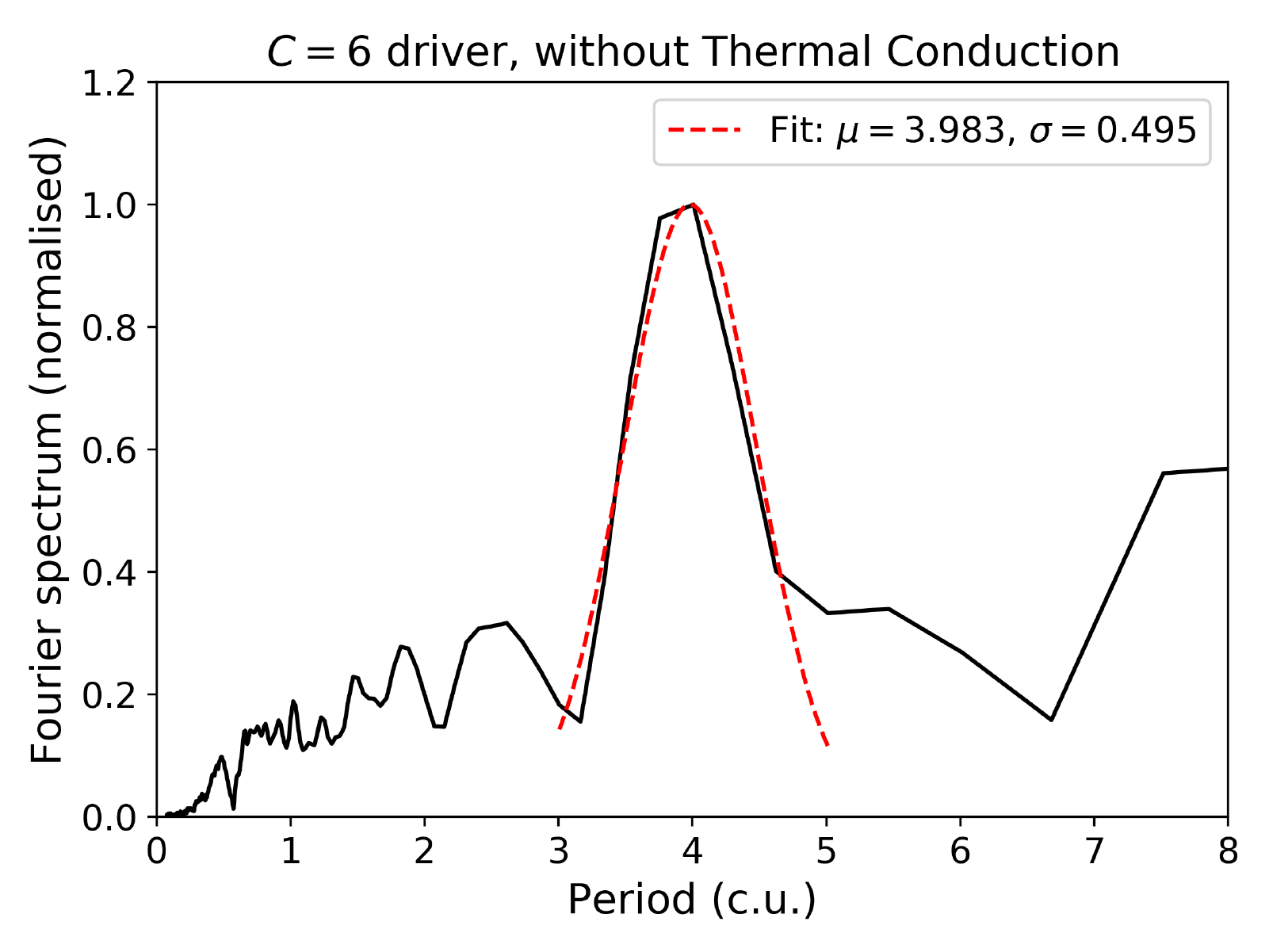}
    \caption{Normalized Fourier spectra for the $J_z$ current density signals of Fig. \ref{fig:ringwavelet} as a function of the period. The red dashed line in each panel shows a Gaussian fit, with the location $\mu$ of the maximum power and standard deviation $\sigma$ given in the legend.}
    \label{fig:ringfourier}
\end{figure*}

\begin{figure*}[t]
    \centering
    \resizebox{\hsize}{!}{
    \includegraphics[trim={0.4cm 0.cm 1.4cm 0.8cm},clip,scale=0.45]{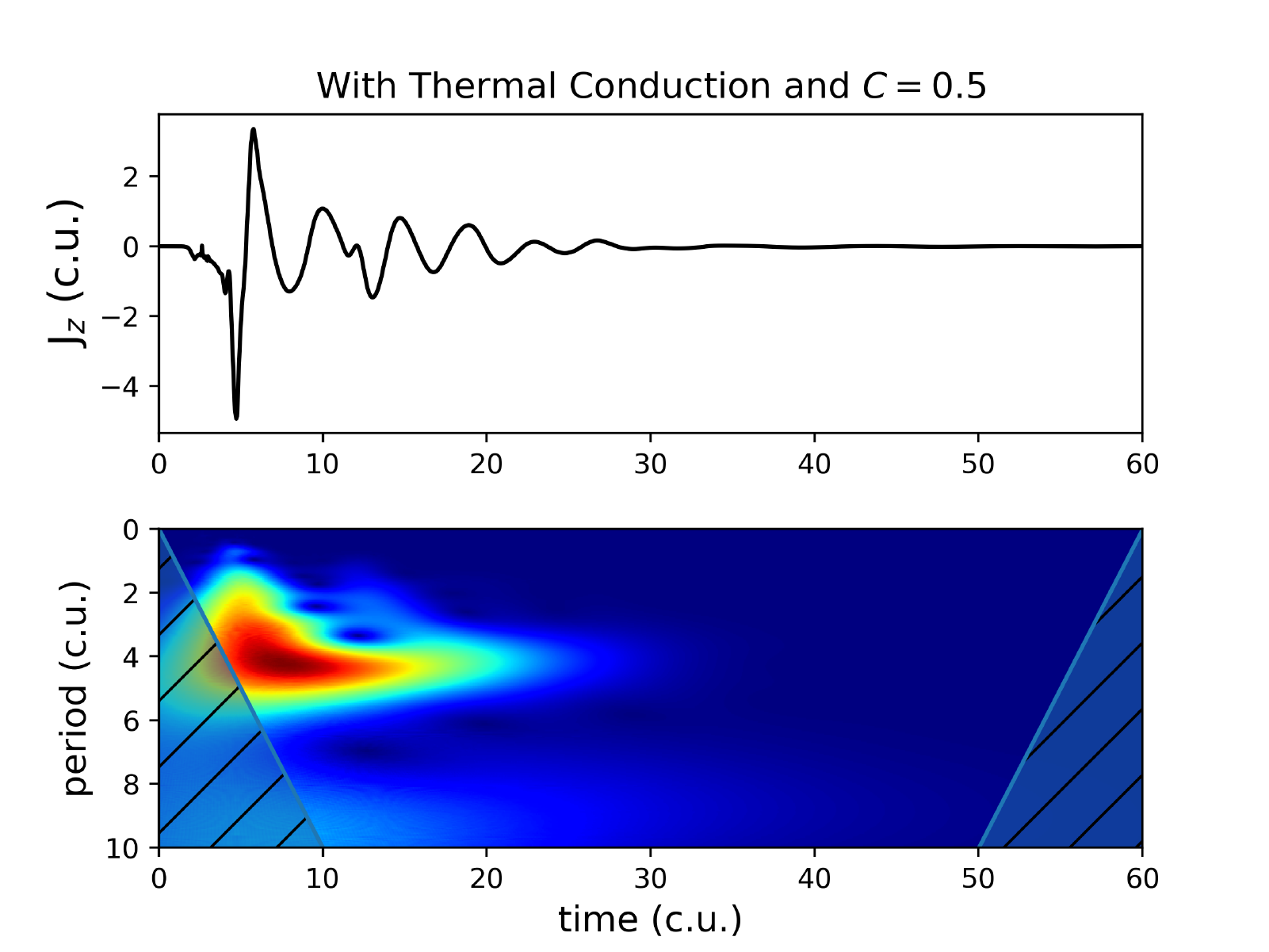}
    \includegraphics[trim={0.4cm 0.cm 1.4cm 0.8cm},clip,scale=0.45]{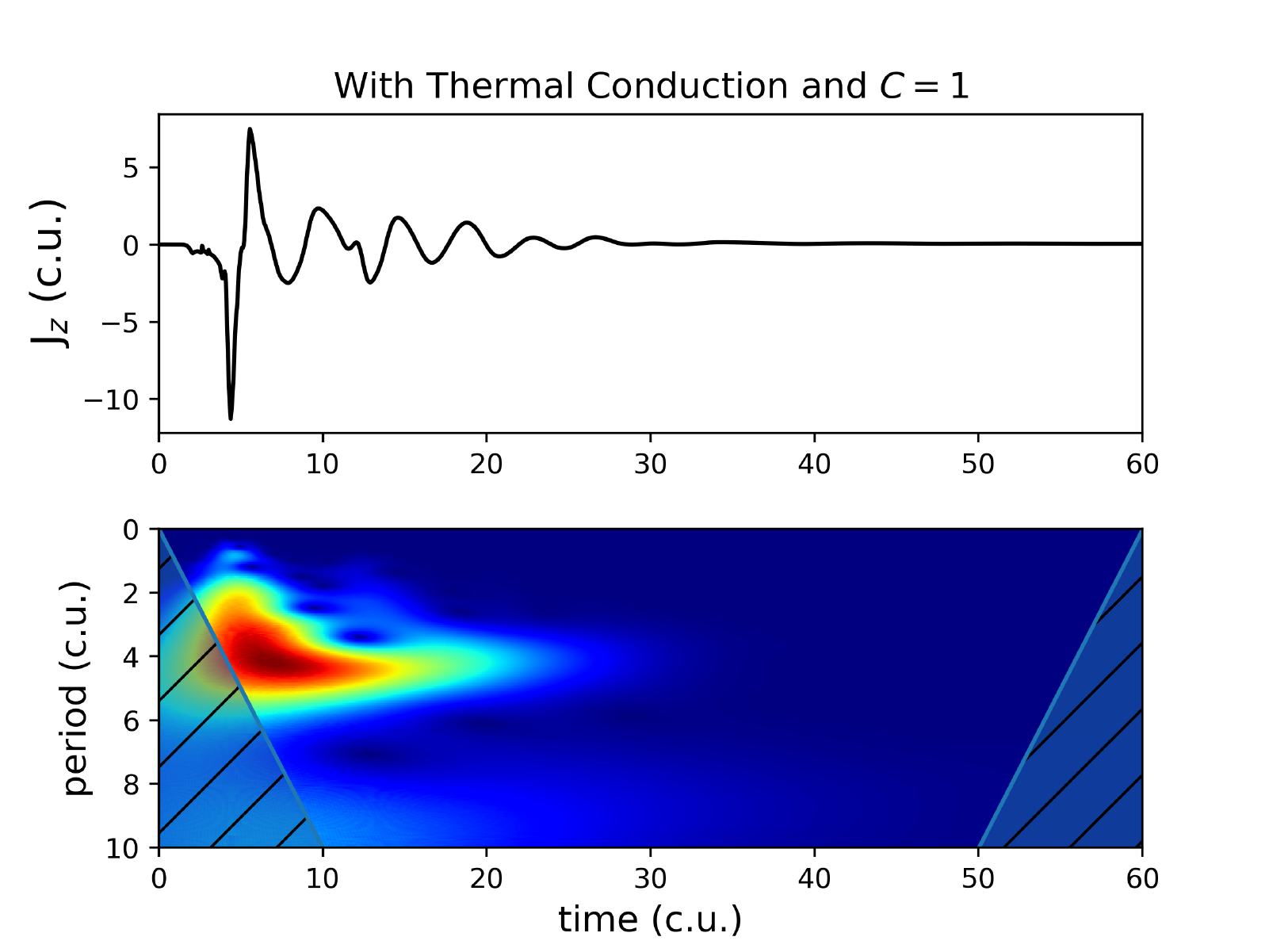}
    \includegraphics[trim={0.4cm 0.cm 1.4cm 0.8cm},clip,scale=0.45]{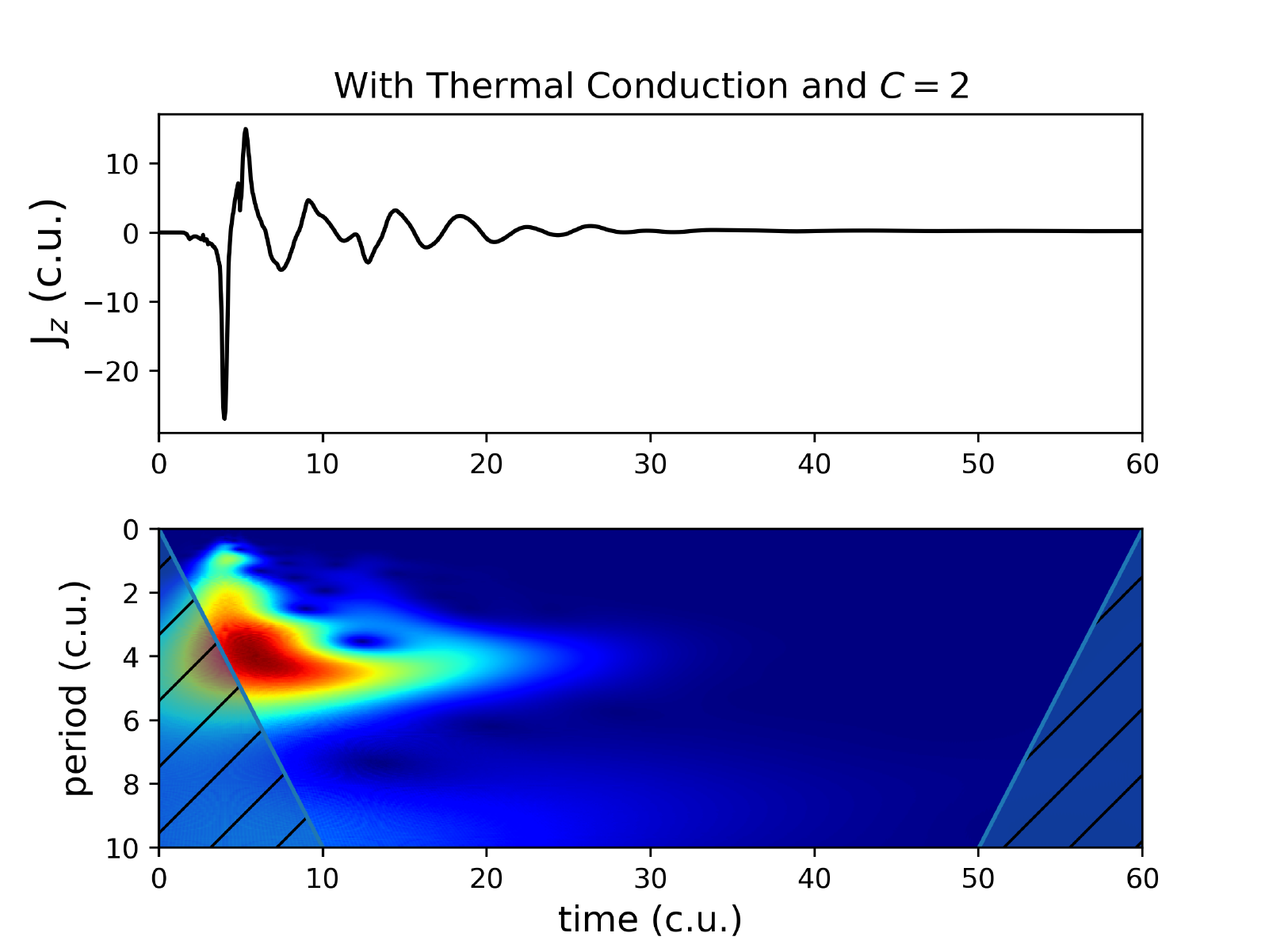}
    }
    \resizebox{\hsize}{!}{
    \includegraphics[trim={0.4cm 0.cm 1.4cm 0.8cm},clip,scale=0.45]{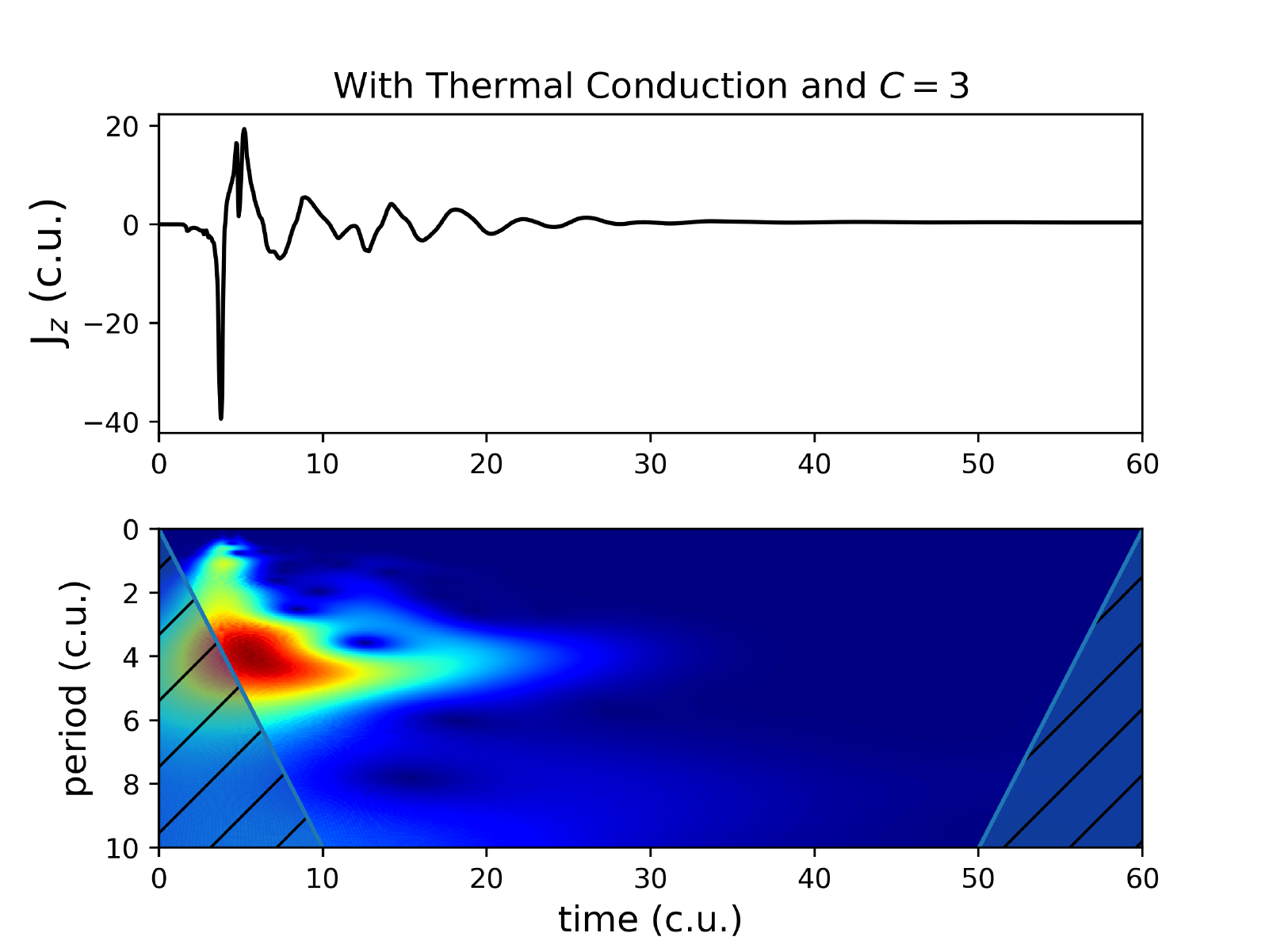}
    \includegraphics[trim={0.4cm 0.cm 1.4cm 0.8cm},clip,scale=0.45]{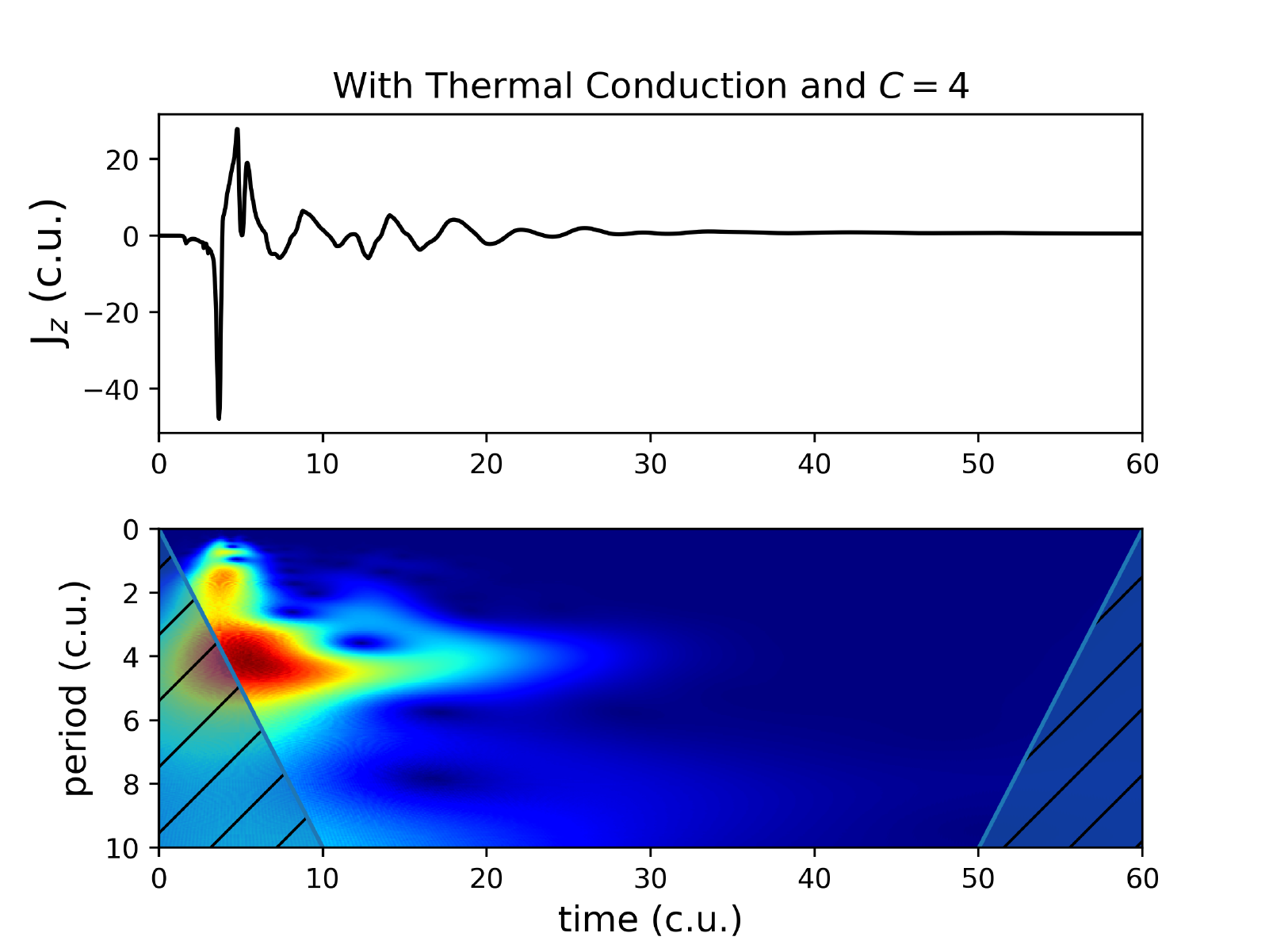}
    \includegraphics[trim={0.4cm 0.cm 1.4cm 0.8cm},clip,scale=0.45]{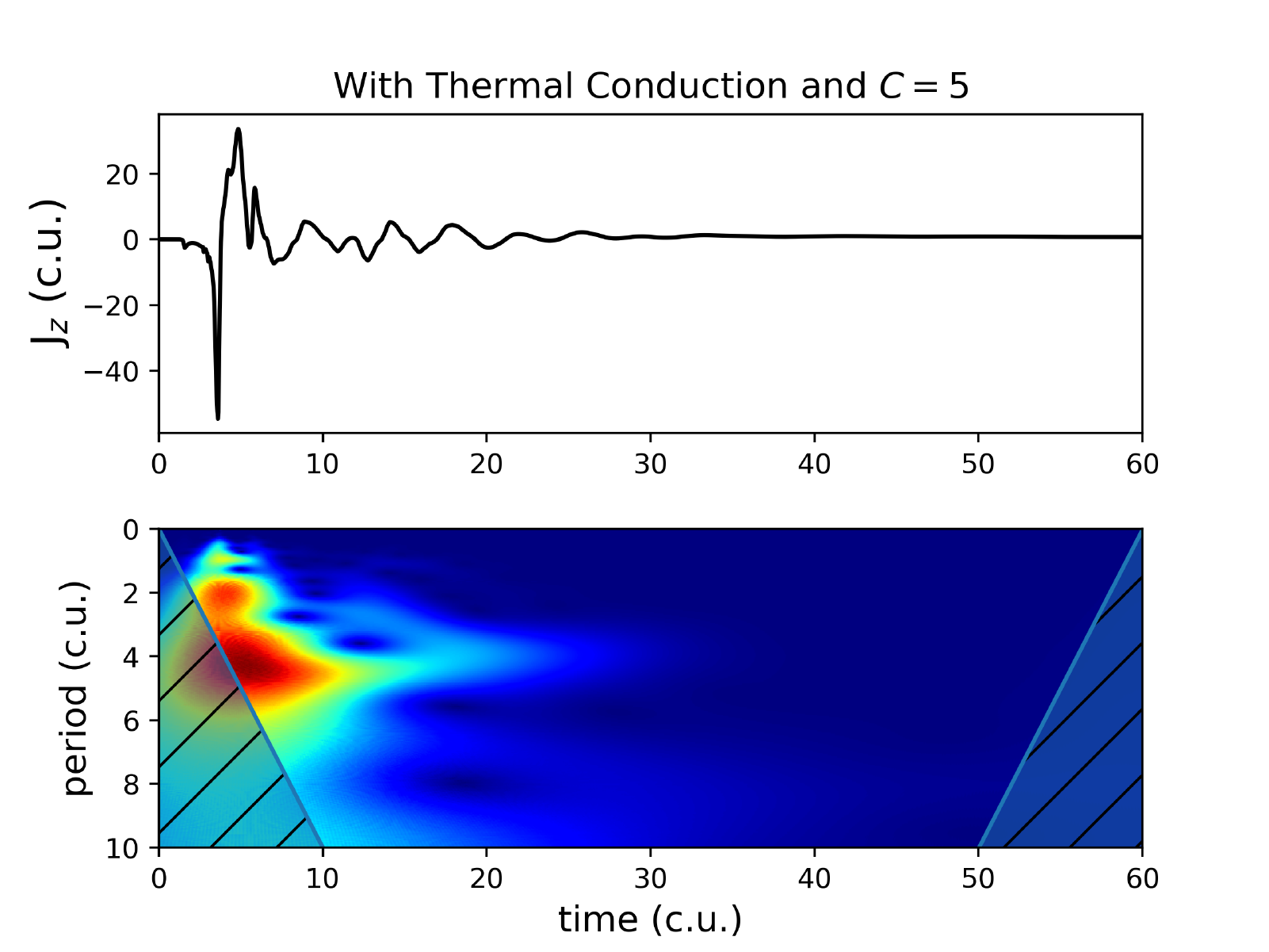}
    }
    \includegraphics[trim={0.4cm 0.cm 1.4cm 0.8cm},clip,scale=0.41]{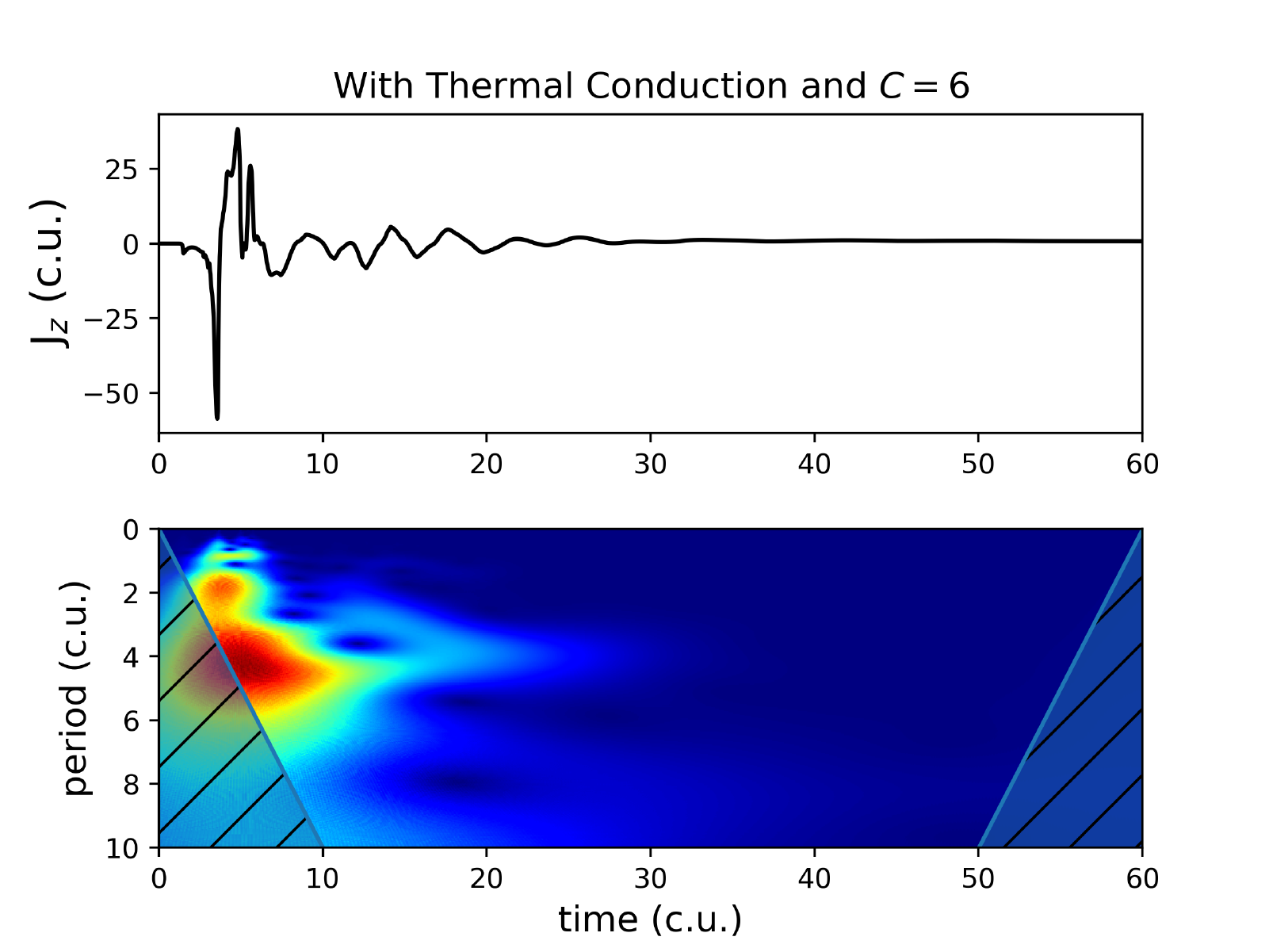}
    \caption{Same as Fig. \ref{fig:ringwavelet}, with the inclusion of anisotropic thermal conduction in the resistive MHD equations.}
    \label{fig:ringwaveletTC}
\end{figure*}

\begin{figure*}[t]
    \centering
    \resizebox{\hsize}{!}{
    \includegraphics[trim={0.cm 0.cm 0.cm 0.cm},clip,scale=0.45]{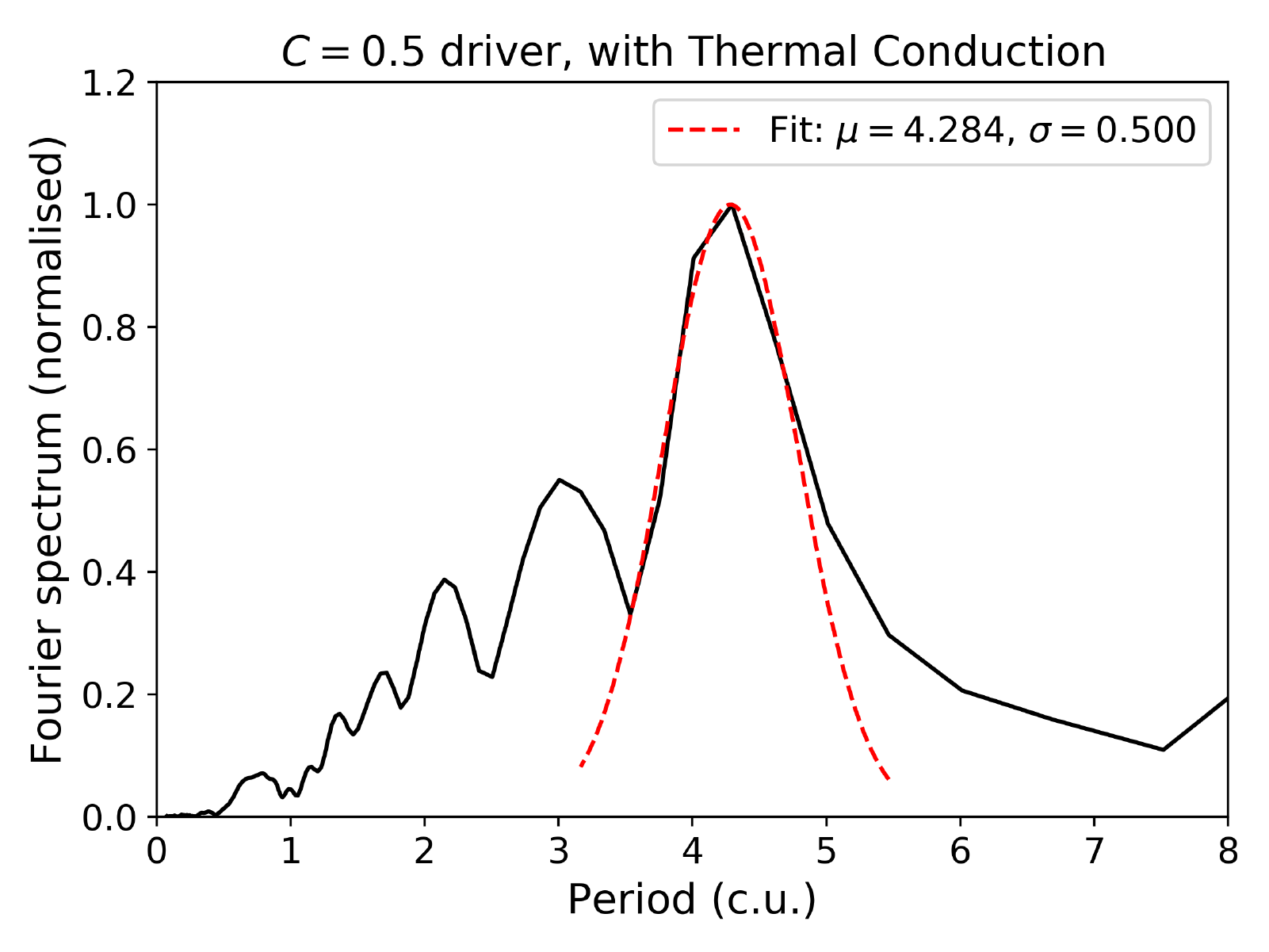}
    \includegraphics[trim={1.cm 0.cm 0.cm 0.cm},clip,scale=0.45]{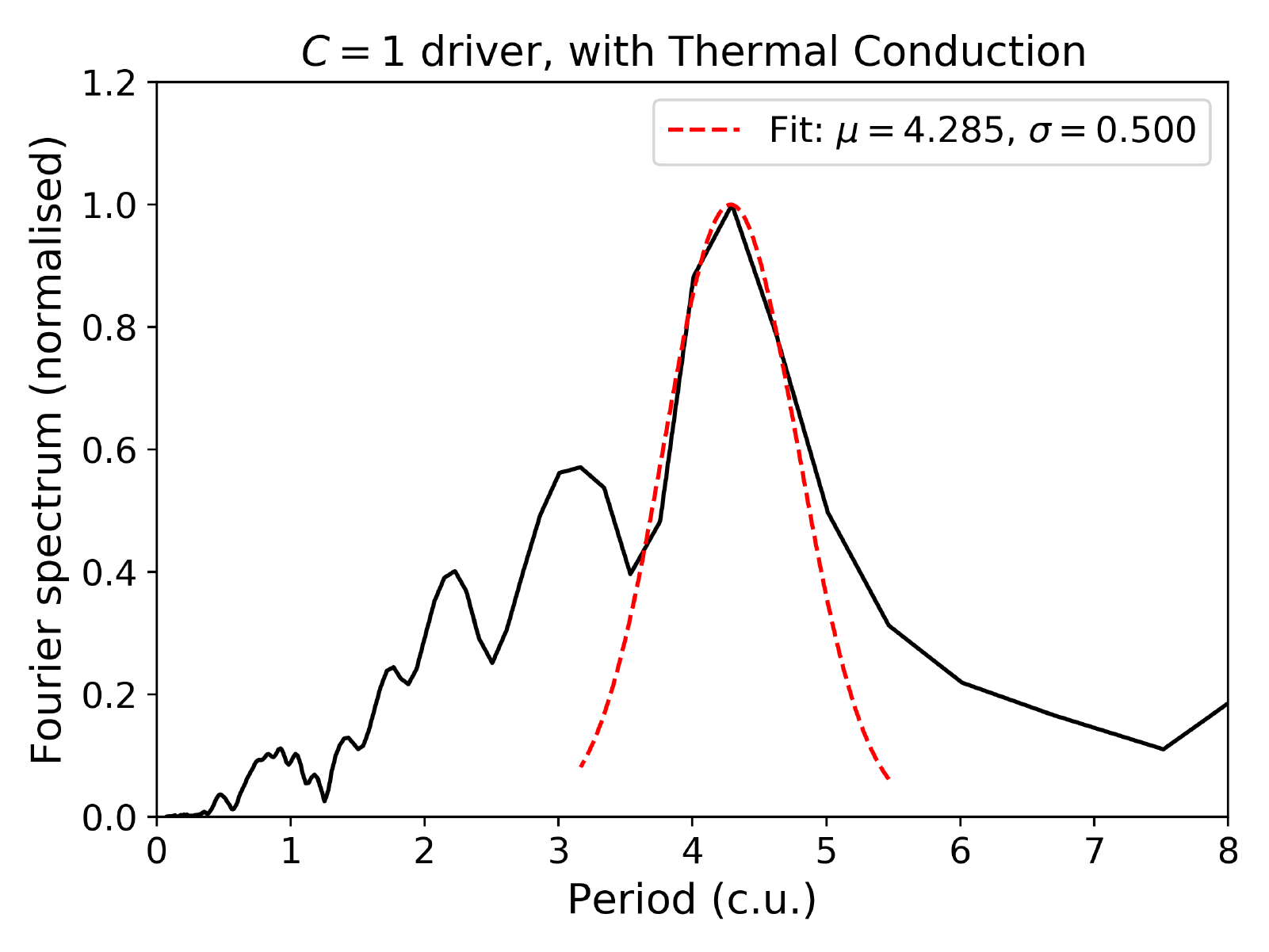}
    \includegraphics[trim={1.cm 0.cm 0.cm 0.cm},clip,scale=0.45]{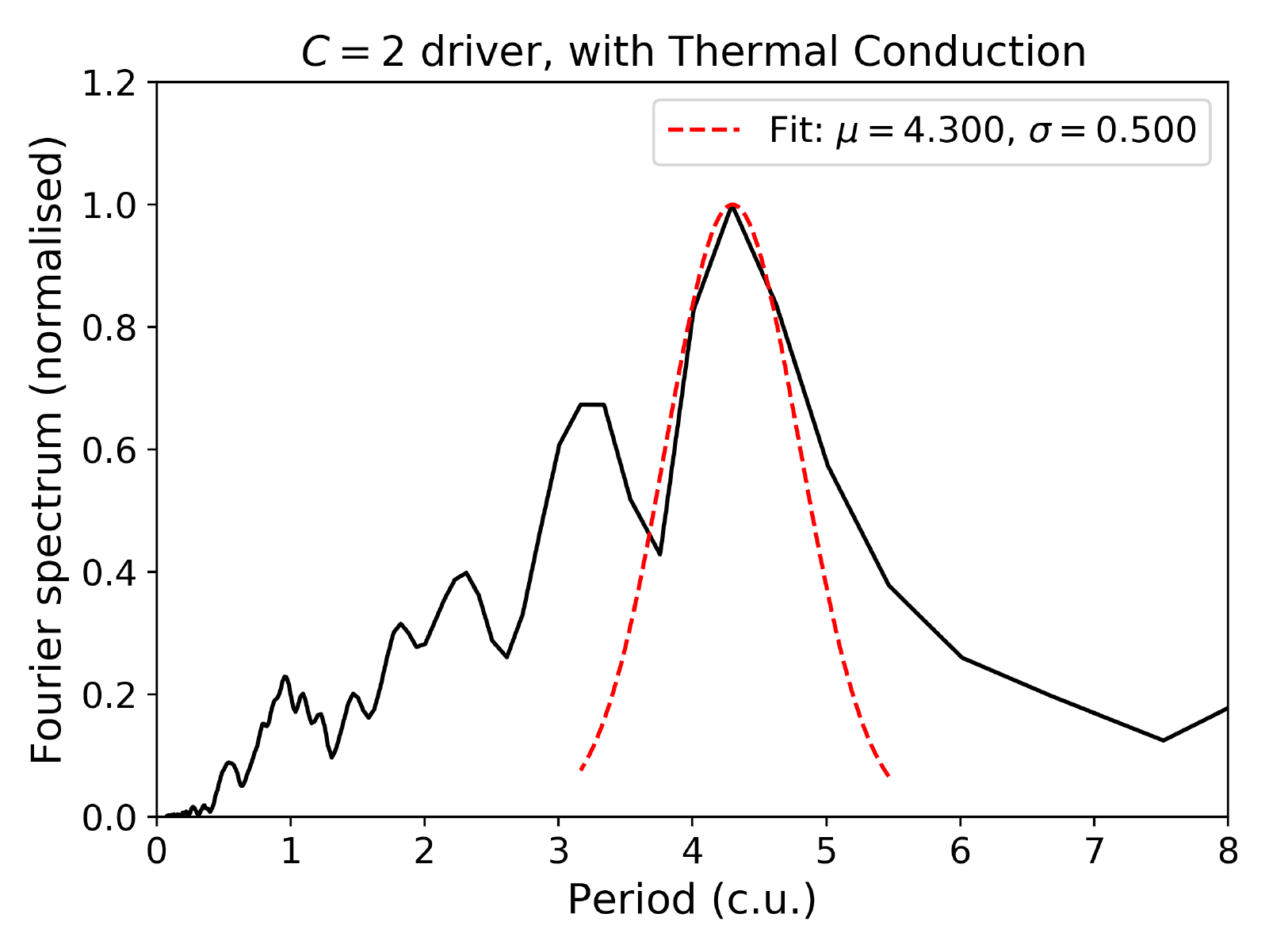}
    }
    \resizebox{\hsize}{!}{
    \includegraphics[trim={0.cm 0.cm 0.cm 0.cm},clip,scale=0.45]{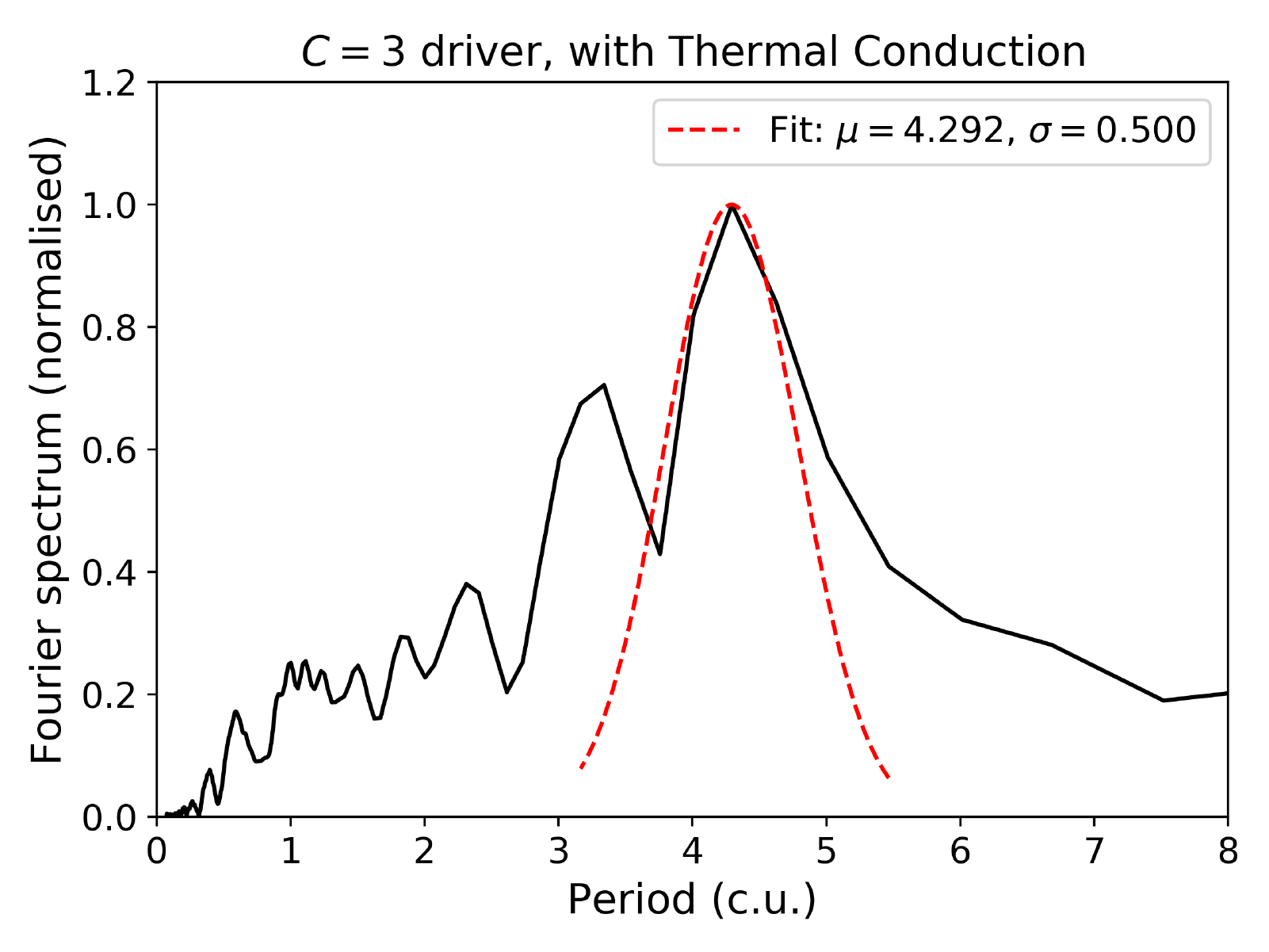}
    \includegraphics[trim={1.cm 0.cm 0.cm 0.cm},clip,scale=0.45]{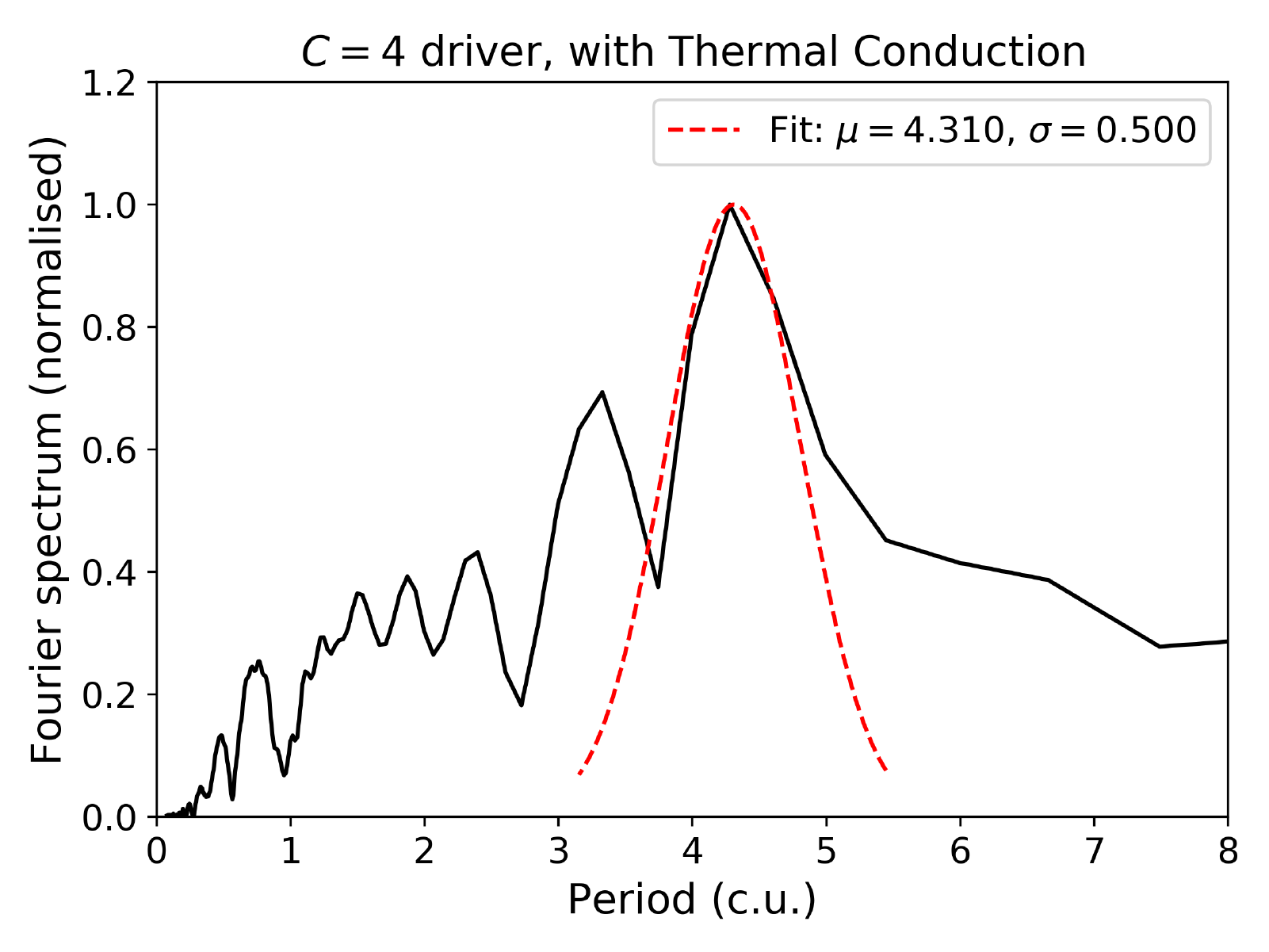}
    \includegraphics[trim={1.cm 0.cm 0.cm 0.cm},clip,scale=0.45]{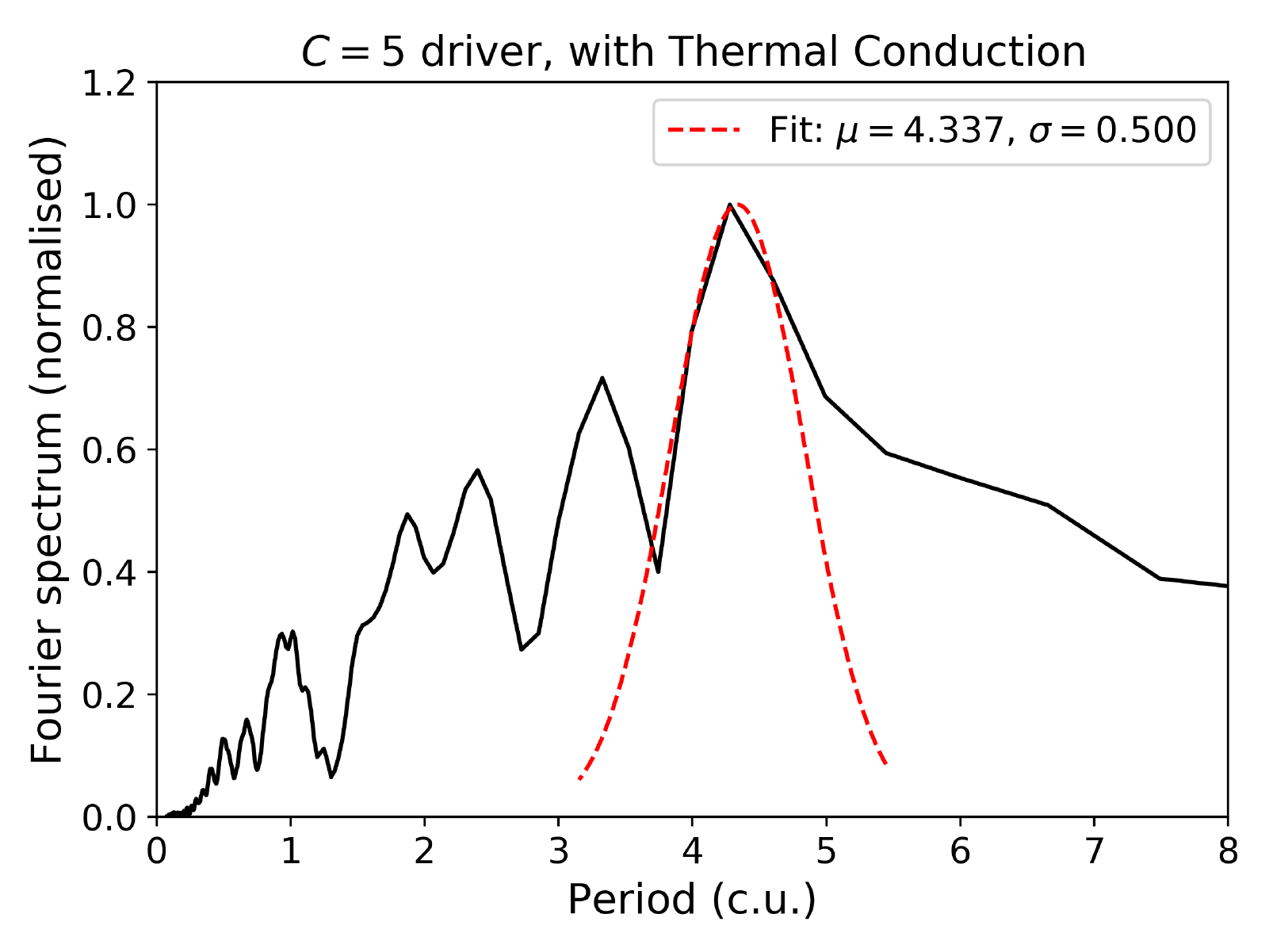}
    }
    \includegraphics[trim={0.cm 0.cm 0.cm 0.cm},clip,scale=0.38]{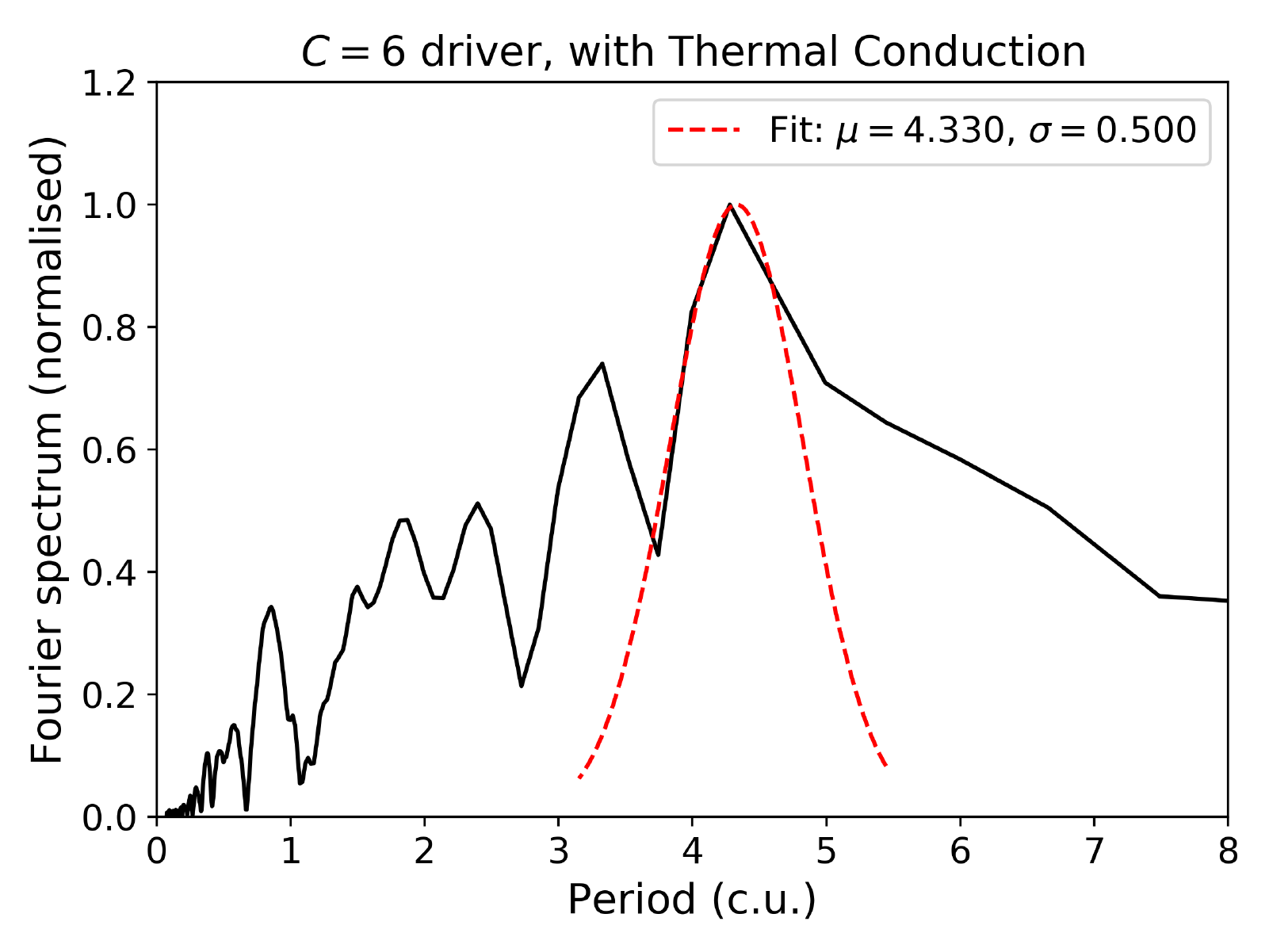}
    \caption{Normalized Fourier spectra for the $J_z$ current density signals of Fig. \ref{fig:ringwaveletTC} as a function of the period. The red dashed line in each panel shows a Gaussian fit, with the location $\mu$ of the maximum power and standard deviation $\sigma$ given in the legend.}
    \label{fig:ringTCfourier}
\end{figure*}

\begin{figure*}[t]
    \centering
    \includegraphics[trim={0.cm 0.cm 0.cm 0.cm},clip,scale=0.42]{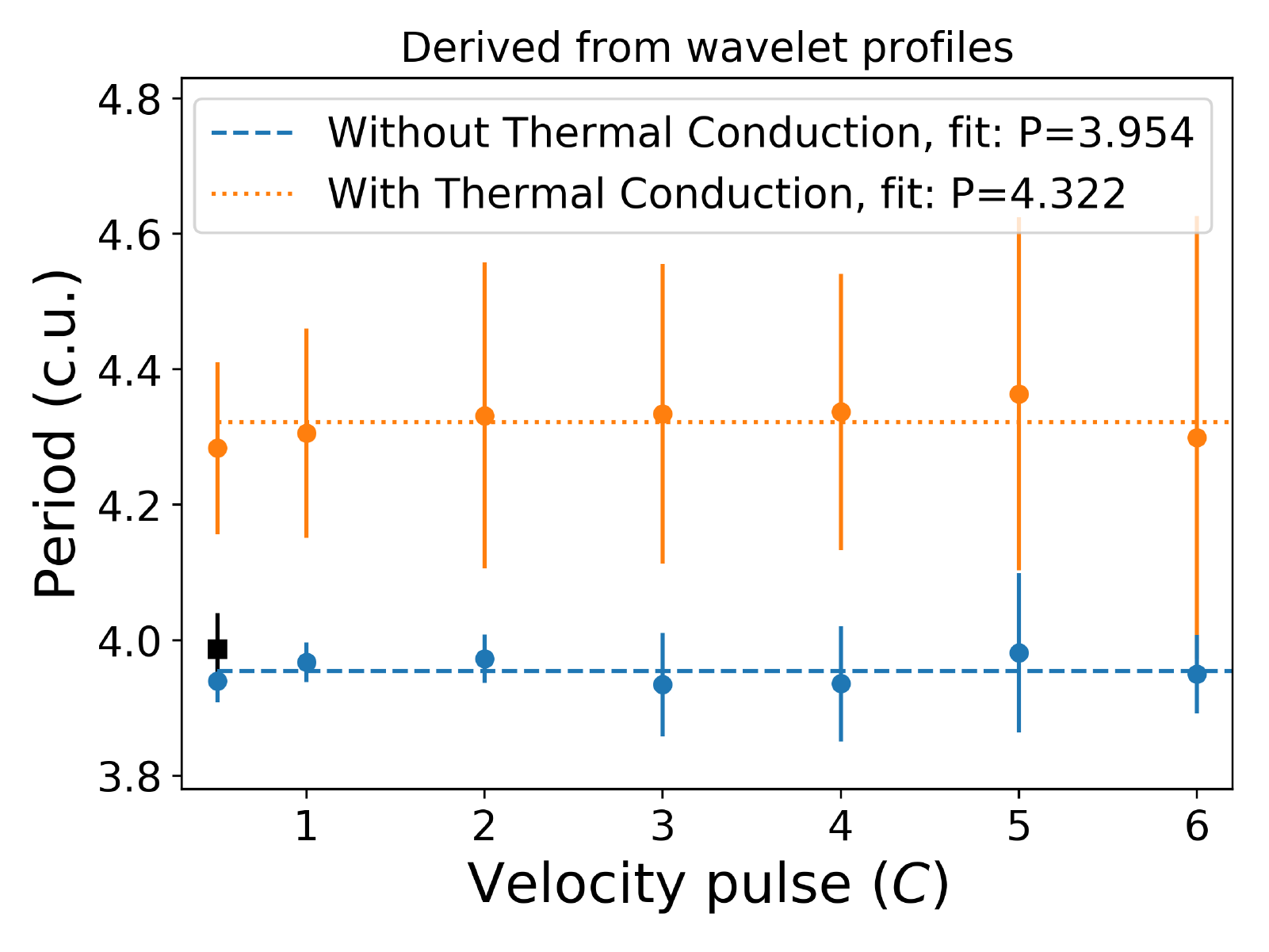}
    \includegraphics[trim={0.cm 0.cm 0.cm 0.cm},clip,scale=0.42] {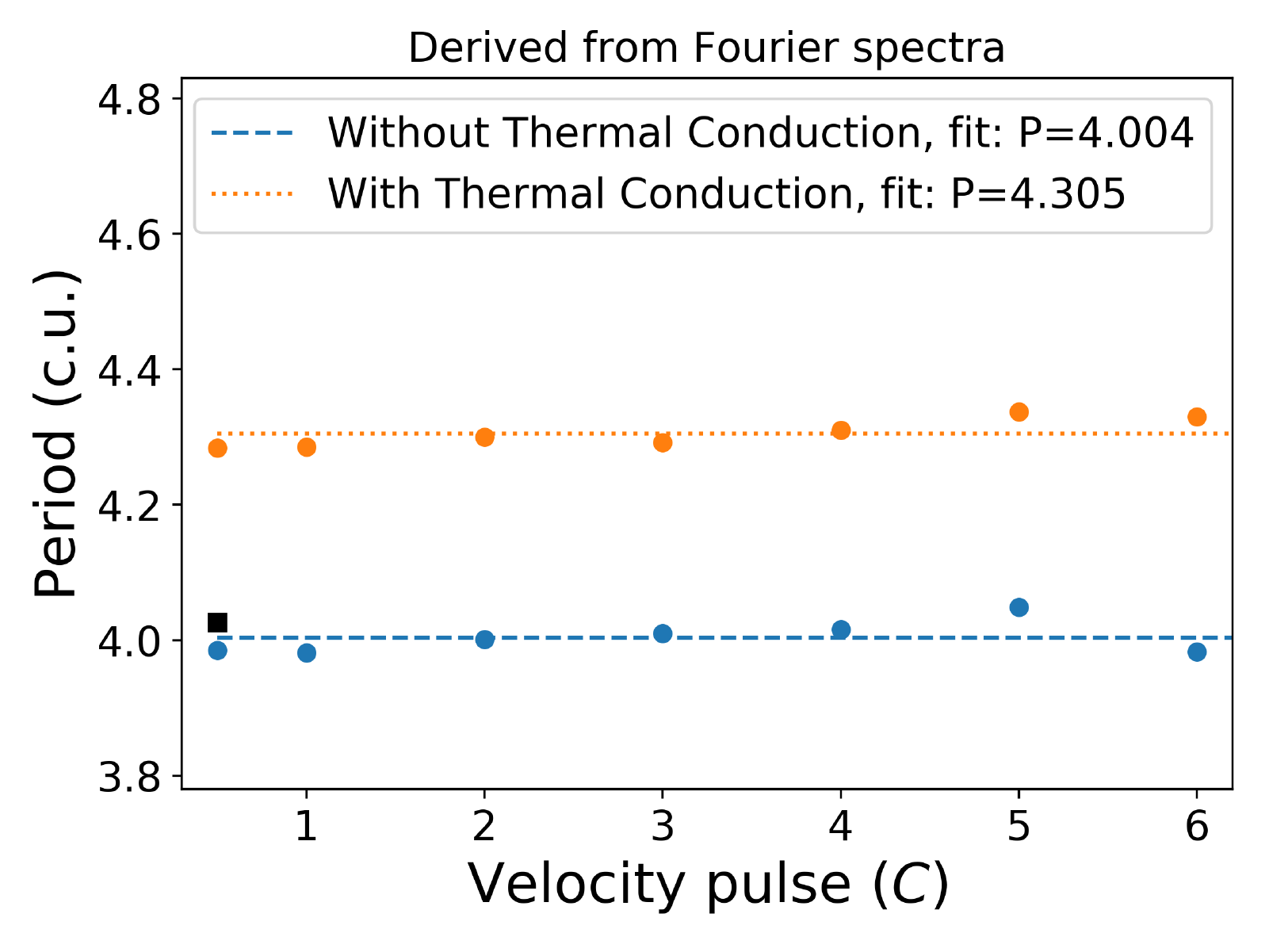}
    \caption{Graphs showing the periods (in code units) of the $J_z$ oscillation profiles with respect to the initial pulse amplitude (in C). The left panel is for the periods calculated from the wavelet profiles of Figs. \ref{fig:ringwavelet},  \ref{fig:ringwaveletTC} and \ref{fig:pinchwavelet}, and the right panel is for the periods from the Fourier spectra of Figs. \ref{fig:ringfourier}, \ref{fig:ringTCfourier} and \ref{fig:pinchwavelet}. The cases without thermal conduction are shown in blue, and with thermal conduction in orange. For both sets on each panel, a constant period $P$ is fitted (blue dashed and orange dotted lines for each case, respectively). The circles plus error bars correspond to the different Ring drivers, and the black square plus error bars to the Pinch driver. The error bars in the right panel have a length of $\sigma \approx 0.5$, as shown in Figs. \ref{fig:ringfourier} and \ref{fig:ringTCfourier}, and are omitted for easier visualization.
    }
    \label{fig:periods}
\end{figure*}

\begin{figure*}[t]
    \centering
    \resizebox{\hsize}{!}{
    \includegraphics[trim={2.cm 1.07cm 0.cm 0.cm},clip,scale=0.45]{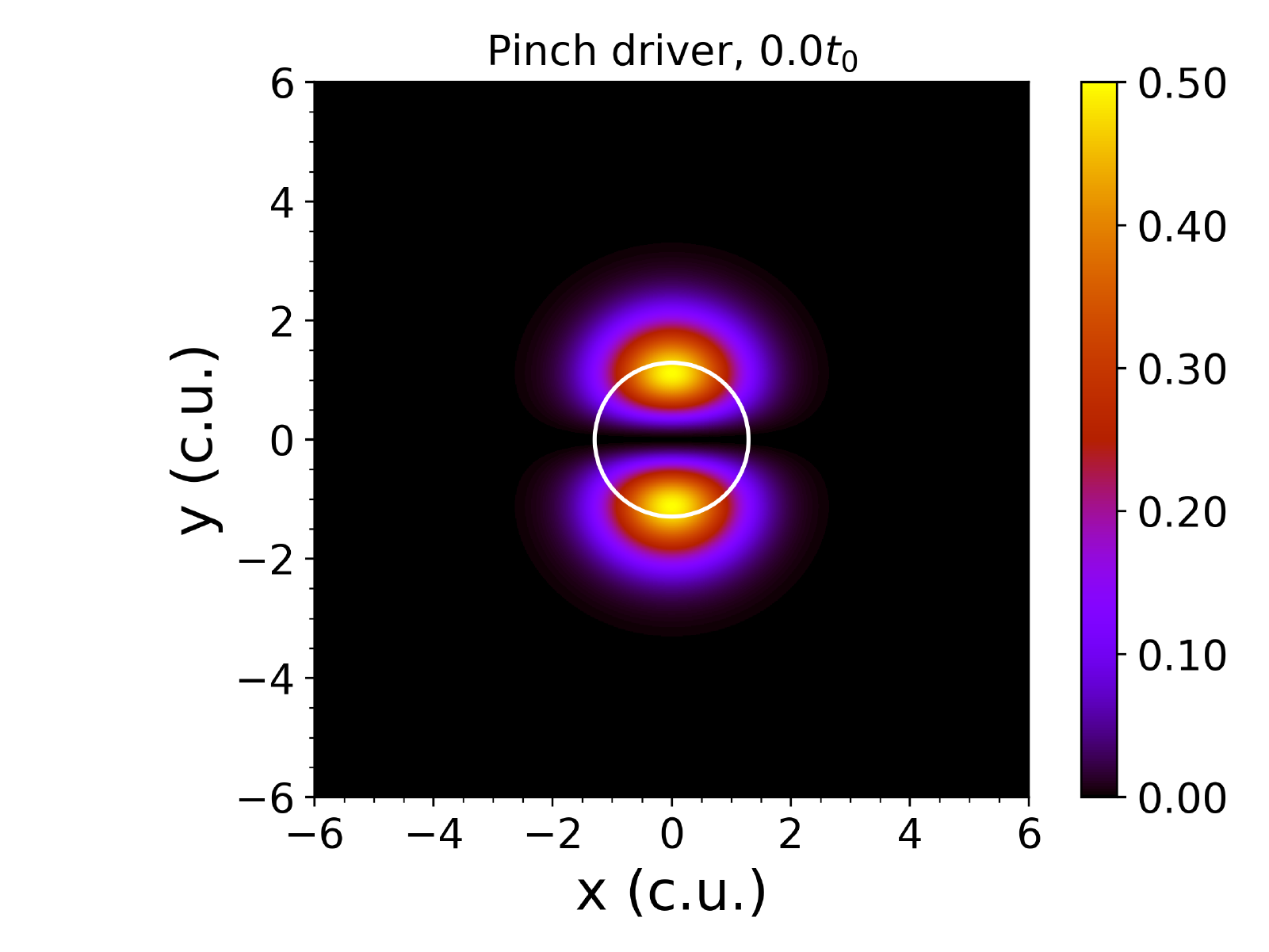}
    \includegraphics[trim={3.cm 1.07cm 0.cm 0.cm},clip,scale=0.45]{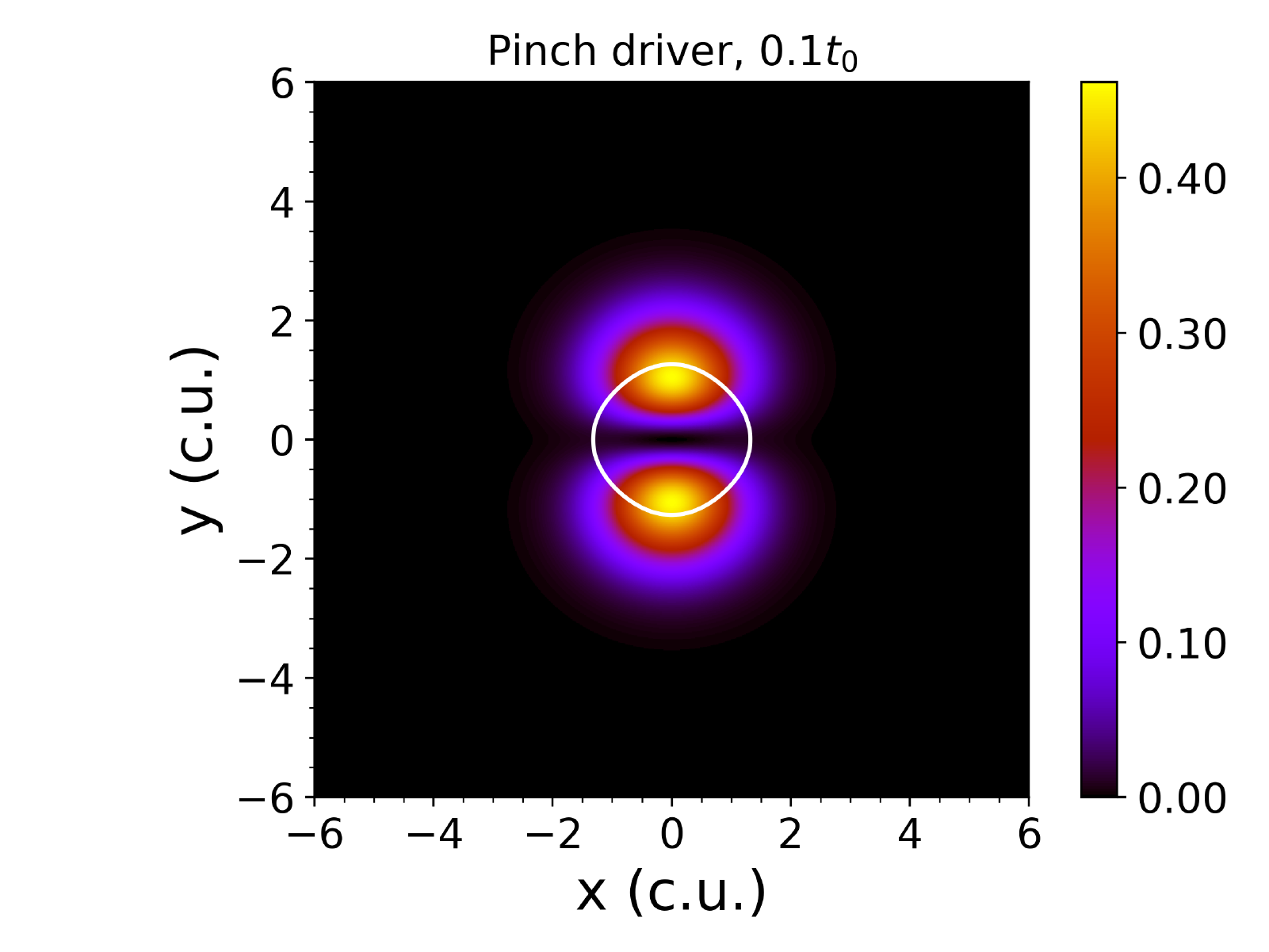}
    \includegraphics[trim={3.cm 1.07cm 0.cm 0.cm},clip,scale=0.45]{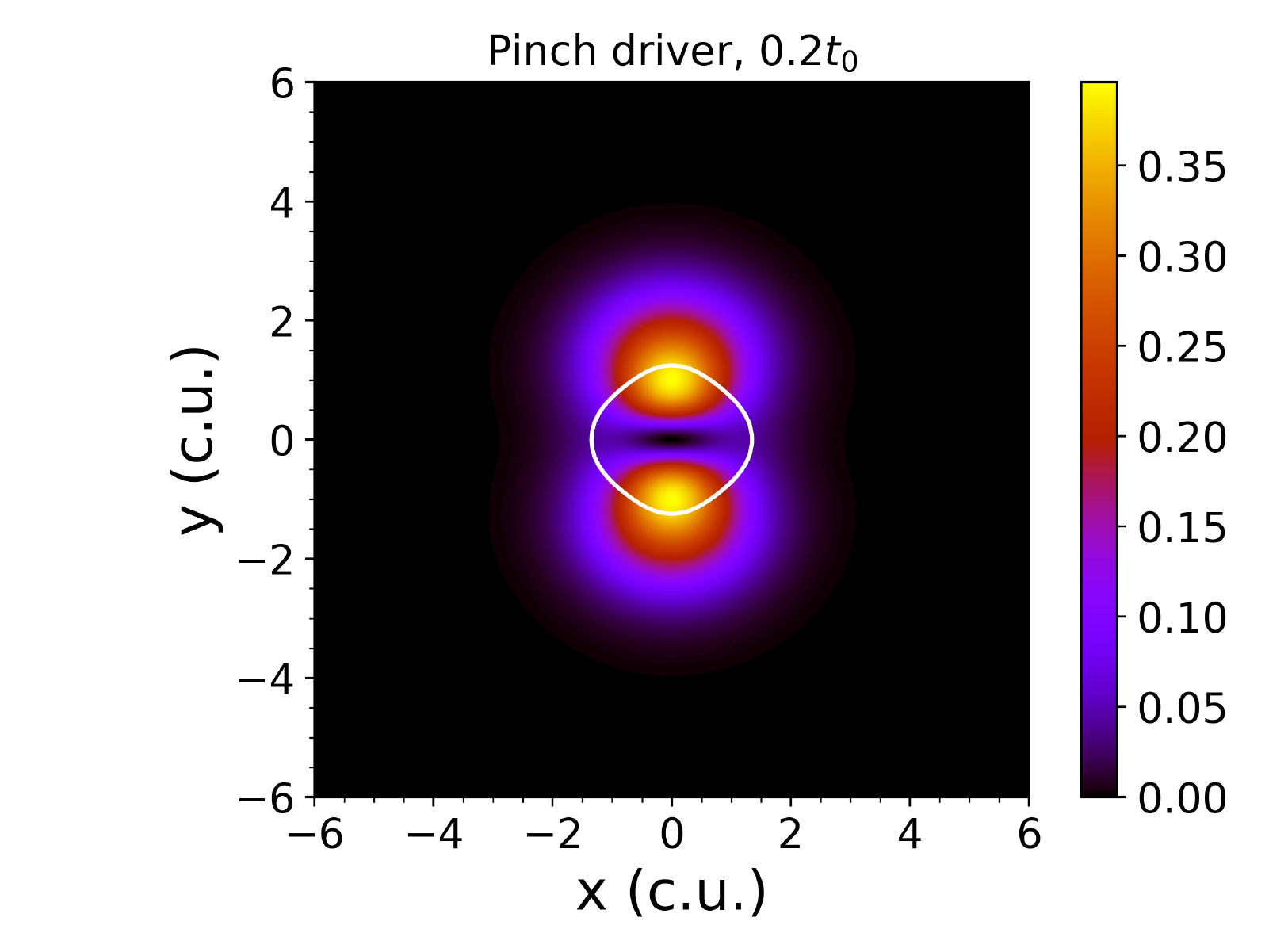}
    \includegraphics[trim={3.cm 1.07cm 0.cm 0.cm},clip,scale=0.45]{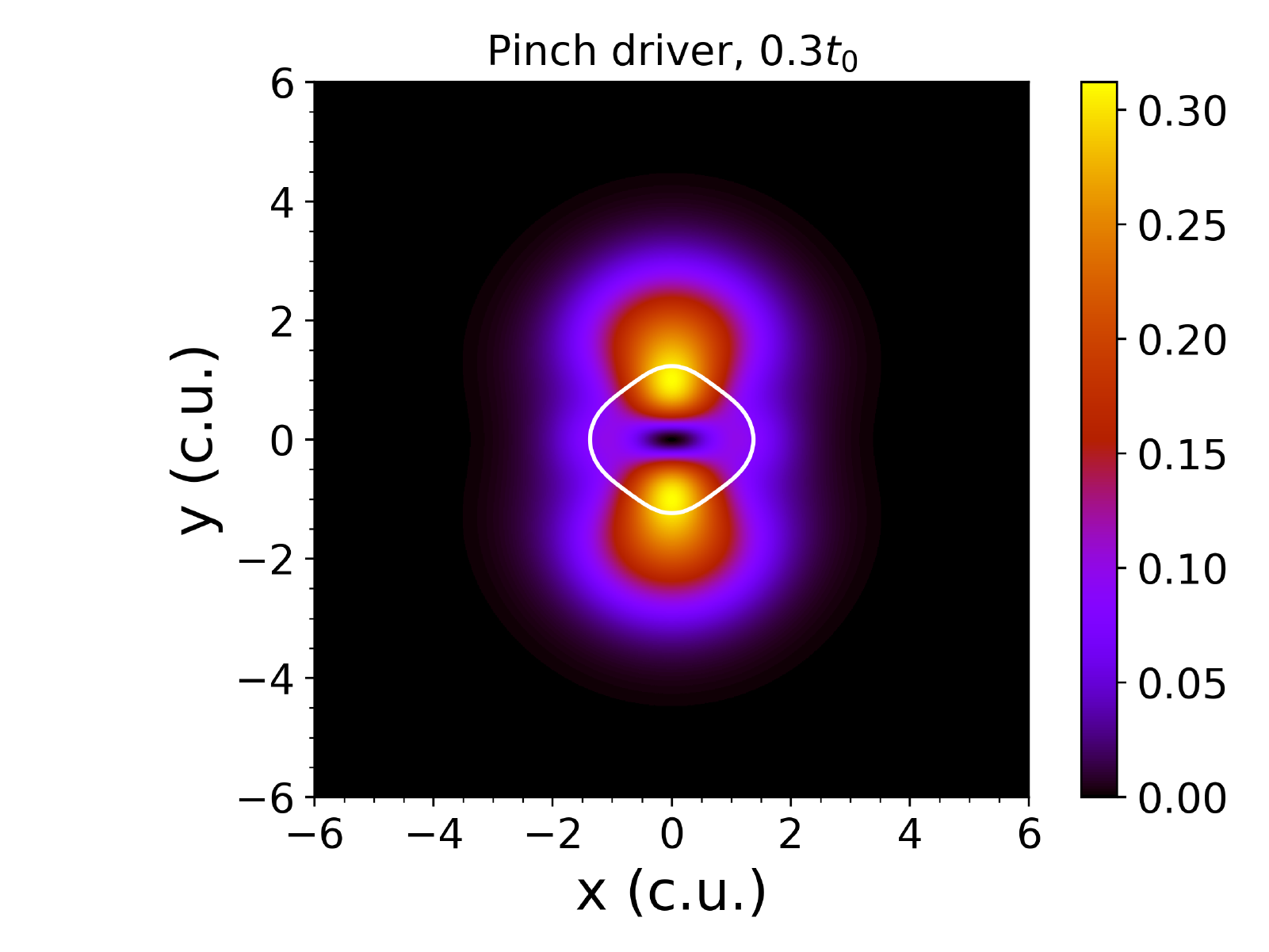}
    }
    \resizebox{\hsize}{!}{
    \includegraphics[trim={2.cm 0.cm 0.cm 0.cm},clip,scale=0.45]{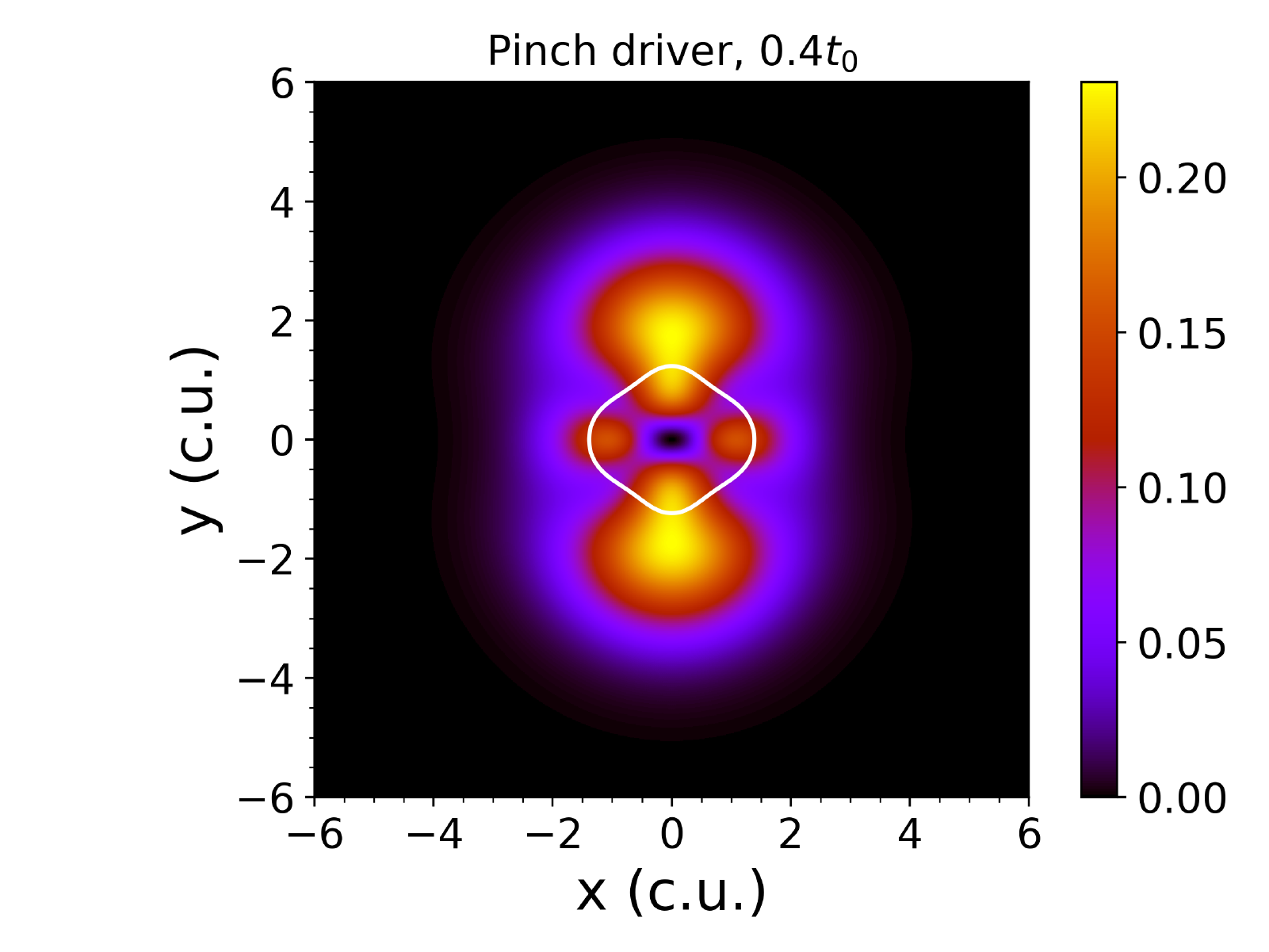}
    \includegraphics[trim={3.cm 0.cm 0.cm 0.cm},clip,scale=0.45]{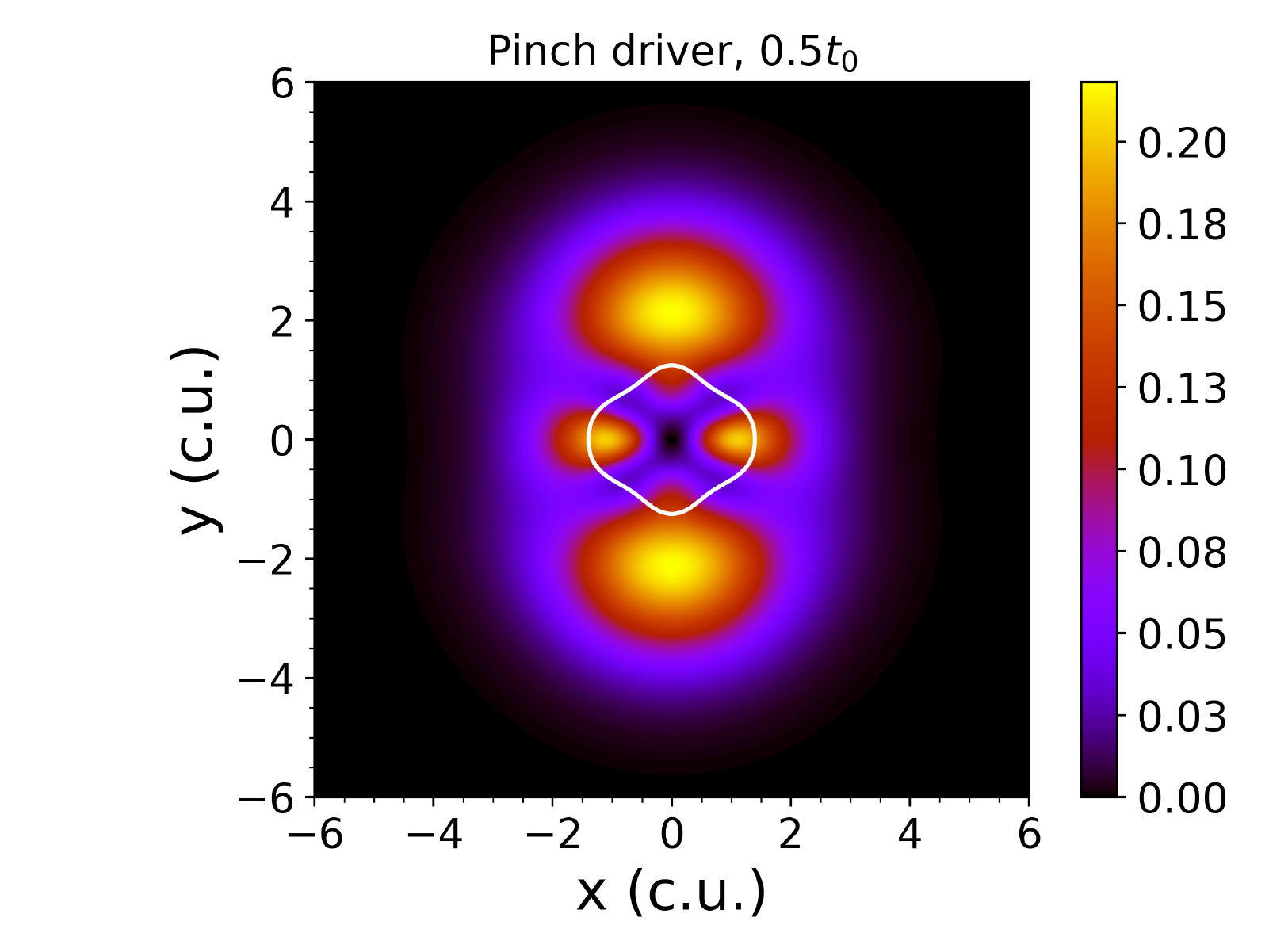}
    \includegraphics[trim={3.cm 0.cm 0.cm 0.cm},clip,scale=0.45]{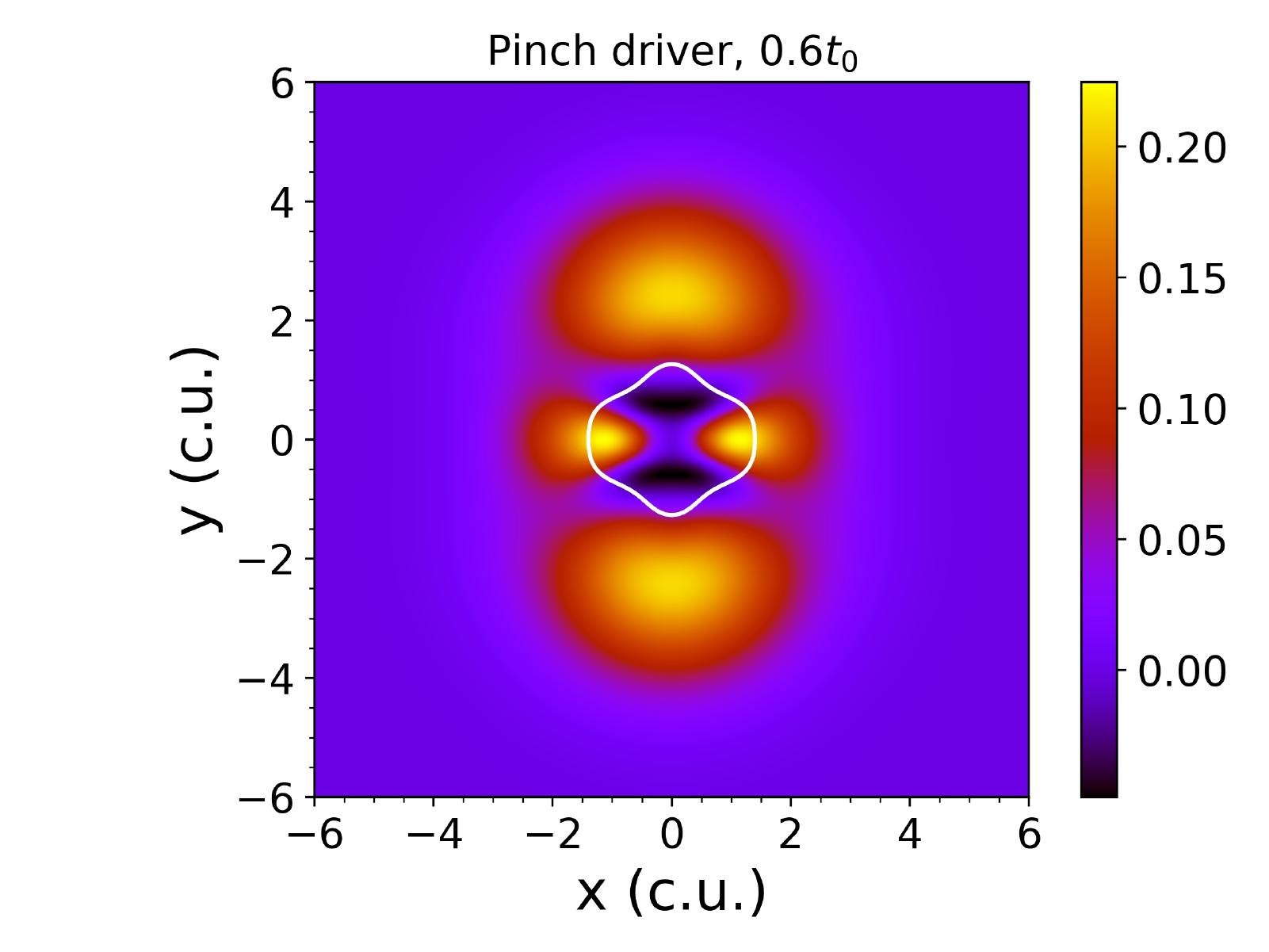}
    \includegraphics[trim={3.cm 0.cm 0.cm 0.cm},clip,scale=0.45]{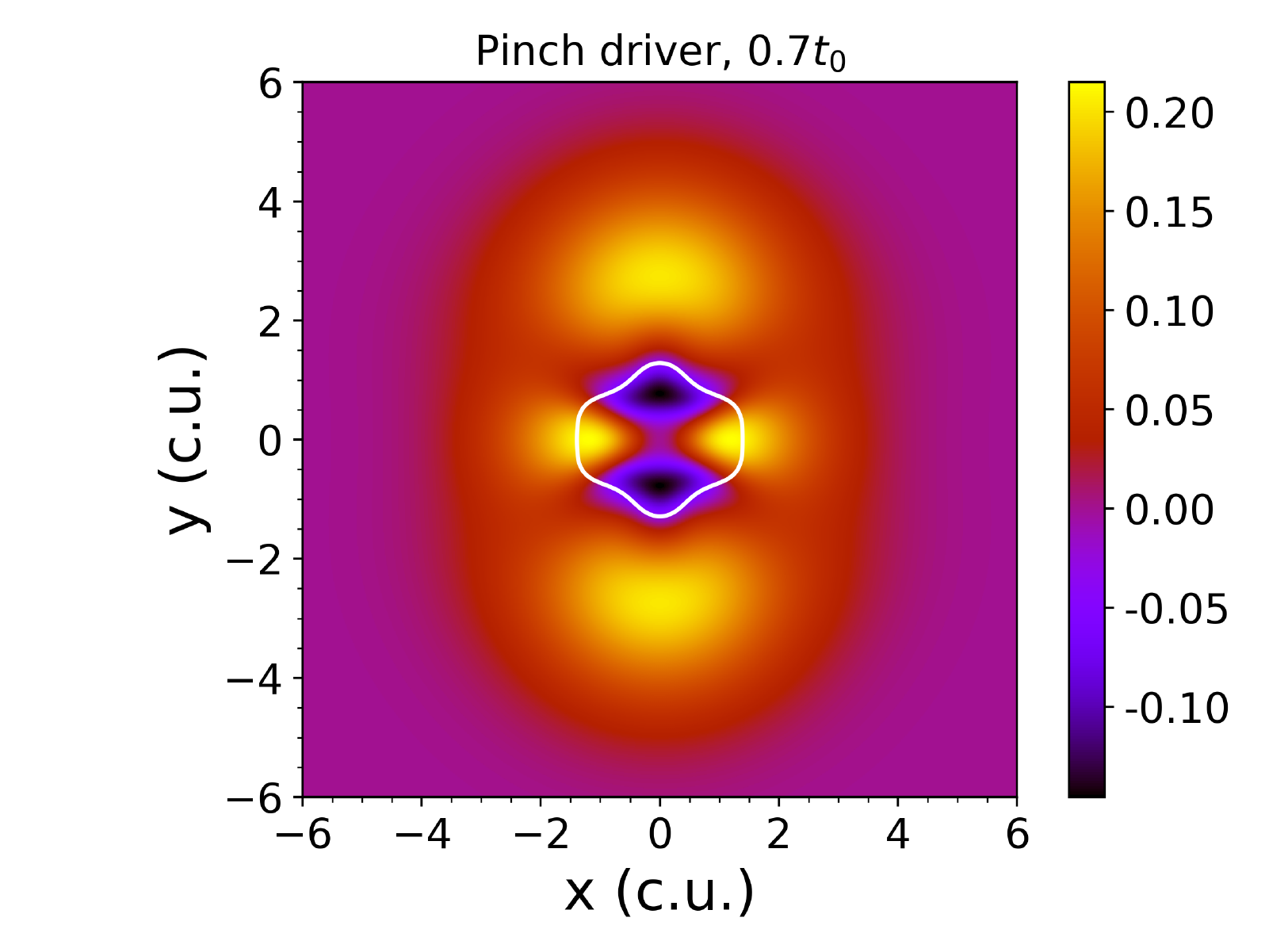}
    }
    \caption{Similar to Fig. \ref{fig:ringdriver} for the initial evolution of the Pinch driver.}
    \label{fig:pinchdriver}
\end{figure*}

\begin{figure}[t]
    \centering
    \includegraphics[trim={0.cm 0.cm 0.cm 0.cm},clip,scale=0.5]{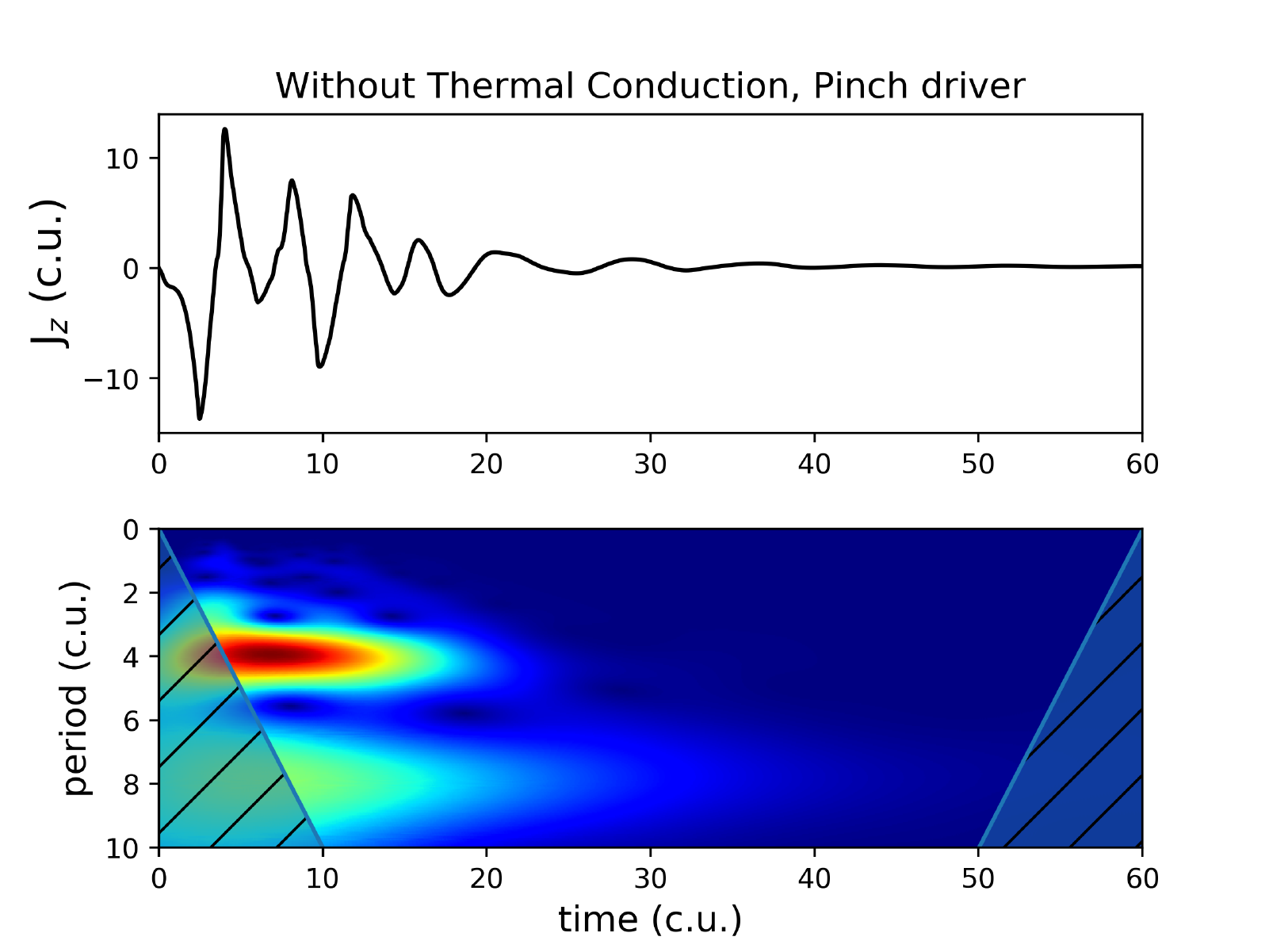}
    \includegraphics[trim={0.cm 0.cm 0.cm 0.cm},clip,scale=0.45]{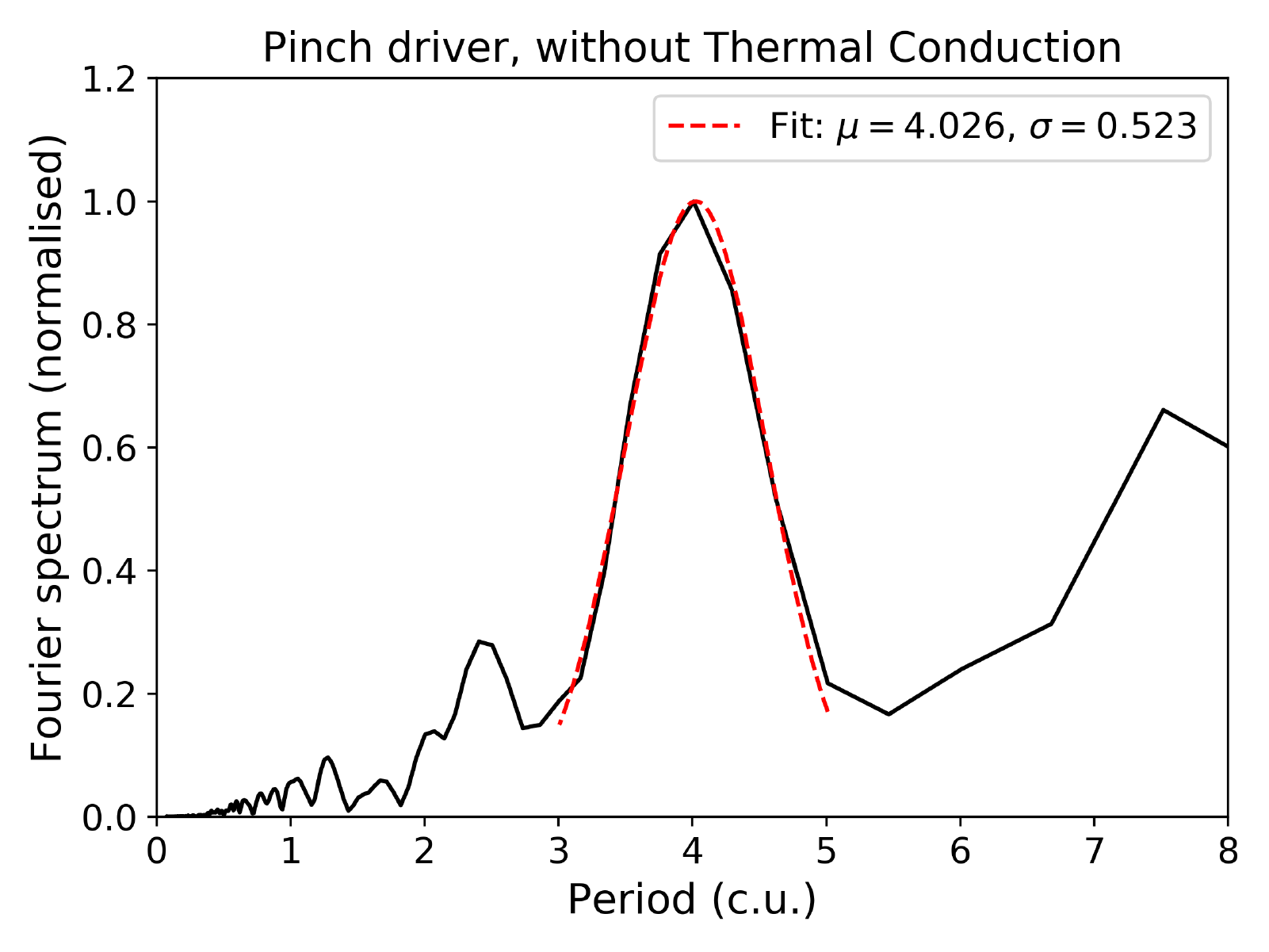}
    \caption{The top two panels show The oscillating $J_z$ signal at the null point and its respective wavelet profile for the Pinch driver. The bottom panel shows the normalized Fourier power spectrum and a Gaussian fit (red dashed line), with respect to the period, for the $J_z$ signal. The location of the maximum $\mu$ and standard deviation $\sigma$ for the period is given in the legend.}
    \label{fig:pinchwavelet}
\end{figure}

\begin{figure*}[t]
    \centering
    \resizebox{\hsize}{!}{
    \includegraphics[trim={1.0cm 1.07cm 0.cm 0.cm},clip,scale=0.45]{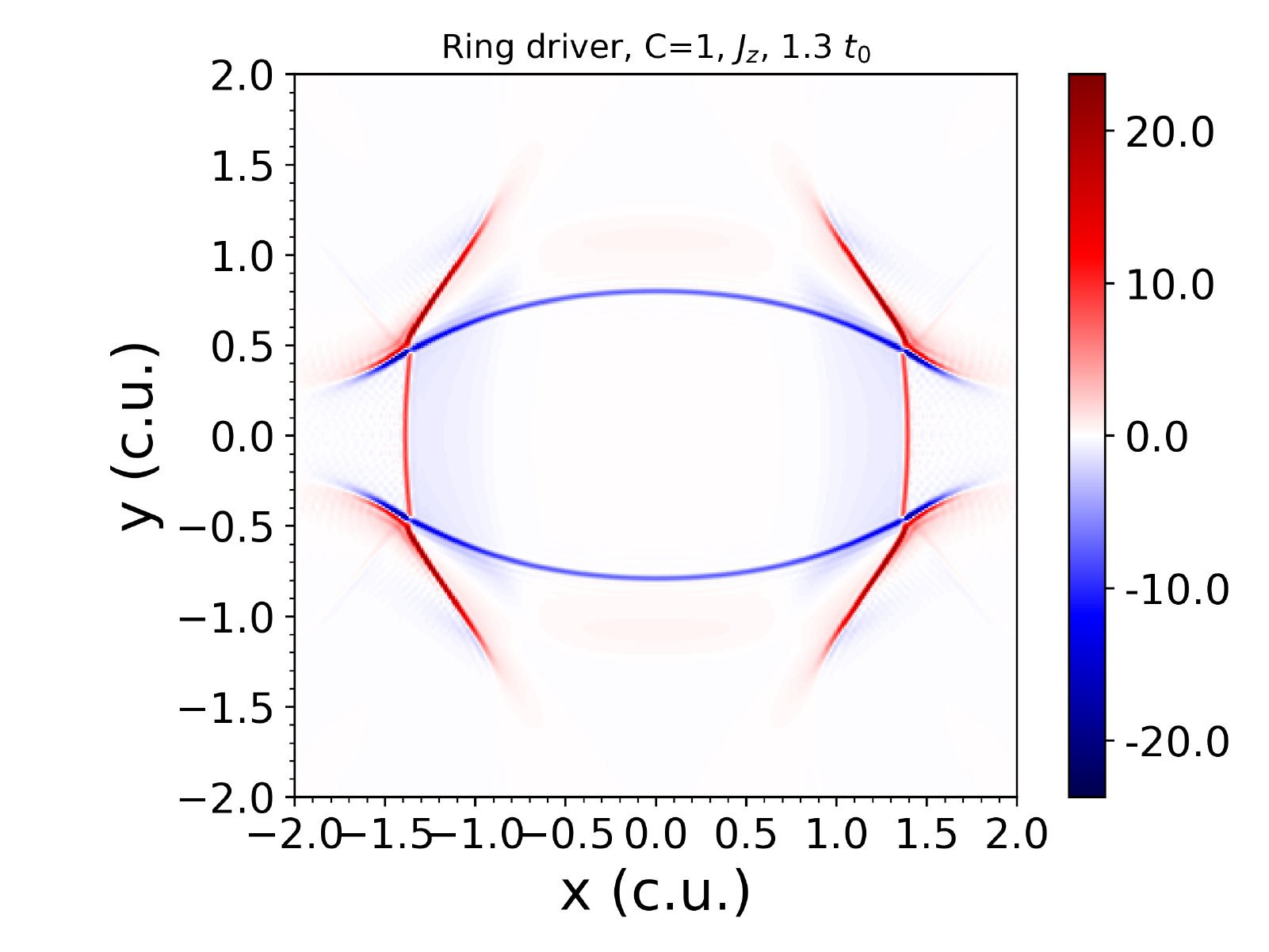}
    \includegraphics[trim={2.3cm 1.07cm 0.cm 0.cm},clip,scale=0.45]{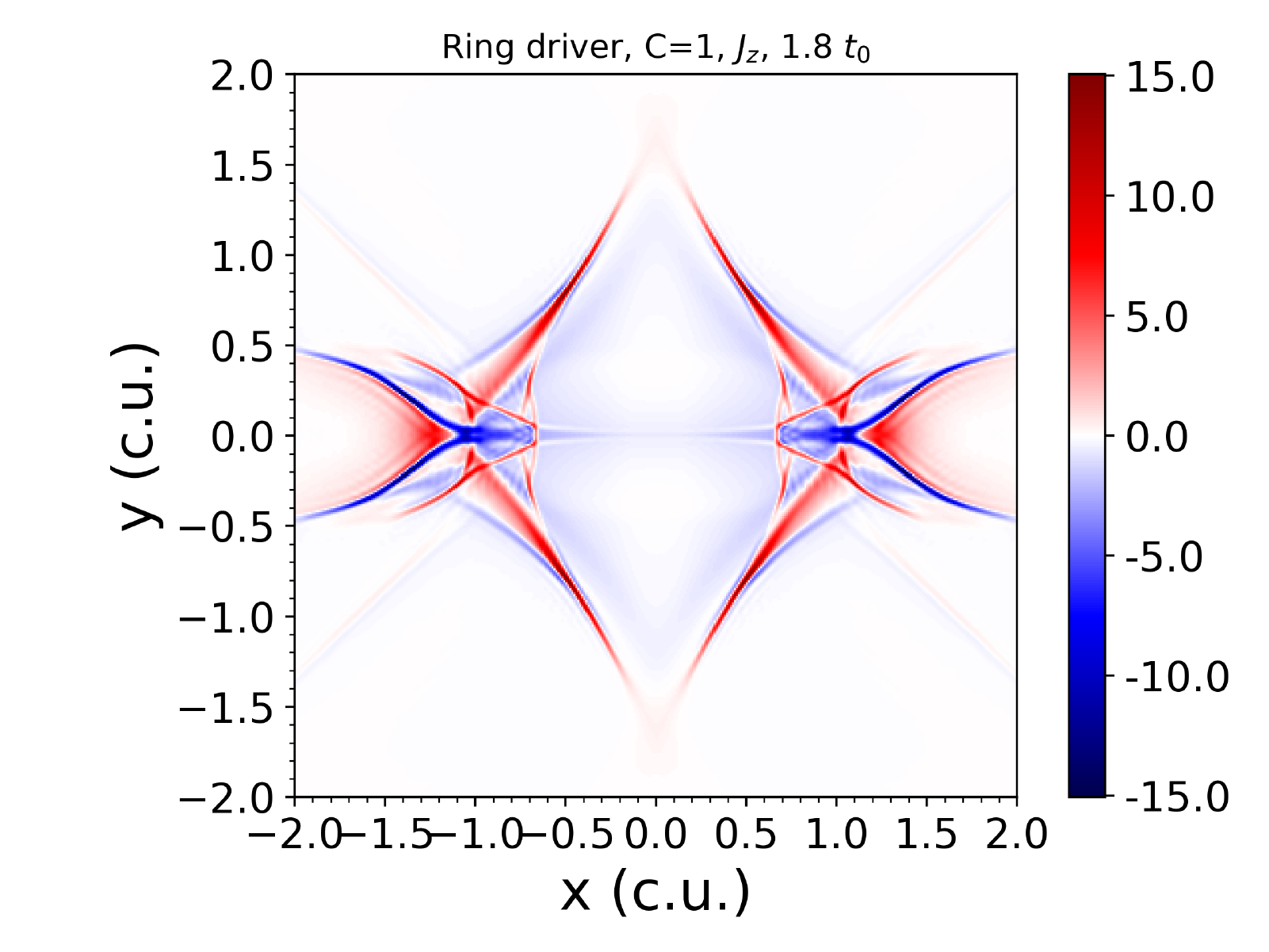}
    \includegraphics[trim={2.3cm 1.07cm 0.cm 0.cm},clip,scale=0.45]{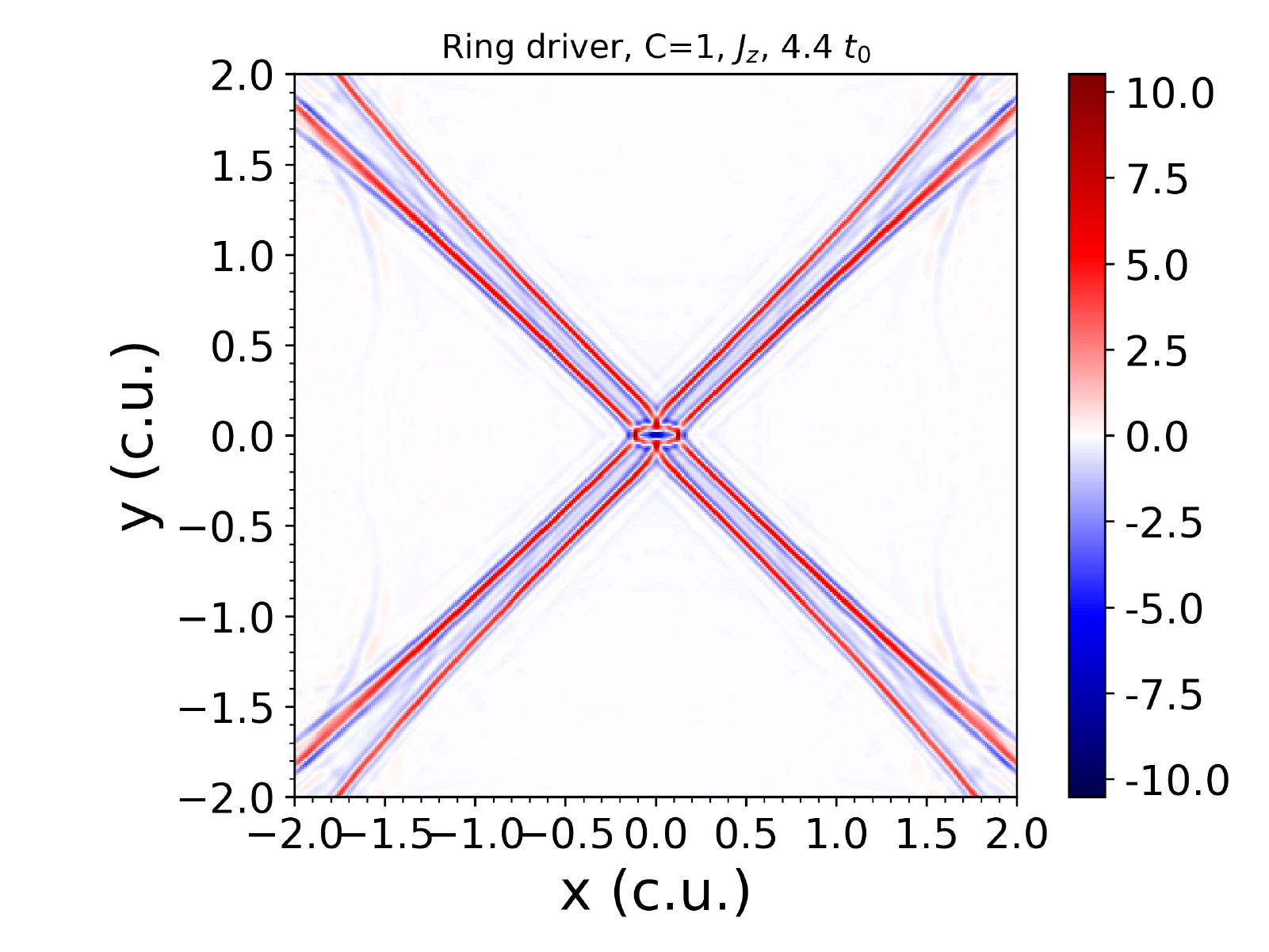}
    \includegraphics[trim={2.3cm 1.07cm 0.cm 0.cm},clip,scale=0.45]{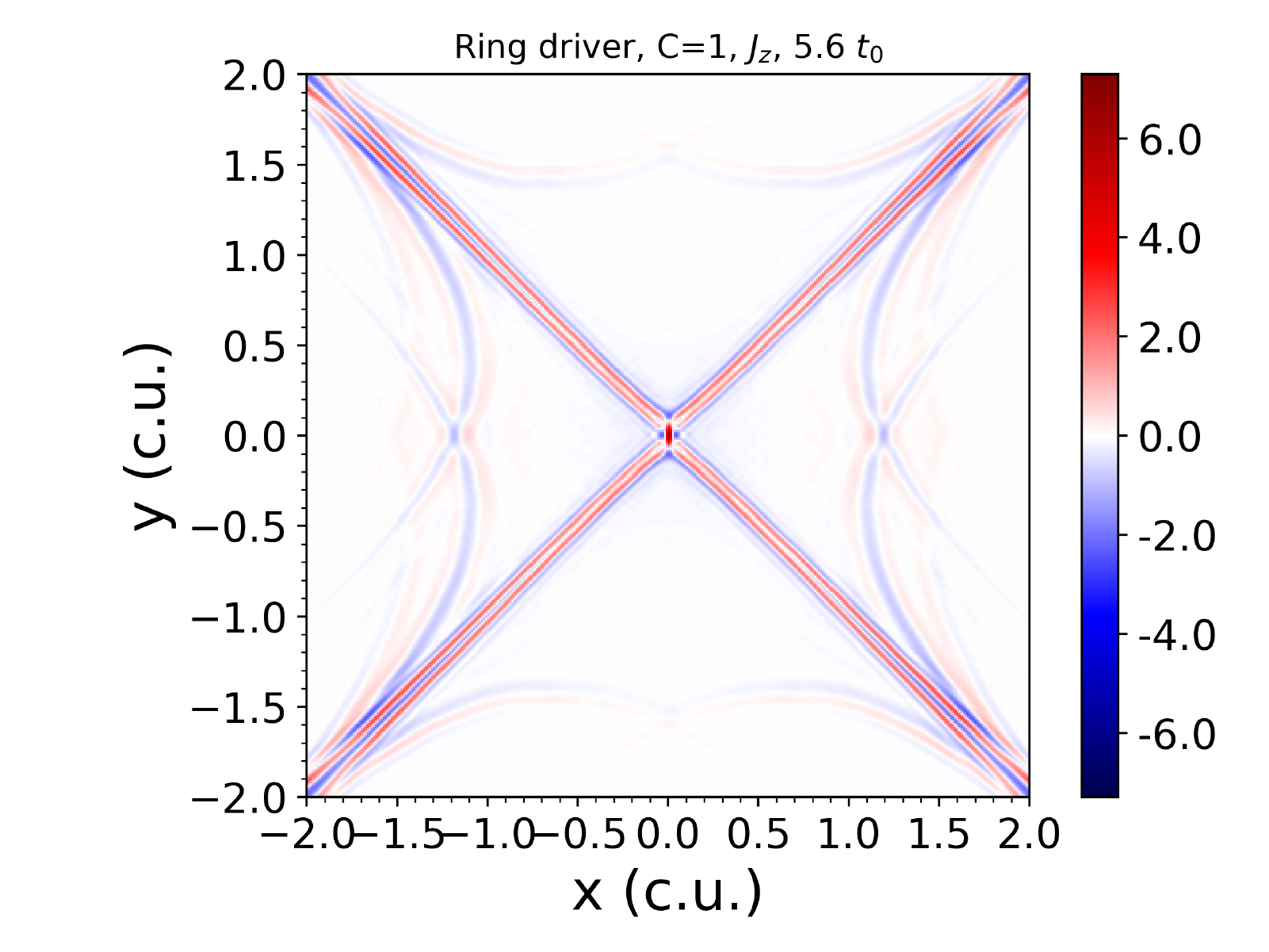}
    }
    \resizebox{\hsize}{!}{
    \includegraphics[trim={1.0cm 0.cm 0.cm 0.cm},clip,scale=0.45]{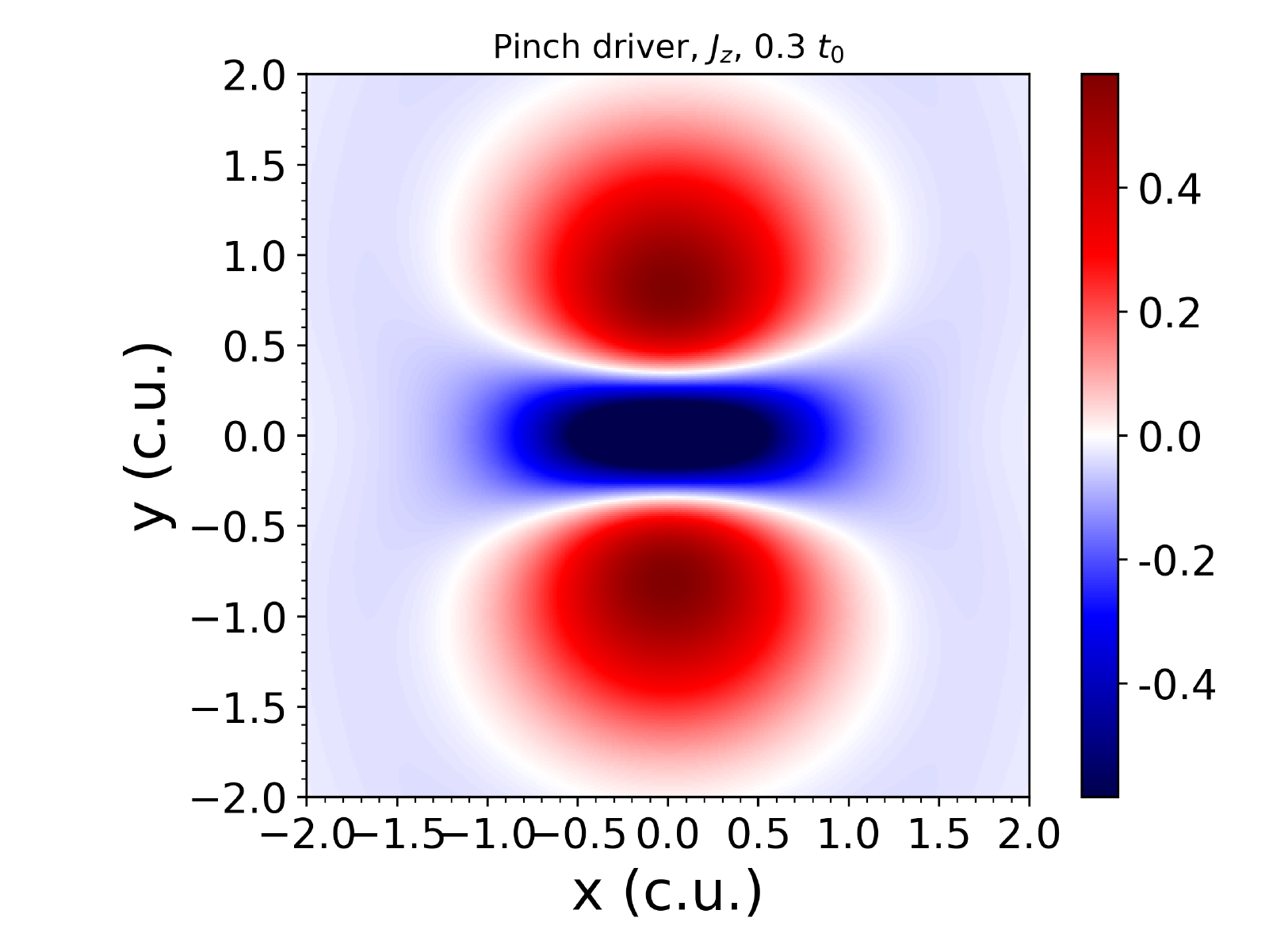}
    \includegraphics[trim={2.3cm 0.cm 0.cm 0.cm},clip,scale=0.45]{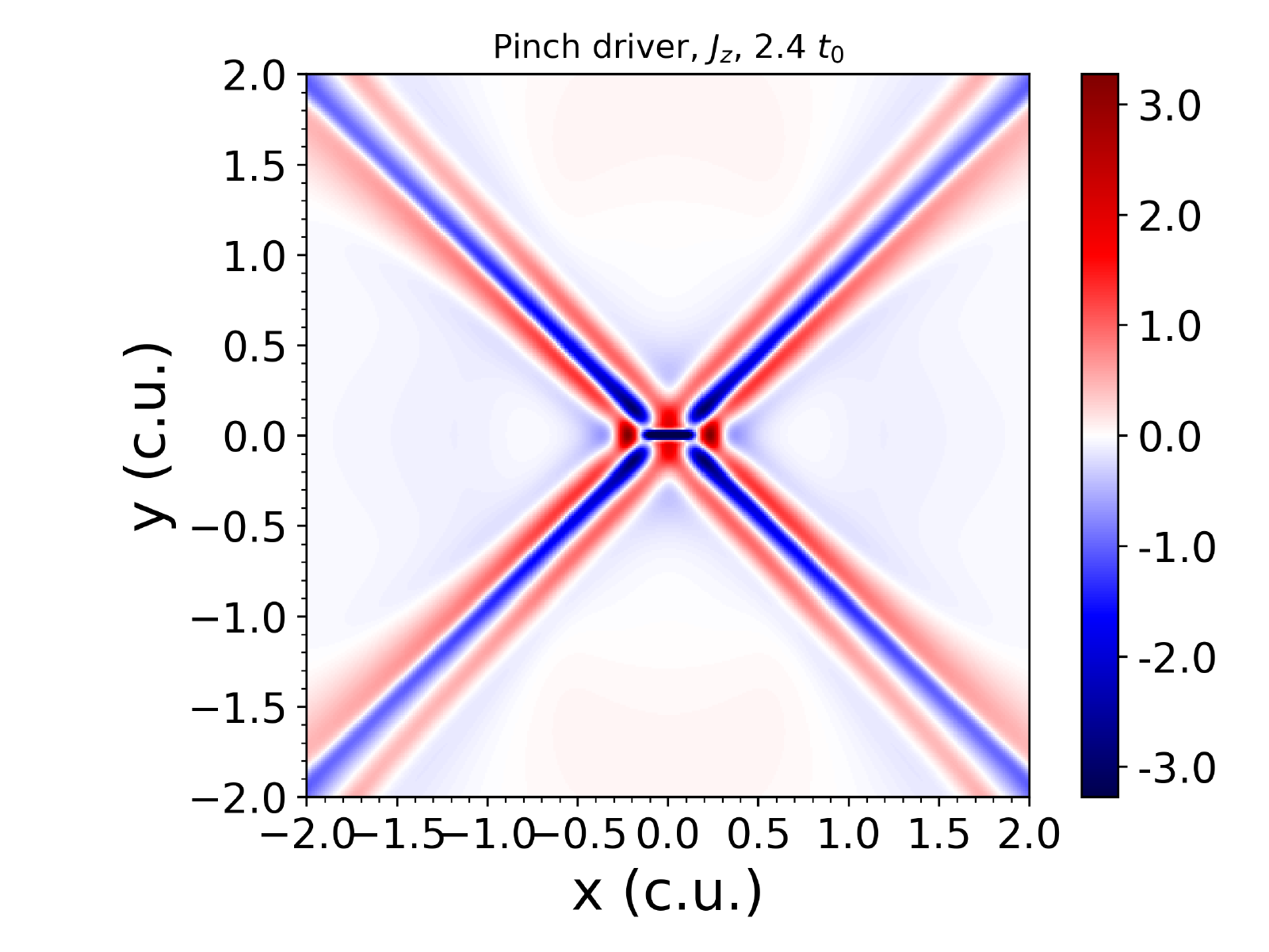}
    \includegraphics[trim={2.3cm 0.cm 0.cm 0.cm},clip,scale=0.45]{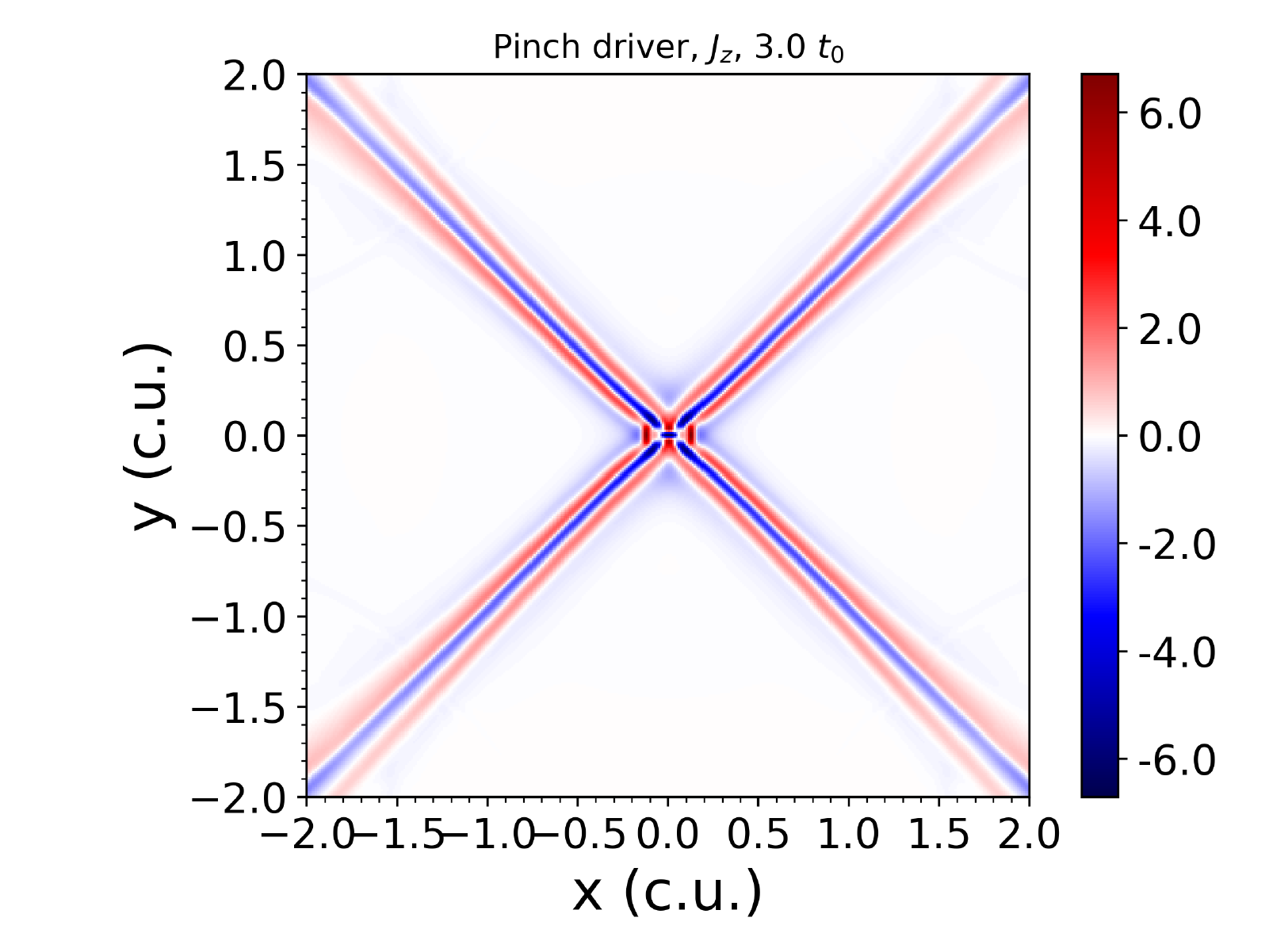}
    \includegraphics[trim={2.3cm 0.cm 0.cm 0.cm},clip,scale=0.45]{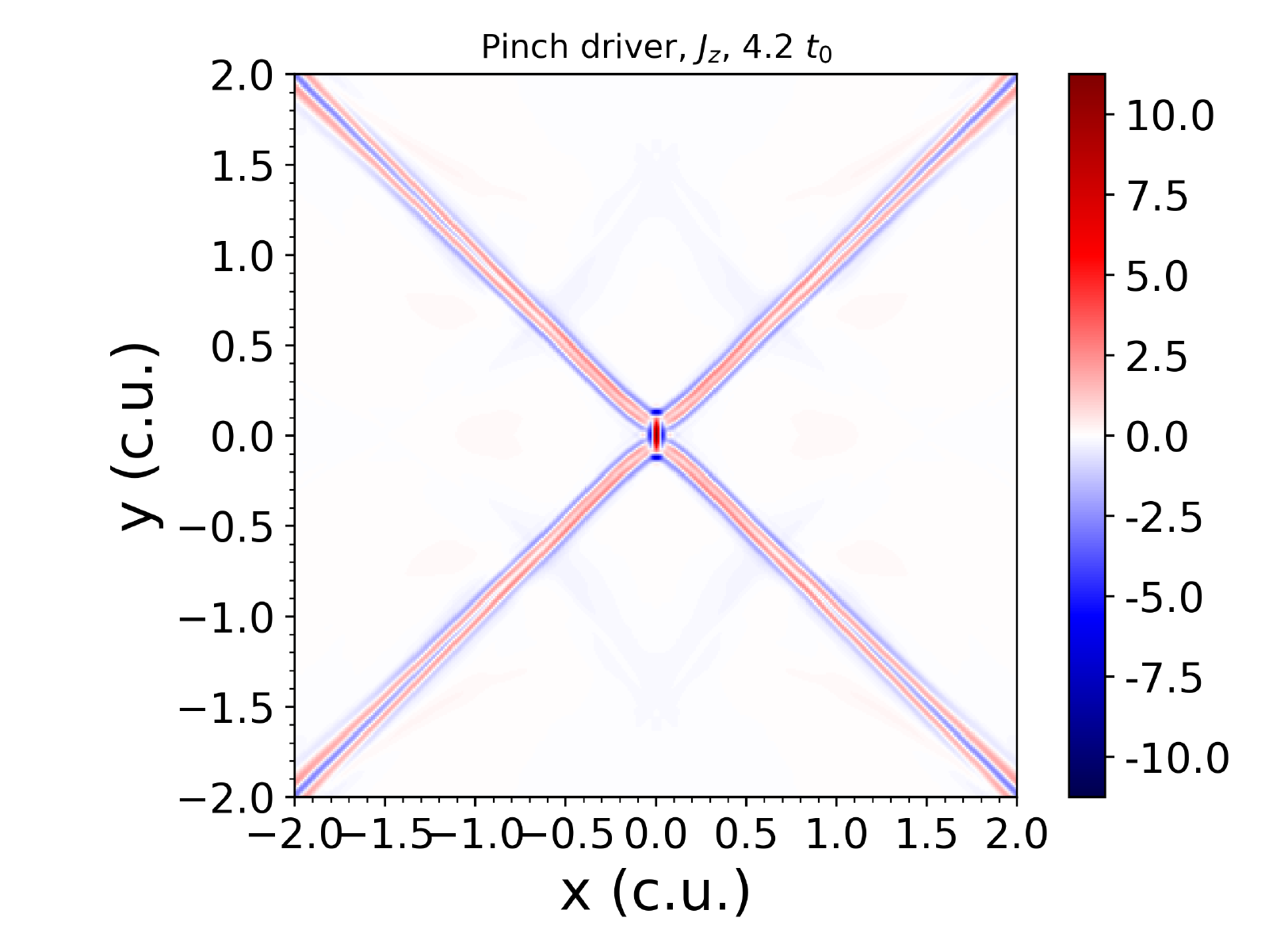}
    }
    \caption{2D profiles of the $J_z$ current density around the null point. The top panels show snapshots for time close to the first horizontal current sheet before ($t=1.8\,t_0$) and after its ($t= 4.4\,t_0$) collapse, and the first vertical current sheet ($t=5.6\,t_0$) for the Ring driver with $C=1$ amplitude. The bottom panels show the corresponding snapshots for a time close to the
    first horizontal ($t=0.3\,t_0$ and $t=3\,t_0$) and vertical current sheet ($t= 4.2\,t_0$), for the Pinch driver.}
    \label{fig:current}
\end{figure*}

\subsection{X-point and initial conditions}
Following the work of \citet{Karampelas2022a}, our basic setup consists of a 2D magnetic X-point, described by the following magnetic field in Cartesian coordinates:
\begin{equation}\label{eq:magneticfield}
    \mathbf{B} = \frac{B_0}{L_0}\left(y,x,0\right),
\end{equation}
where $B_0$ is the characteristic field strength and $L_0$ is the length scale for magnetic field variations. Our setup has a uniform initial density equal to $\rho_0 = 10^{-12}$\,kg m$^{-3}$ and a uniform initial temperature of $T_0 = 1$\,MK. A uniform equilibrium density was chosen in order to prevent the effects of phase mixing (e.g. \citealt{heyvaerts1983}), while the uniform density and temperature profiles lead to a uniform initial sound speed ($V_S$). The uniform $V_S$ contrasts with the Alfv\'{e}n speed $V_A$, which increases away from the X-point. 
The two different profiles for the Alfv\'{e}n and sound speeds lead to an area where $V_A = V_S$, which is the location of the equipartition layer. The importance of this layer lies in that the coupling between fast and slow waves is most efficient there. In our setup, this layer is initially a circle, with a radius $\sqrt{2\,/\,\gamma}\approx 1.095$ times smaller than that of the plasma-$\beta=1$ layer. Thus, the equipartition layer describes the limit between a low-$\beta$ environment, as in most parts of the solar corona and a high-$\beta$ environment, as is the case in the vicinity of null points. Fig. \ref{fig:profileB} depicts the initial magnetic field lines (in gray) of the X-point, alongside the initial equipartition layer (magenta line) and the separatrices (in black) that separate the regions with opposing magnetic field polarity.

As we aim to study the phenomenon of oscillatory reconnection, we need to perturb our system from its equilibrium state, in order to start the process. To that end, we consider an initial fast magnetoacoustic wave pulse of the form:
\begin{eqnarray}
v_{\perp}(t=0) &=& \frac{C}{0.2 \sqrt{2\pi}}\exp\left( -0.5 \frac{(r-5)^2}{0.2}\right),\label{eq:ringvp} \\
v_{\parallel}(t=0)&=&0,\label{eq:ringvl}
\end{eqnarray}
where $v_{\perp} = (\mathbf{v}\times \mathbf{B})\cdot\hat{\mathbf{z}}$ is a quantity related to the velocity component perpendicular to the magnetic field lines, $v_{\parallel}=\mathbf{v}\cdot\mathbf{B}$ is related to the velocity component parallel to the magnetic field lines, and $C$ (in code units) is the amplitude of the perturbation. This annulus-shaped velocity perturbation, or Ring driver, has been used in the past as a triggering mechanism for oscillatory reconnection \citep[e.g.][]{McLaughlin2009,Karampelas2022a}. In this work we will focus on this driver, by performing a parameter study by employing initial perturbations of different amplitudes (i.e., $C=0.5,\,1,\,2,\,3,\,4,\,5$, and $6$). Note that the case with an amplitude of $C=1$ is equivalent to that used in \citet{Karampelas2022a}.

We will also consider a second type of velocity pulse, to perturb our system in a different way. To that end, we devise what can be described as a pinch-like velocity pulse, or Pinch driver, by introducing a $v_y$ velocity:
\begin{equation}\label{eq:pinchvy}
v_y = -1.084\,y\,\exp{\left( \frac{r^2}{0.5\sqrt{2\pi}} \right)}\;.
\end{equation}
The initial 2D profiles of the Ring driver for $C=1$ and the Pinch driver are shown in Fig. \ref{fig:profiledriver}, where the $v_{\perp}$, $v_{\parallel}$, $v_x$ and $v_y$ components of each driver are shown.


\subsection{Numerical Scheme}
We solve the 2D compressible resistive MHD equations in cartesian coordinates with the use of PLUTO \citep{mignonePLUTO2007, mignonePLUTO2012}, a finite-volume, shock-capturing code. We perform the time integration using the third-order Runge-Kutta method and the spatial integration using the fifth-order monotonicity preserving scheme (MP5) and the total variation diminishing Lax–Friedrich (TVDLF) solver. To treat the nonzero divergence of the magnetic field, we employ the Constrained Transport method.

The 2D compressible and resistive MHD equations, in the absence of gravity, are given below:
\begin{equation}
  \frac{\partial\rho}{\partial t} + \nabla \cdot (\rho \mathbf v) = 0\,,
\end{equation}
\begin{equation}
 \rho \left[ \frac{\partial\mathbf v }{\partial t} + (\mathbf v \cdot \nabla)\mathbf v \right] = - \nabla p + \frac{1}{\mu}\left(\nabla\times\mathbf{\mathbf B}\right) \times \mathbf B,
\end{equation}
\begin{equation}
    \rho \left[ \frac{\partial \epsilon}{\partial t} + (\mathbf v \cdot \nabla)\epsilon \right] = -p \nabla \cdot \mathbf v + \frac{1}{\sigma} |\mathbf J|^2 + \nabla \cdot \mathbf{F}_c,
\end{equation}
\begin{equation}
  \frac{\partial\mathbf B}{\partial t} = \nabla \times (\mathbf v \times \mathbf B) + \eta\nabla^2 \mathbf B
\end{equation}
where $\rho$, $p$, and $\mathbf v$ are the density, plasma pressure, and velocity, respectively. The electric current is defined as $\mathbf J = \nabla\times\mathbf B /\mu$, while for the magnetic field the solenoidal condition $\nabla\cdot\mathbf B = 0$ must also hold. In the above equations, we can also see the magnetic permeability $\mu = 4 \pi \times 10^{-7}$\,H m$^{-1}$, the electrical conductivity $\sigma = 1/(\mu\,\eta)$ ($\eta$ is the magnetic diffusivity),  $\epsilon = p/\left[\rho (\gamma-1)\right]$ is the specific internal energy density, and $\gamma = 5/3$ is the ratio of the specific heats.  $\nabla \cdot \mathbf{F}_c$ is a source term that introduces anisotropic thermal conduction into our set of MHD equations. $\mathbf{F}_c$ is an expression that varies between the classical ($\mathbf{F}_{class}$) and saturated thermal conduction ($F_{sat}$), with the effects of saturation being considered for very large temperature gradients. 

After defining the unit vector in the direction of the magnetic field $\mathbf{\hat{b}} = \mathbf{B}/|\mathbf{B}|$, the expressions for $\mathbf{F}_c$, $\mathbf{F}_{class}$  and $F_{sat}$ are calculated as:
\begin{equation}
    \mathbf{F}_{c} = \frac{F_{sat}}{F_{sat}+|\mathbf{F}_{class}|} \mathbf{F}_{class}   
\end{equation}
\begin{equation}
    \mathbf{F}_{class} = \kappa_{\parallel} \mathbf{\hat{b}} \left(\mathbf{\hat{b}} \cdot \nabla T \right) + \kappa_{\perp} \left[ \nabla T - \mathbf{\hat{b}}\left( \mathbf{\hat{b}} \cdot \nabla T \right) \right]
\end{equation}
\begin{equation}
    F_{sat} = 5\,\phi\,\rho\, V_{S,iso}^3,
\end{equation}
where $V_{S,iso}$ is the isothermal sound speed and $\phi$ is a free code parameter (whose default value is $0.3$). In the hydrodynamical limit (zero magnetic field), $\mathbf{F}_{c}$ reduces to $\mathbf{F}_{c}= \kappa_{\parallel} \, \nabla T$. The values for the parallel and perpendicular coefficients (in J\,s$^{-1}$\,K$^{-1}$\,m$^{-1}$) are calculated from the Spitzer conductivity \citep{Orlando2008ApJ} and are given below:
\begin{eqnarray}
\kappa_{\parallel,ph} &=& 5.6 \times 10^{-12}\, T_{ph}^{\frac{5}{2}},\label{eq:kpar}\\
\kappa_{\perp,ph} &=& 3.3 \times 10^{-21}\, \frac{n_{H,ph}^2}{\sqrt{T_{ph}}B_{ph}^2},\label{eq:kprp}
\end{eqnarray}
where $\kappa_{\parallel,ph}$, $\kappa_{\perp,ph}$ and the hydrogen number density $n_{H,ph}$, temperature $T_{ph}$ and magnetic field $B_{ph}$ are all given in physical units. For a more detailed description of the implementation of anisotropic thermal conduction, the readers are referred to \citet{mignonePLUTO2012}. 

In the rest of the manuscript, all the quantities are expressed in code units $U = U_{ph}\, U_0^{-1}$, derived from the physical quantities divided by the normalization units $U_0$. The constants $U_0$ are characteristic values, appropriate for the conditions in the solar corona. We consider the unit length $L_0=1$\,Mm, the unit density $\rho_0 = 10^{-12}$\,kg m$^{-3}$, the unit velocity $v_0$ (equal to $V_S / \sqrt{\gamma}$ for coronal plasma at $1$\,MK), and the unit temperature $T_0=1\,$MK. The characteristic magnetic field and unit time are respectively $B_0 = \sqrt{\mu \rho_0 v_0^2} = 1.44$\,G and $t_0 = L_0 / v_0 = 7.78$\,s.

We consider the magnetic diffusivity in code units $\eta = R_m^{-1} = 10^{-5}$, where $R_m$ is the magnetic Reynolds number, assuming that the typical length and velocity scales of our system are respectively $L_0$ and $v_0$. Present in our code is also the `effective' numerical diffusivity, which has been estimated to be between $10^{-6}$ and $10^{-5}$ in code units, through a parameter study. The effects of numerical dissipation for our resolution prevent us from considering higher and more realistic values of $R_m$, thus making it harder to get quantitatively accurate results.

\subsection{Boundary Conditions, Damping Regions, and the Solenoidal Constraint}\label{sec:scheme}
Our setup consists of a square domain containing the 2D magnetic X-point. A structured uniform grid is used, extending to $(x,y)\in (-10,10)$\, in code units, with a resolution of $1801\times 1801$ grid points.

We consider effectively reflective boundaries for the velocity components across the boundaries, by fixing them to zero there, ensuring that we have no flows across the boundary that would disrupt the initial equilibrium. We also take gradient boundary conditions for the magnetic field (of the form $d_i - d_{i-1} = d_{i-1} - d_{i-2}$), in order to prevent any artificially developed currents at the boundaries. Finally, we fix the values for the pressure and density  to their initial conditions at the boundaries, which prevents the accumulation of heat at the boundaries when thermal conduction is switched on. 

In order to prevent the waves reflected at the boundaries from returning and affecting the solution at the null point, we employ a few numerical strategies to dampen these unwanted perturbations. Firstly, and following the approach used in \citet{Karampelas2022a}, we reset the velocity components to their initial (zero) values for the region with radius of $r\geqslant 7$, at time $t_C=0.6\,t_0$ for the Ring drivers and time $t_C=2.5\,t_0$ for the Pinch driver. This allows us to deal quickly with the initial strong velocity perturbations that move away from the null point and toward the boundaries. 

The second step is to then define a region with radius $r\geqslant 3\,L_0$ where we use a combination of viscous and numerical dissipation in order to dampen the rest of the perturbations escaping the vicinity of the null point. For the viscous dissipation, the viscosity coefficient in code units is given by the equation:
\begin{equation}\label{eq:nd}
    n_{visc}= A_{visc} +  A_{visc}\tanh(r - 5), \qquad t>t_C,
\end{equation}
where $A_{visc}=0.1$ and $t_C=0.6$ for all the Ring drivers and $A_{visc}=0.25$ and $t_C=2.5$ for the Pinch driver. For the numerical dissipation scheme, we take away kinetic energy from the waves by dividing the velocity components by the dissipation coefficient $n_d$ per each iteration. The coefficient is defined in code units as:
\begin{equation}\label{eq:nv}
    n_{d}= 1.0005 +  0.0005\tanh(r - 5), \qquad t>t_C,
\end{equation}
where $t_C=0.6$ and $t_C=2.5$ for the Ring and Pinch drivers, respectively. This new definition of the coefficients for the viscous and numerical dissipation leads to a weaker but more gradual dissipation of waves away from the null. This reduces the reflection from waves entering this artificial region, leading to a cleaner oscillating signal. These three different cases of the coefficients are presented in the two panels of Fig. \ref{fig:profilecoeff}.


\section{Results} \label{sec:results}

\subsection{Fast Magnetoacoustic Wave Pulses without Thermal Conduction} \label{sec:NoTC}
The main goal of this study is to derive the relation between the kinetic energy of the initial oscillation and the resulting oscillation period. The first step to achieve that is by expanding the results of \citet{Karampelas2022a} by performing a parameter study on the effects of the kinetic energy, or in our case the amplitude of the initial velocity pulse, on the oscillation of the $J_z$ current density. To that end, we start by taking the ring-type fast magnetoacoustic wave pulse, that was first introduced in \citet{McLaughlin2009}. As was already described in the previous section, we will be considering a range of amplitudes for that driver (i.e., $C=0.5,\,1,\,2,\,3,\,4,\,5$ and $6$), with the case of $C=1$ amplitude being equivalent to the one used by \citet{Karampelas2022a}. Note that in this first step, no thermal conduction is considered and also that the initial magnetic field, pressure, and density are kept the same across the different cases.

The initial velocity perturbation described by Eqns. (\ref{eq:ringvp}) and (\ref{eq:ringvl}) splits into two counter-propagating pulses, each with an amplitude of $0.5\,C$ and that travel in opposing directions from one another, toward and away from the null point. Our previously discussed dissipation scheme is used in order to remove the outwardly propagating pulse and then dissipate the artificially generated secondary perturbations caused by this resetting before they can reach the area of the null point. This process can be seen in the 2D profiles of the $v_{\perp}$ velocity in Fig. \ref{fig:ringdriver}. The inwards propagating pulse focuses at the X-point, due to refraction, forming compression and rarefaction shocks in the $y$-direction and $x$-direction respectively \citep[see also][for an analysis]{Gruszecki2011null}. The heating caused by the incoming shock fronts disturbs the surrounding medium and deforms the initially circular equipartition layer, while mode conversion takes place as the low-$\beta$ fast magnetoacoustic wave ($v_{\perp}$ pulse) turns into high-$\beta$ fast and slow waves ($v_{\perp}$ and $v_{\parallel}$ components, respectively) while crossing this layer \citep[e.g.][]{McLaughlin2006b,Karampelas2022a}.

Once the incoming pulse reaches the X-point, it perturbs it from its equilibrium, leading to an initial collapse and then to the creation of horizontal and vertical current sheets, corresponding to negative and positive peak values for the $J_z$ current density at the null point, which we describe as oscillatory reconnection \citep[][]{McLaughlin2009}. Focusing on the oscillating $J_z$ signal for the different setups, we perform a wavelet analysis \citep[see][]{1992AnRFM..24..395F,1998BAMS...79...61T} in order to study the periodicity of this signal, as seen in Fig. \ref{fig:ringwavelet}. For this analysis, we employ the Morlet wavelet for calculating the spectra. The computations were done with the use of the PyWavelets tool, which is an open-source wavelet transform software for Python.

The periodicity and decay rate of oscillatory reconnection for an X-point configuration in hot coronal plasma has already been studied for the Ring driver (with $C=1$ amplitude) in \citet{Karampelas2022a}. One of the main observations was the existence of two period bands in the wavelet spectra, for the cases without anisotropic thermal conduction. The origin of the second periodic signal that was superimposed on top of the oscillatory reconnection signal was not clear in the previous paper. The two suggestions were that the waves were either the result of mode conversion at the equipartition layer, or reflections from outward propagating waves while entering the artificial dissipation region described in \citet{Karampelas2022a}. To address this issue, we have modified the dissipation scheme used in \citet{Karampelas2022a}  leading to the one described in \S\ref{sec:scheme}. This new scheme allows for a weaker but more gradual dissipation of waves away from the null, reducing the reflection from waves entering this artificial region. Indeed, the resulting wavelet profile for the driver with $C=1$ amplitude now only shows one prominent period band, with a secondary one barely visible. This is proving two things. Firstly, that  the superimposed periodic signal was caused by waves reflected in the edges of that region, and second, that what we had previously assumed to be the period band associated with oscillatory reconnection is in fact the secondary superimposed signal and vice versa. 

Returning to the task at hand, we have performed additional simulations, with the new dissipation scheme, while changing the amplitude of the Ring driver. A comparison of the different cases is shown in the $J_z$ time series and their corresponding wavelet profiles in Fig. \ref{fig:ringwavelet}. The first that we see is that oscillatory reconnection takes place for all the cases within the range of initial velocities that we have chosen. Regarding the period of this oscillation, the power of the secondary oscillating signal from the reflected waves is gradually reinforced as we consider cases with a stronger initial velocity pulse. This is in agreement with our previous discussion on the nature of the secondary periodic signal found in \citet{Karampelas2022a}. Stronger perturbations will be harder to dissipate, leading to a more noisy $J_z$ signal. We can also see that for stronger initial Ring drivers, the amplitude of the initial perturbation of the X-point gets larger with respect to the rest of the oscillating signal. Due to the increased contribution from this initial perturbation, the maximum values of the wavelet power spectra appear earlier in the wavelet profiles.

Our primary concern however is the location of the main period band in the wavelet profiles. A first visual inspection shows that all of these narrow period bands are located around a period of $\approx 4\,t_0$, which also happens to be the value for the corresponding band of periods in the simulations without thermal conduction in \citet{Karampelas2022a}. The same conclusions are reached once we look at the Fourier profiles for the $J_z$ signal of each setup in Fig. \ref{fig:ringfourier}. Fitting a Gaussian profile on the highest peak of each spectrum, we see that the respective values of the period only show minor fluctuations from the $\approx 4\,t_0$ value. This consistency is important, as an independence of the oscillation frequency from the characteristics of the driver could be very important for future coronal seismology studies. This is the first time that this independence from the kinetic energy is explicitly shown in the context of oscillatory reconnection.


\subsection{Fast Magnetoacoustic Wave Pulses with Thermal Conduction}\label{sec:TC}

Our results so far have given us the first clues regarding the relationship between the strength of the driver and the characteristics of the resulting oscillatory reconnection. Our next step is to create a setup closer to the conditions of the solar corona. To that end, we will add anisotropic thermal conduction in our setups, and repeat the previous parameter study. As expected for coronal conditions, anisotropic thermal conduction will dissipate heat along the magnetic field, since the parallel to the field conduction coefficient is much larger than the perpendicular one. Our boundary conditions allow us to dissipate this heat away from our X-point, once it reaches the edges of our computational domain.

Regarding the period of the oscillatory reconnection, the wavelet profiles of Fig \ref{fig:ringwaveletTC} show a dominant value around $\approx 4.3\, t_0$, which is again, on average, constant between the different cases considered. The wavelet profiles get more complicated for stronger initial drivers, due to the stronger secondary perturbations generated by the reflections, which cannot be resolved as effectively by our treatment, as in the cases of smaller amplitudes. However, in all cases we see the same main value for the period, which corresponds to the main signal due to oscillatory reconnection. Similar to the case without thermal conduction, we see that the increased contribution from the initial perturbation for the stronger drivers shifts the maximum values of the wavelet power spectra toward earlier times. From the wavelets spectra for the $C=5$ and $C=6$ cases, we also detect this additional signal for periods around $2\,t_0$, which is confined at the very initial stages of the oscillation. Due to its short duration, this period band does not affect the rest of the signal and it can also be seen detected in the spectra of all of the studied cases, with and without thermal conduction, albeit less pronounced. Therefore, it is assumed that it is associated with the initial perturbation of the X-point.

The fact that this visually derived value for the oscillation period is larger than the respective value for the cases without anisotropic thermal conduction is in agreement with the findings of \citet{Karampelas2022a}, where it was shown that the addition of thermal conduction leads to both higher periods of oscillation, as well as a faster decay of the oscillating signal, which we also see in our current results. Similar to the cases without thermal conduction, the same trend is shown at the Fourier profiles for the $J_z$ signal of each setup in Fig. \ref{fig:ringTCfourier}. Again, fitting a Gaussian profile on the highest peak of each spectrum, we see that the respective values of the period only show minor fluctuations from the $\approx4.3\,t_0$ value. In the same figure, we see that the standard deviation is constrained with an upper value of $\sigma=0.5\,t_0$. We chose this value in order to better fit the Gaussian to the peak of the dominant period for each case, by reducing the influence of the nonsymmetrical wings of those peaks.


\subsection{Calculating Oscillation Periods for different Parameters Using Wavelet Analysis and Fourier Spectra}\label{sec:periods}

In order to quantify the relation between the amplitude of the initial velocity pulse, and the period of the resulting oscillating signal for the $J_z$ current density, we need to calculate more accurately the period for each one of our simulations. In this study, we have developed a new method to calculate the period, different from the one used in \citet{Karampelas2022a}. In particular, we go back to the wavelet profiles of Figs. \ref{fig:ringwavelet} and \ref{fig:ringwaveletTC}, and we focus on a band of periods between $3.8\,t_0$ and $5.7\,t_0$. This band of periods is where we get the highest values of the spectral power in both categories of setups. By focusing on this band of periods, we exclude from our analysis the overlapping signals from the secondary perturbations due to the reflections and the mode conversion, that become stronger as the amplitude of the initial pulse increases. 

Within this narrow band of frequencies, we want to get the values that will lie outside of the cone of influence, while at the same time being representative of the oscillating signal, before its decay. We also want to use as many of the data points of the signal as possible, in order to get a more representative average value, while also accounting for the standard deviation from each set of values. For the setups of Fig. \ref{fig:ringwavelet}, we consider the values within the time window of $t\in \left[ t_{p-max}-2, t_{p-max}+7 \right]\,t_0$, where $t_{p-max}$ is the time when the power of the wavelet signal takes its maximum value. Similarly, for the setups of Fig. \ref{fig:ringwaveletTC}, we take the values within the time window of $t\in \left[ t_{p-max}, t_{p-max}+15 \right]\,t_0$. More points were chosen for the cases with thermal conduction, because their wavelet profiles exhibit a relatively stronger gradual swift over time, and thus more points were needed for more reliable results.

Placing these average values on a graph, while using the calculated standard deviation as error values, we get the period (in code units) versus velocity pulse amplitude (in $C$) graph of Fig. \ref{fig:periods}. The blue circular disks and their error bars correspond to the average values and standard deviation for the cases without thermal conduction, and the orange ones correspond to the cases with the added anisotropic thermal conduction. We see that the cases with thermal conduction exhibit higher periods than the cases without, as was also reported in \citet{Karampelas2022a}. Additionally, they also exhibit values with longer error bars from the cases without anisotropic thermal conduction. This is the result of the wavelet profiles showing a gradual shift in the values of the periods over time, a shift that can be attributed to nonlinear effects like the slight redistribution of the mean magnetic field and the gradual loss of energy from the system due to thermal conduction. Repeating the same process but using the period values from the Fourier spectra of Figs. \ref{fig:ringfourier} and \ref{fig:ringTCfourier}, we again see two distinct groups for the period, corresponding to the cases with and without thermal conduction. Following the same convention as before, we plot these on the right panel of Fig. \ref{fig:periods}. Here we have omitted the error bars, because the respective standard deviation values are too big, and they would visually clutter the graph. 

Finally, we have fitted two constant values for the period ($P$) in these two data sets, shown in Fig. \ref{fig:periods} with the same respective colors for each case (see the dashed line for the cases without thermal conduction, and the dotted line for those with). From this graph, it becomes clear that the resulting average period of oscillatory reconnection for these setups is practically independent of the amplitude of the initial perturbation, as was already hinted by the visual inspection of the wavelet profiles. Repeating this for the graph on the right, we again see the same independence, for very similar constant values for the period $P$. This is the first time that this independence is shown for a 2D null point in hot plasma, undergoing oscillatory reconnection, excited by an external wave pulse. This independence suggests important implications for the seismological prospects of this mechanism.

\subsection{Alternative driver for Oscillatory Reconnection}\label{sec:pinch}

Finally, we have decided to test the sensitivity of oscillatory reconnection by using an initial perturbation of a different nature altogether, than just changing its amplitude. We take the so-called pinch-like driver that was described with eqn. (\ref{eq:pinchvy}), for a setup without thermal conduction. Unlike the Ring driver, this pulse is of a mixed nature, with velocity components both perpendicular and parallel to the magnetic field. The spatial characteristics of this driver are also different from those used in the past, within the context of this $2D$ magnetic X-point. The initial evolution of this new type of driver is depicted in Fig. \ref{fig:pinchdriver} for the $v_{\perp}$ pulse component, where we see that this driver does not produce such strong shock fronts as the Ring driver does. 

Comparing the $J_z$ current density signal at the null point, and its respective wavelet profile for the Pinch driver in Fig. \ref{fig:pinchwavelet}, with the equivalent Ring drivers from Fig. \ref{fig:ringwavelet}, we can see that this new driver also leads to the development of oscillatory reconnection, once the velocity pulse perturbs the null point from its initial equilibrium. The different spatial profile of the driver leads to an earlier and more gradual perturbation of the null, which translates into an earlier and more gradual initial $J_z$ negative peak in the profile. After that, however, we see that the signal behaves in a similar fashion to the rest of the cases, oscillating between positive and negative values, before its decay. From the wavelet profile, it is obvious that we obtain some additional periodicity in our signal. The origin of this is again due to the wave reflections from the artificial dissipation scheme and the boundaries.

Looking at its wavelet profile, the oscillating signal of this new initial pulse is not as clean as the Ring driver producing $J_z$ values of equivalent amplitude. Instead, we have additional period bands that indicate an overlapping oscillating signal. The more spatially extended nature of this new style of driver makes the dissipation of the reflections similarly challenging as in the case of the higher amplitude Ring drivers. However, both from the visual inspection of the wavelet profile and from the  Fourier spectrum on the bottom panel of Fig. \ref{fig:pinchwavelet}, we again find a similar period as for the Ring drivers without thermal conduction ($\approx 4\,t_0$). This stresses the consistency of the periodicity of oscillatory reconnection.

We follow the same method of calculating the oscillation period from the wavelet profile for the new driver, i.e. focusing on the period band between $3.8\,t_0$ and $5.7\,t_0$, for time $t\in \left[ t_{p-max}-2, t_{p-max}+7 \right]\,t_0$, where $t_{p-max}$ is the time when the power of the wavelet signal takes its maximum value. The resulting average value for the period, with the standard deviation added as error values, we get the black square point in Fig. \ref{fig:periods}. This new point sits very well alongside the rest of the cases for the Ring driver, without thermal conduction. This result expands on our previous findings and for the first time shows that the period of the resulting oscillatory reconnection signal is independent of the type of the external pulse driver used to perturb the null point, and is instead affected by the local characteristics of the plasma, such as the magnetic field \citep{Karampelas2022a}, highlighting  again the potential usefulness of this mechanism as a tool for coronal seismology.


\section{Discussion and Conclusions} \label{sec:discussions}

In this paper we revisit the mechanism of oscillatory reconnection of a 2D null point in hot coronal plasma. The large number of observables that can be attributed to oscillatory reconnection stresses the need for a more complete understanding of the underlying physics of this mechanism. In \citet{Karampelas2022a}, a first step was taken toward that direction, with the decay rate and periodicity of this mechanism being studied in the presence of anisotropic thermal conduction in coronal conditions. In that study, a clear connection between the strength of the underlying magnetic field and the frequency of the oscillation was revealed. In this present work, we examined whether the amplitude of the initial pulse perturbing the 2D null point affects the period of the resulting oscillatory process.

Using the PLUTO code, we have solved the compressible and resistive 2D MHD equations, for cases with and without anisotropic thermal conduction. The main part of this work consists of a parameter study using an initially circular fast wave velocity pulse, which we refer to as the Ring driver. This is the same type of wave pulse used in past studies such as \citet{McLaughlin2009}. For our parameter study, we have considered a range of amplitudes for that driver (i.e., $C=0.5,\,1,\,2,\,3,\,4,\,5$ and $6$), with the case of $C=1$ amplitude being the one used in \citet{Karampelas2022a}. For each simulation, we calculated the wavelet profiles for the oscillating $J_z$ current density signal at the reconnecting null point and derived the average period of oscillation for each case. 

Gathering together the average values for the period and its error margins for each simulation from the wavelet and Fourier spectra, we made a period - velocity amplitude graph, shown in Fig. \ref{fig:periods}, where two different groups of points have formed, corresponding to the cases with and without thermal conduction. For each of these groups, the points fluctuate around a constant value for the period ($P$), which has a different value for the cases with and without thermal conduction. As expected from \citet{Karampelas2022a}, the respective value for the setups with thermal conduction is higher than the equivalent for those without. The points for the period, when thermal conduction is considered, exhibit large error bars due to the gradual shift in the values of the periods over time for these setups. However, the average values are clearly near the fitted constant function, in agreement with the trend for the cases without thermal conduction.

Finally, performing a similar simulation for a completely new type of a velocity driver, for a setup without anisotropic thermal conduction has led to an average period that also follows the trend of the previously considered cases, as shown in Fig. \ref{fig:periods}. This provides additional evidence that the period of oscillation for this mechanism is independent of not only the strength of the initial wave pulse but also of the type of the initial driver. 

As was previously mentioned in  \S\ref{sec:introduction}, oscillatory reconnection has been considered as one (of many) possible underpinning mechanisms that are responsible for the manifestation of quasi-periodic pulsations (QPPs) \citep[see][for reviews on the different mechanisms]{McLaughlin2018SSRv,Kupriyanova2020STP,Zimovets2021SSRv}. Our study has revealed period values for oscillatory reconnection between $3.9\,t_0\approx30.3$\,s and $4.4\,t_0\approx34.2$\,s, which are within the range found in QPP observations over recent years. In particular, \citet{McLaughlin2018SSRv} created a histogram of the statistics of 278 QPP events reported in the literature (correct as of 2017, and that paper also links to a continuously updated online catalog). In that histogram, more than 120 QPP events were reported to have periods in the 10-60 second band. In addition, there have been many examples where reported periods were $\approx 30$\,s \citep[e.g.][]{2016ApJ...830..110C,2016ApJ...833..284I,2016ApJ...832...65Z}. However, caution is advised when looking at this good agreement: the resultant period of oscillatory reconnection comes from the local properties of the null point itself, and that means this is a robust mechanism that can be tailored to produce different period results and not just the periodicity reported here. However, the fact that our results are of the correct order for coronal conditions highlights the potential of this mechanism for coronal seismology.

Coming back to the previously mentioned independence between the velocity driver and the oscillation period, this has to be discussed while taking into account the results of past studies. In particular, in \citet{McLaughlin2012A&A} a correlation was found between the strength of the wave pulse and the period of oscillatory reconnection. This seems to contradict our present results; however, the strength of the driving pulse was also connected to the length of the initial current sheet formed from the collapse of the null point. The conclusion reached by \citet{McLaughlin2012A&A} was that the period of the oscillation seems to be related to the length of the first current sheet, with longer current sheets resulting in higher periods for the oscillating $J_z$ current density signal. This result was further reinforced by the findings of \citet{Thurgood2019A&A} for a 3D null point. In their work, \citeauthor{Thurgood2019A&A} have introduced a magnetic field perturbation around the null, which in turn collapsed into a current sheet, leading to oscillatory reconnection. In the nonlinear regime, which is the relevant one in coronal conditions, the period has been associated with the amplitude of the initial $J_z$ current density perturbation, and the subsequent length of the first current sheet. We note that \citet{McLaughlin2012A&A} and \citet{Thurgood2019A&A} each considered an initially cold plasma.

From the current density 2D profiles of Fig. \ref{fig:current} for the Ring driver, we see that the first current sheet from the collapse of a null point due to an incoming compression pulse does not exhibit the strongest negative $J_z$ values, as shown from the oscillating signal. Instead, these values take place as this first current sheet collapses further and focuses around the null point. This secondary collapse is in part reinforced by the strong rarefaction pulses that reach the solution after the compression pulses in the case of hot plasma, as was also shown in $J_z$ profiles in \citet{Karampelas2022a}. A similar collapse for the current density can be found when we use the new Pinch driver, also shown in Fig. \ref{fig:current}. In that case, the initial current sheet, formed at the start of the simulation, collapses and focuses around the null point. Due to this secondary collapse, the length of the initial current sheet seems to play only a secondary role in the resulting period of the oscillation, for the range of pulse amplitudes considered in this study. This explains why varying the amplitude of the initial wave pulse does not alter the period of oscillatory reconnection in the case of a hot plasma.

Using stronger pulses, or driving the null point collapse by moving the magnetic field footpoints, will potentially lead to longer current sheets with higher values of $J_z$ current density. However, as was the case with \citet{Sabri2020ApJ} for a null point collapse in hot plasma due to an external pulse, we would then face the prospect of initiating the plasmoid instability in this high, overly long sheet. In \citet{2018ApJ...855...50T} it was shown that high aspect ratio current sheets can be subjected to nonlinear tearing in coronal conditions. The formation of the overly long current sheet, accompanied by an emerging O-point magnetic field configuration, and the resulting tearing of the current sheet in \citet{Sabri2020ApJ} blocked the evolution of oscillatory reconnection after the first two reversals in magnetic connectivity.

\begin{figure}[t]
    \centering
    \includegraphics[trim={0.cm 0.cm 0.cm 0.cm},clip,scale=0.4]{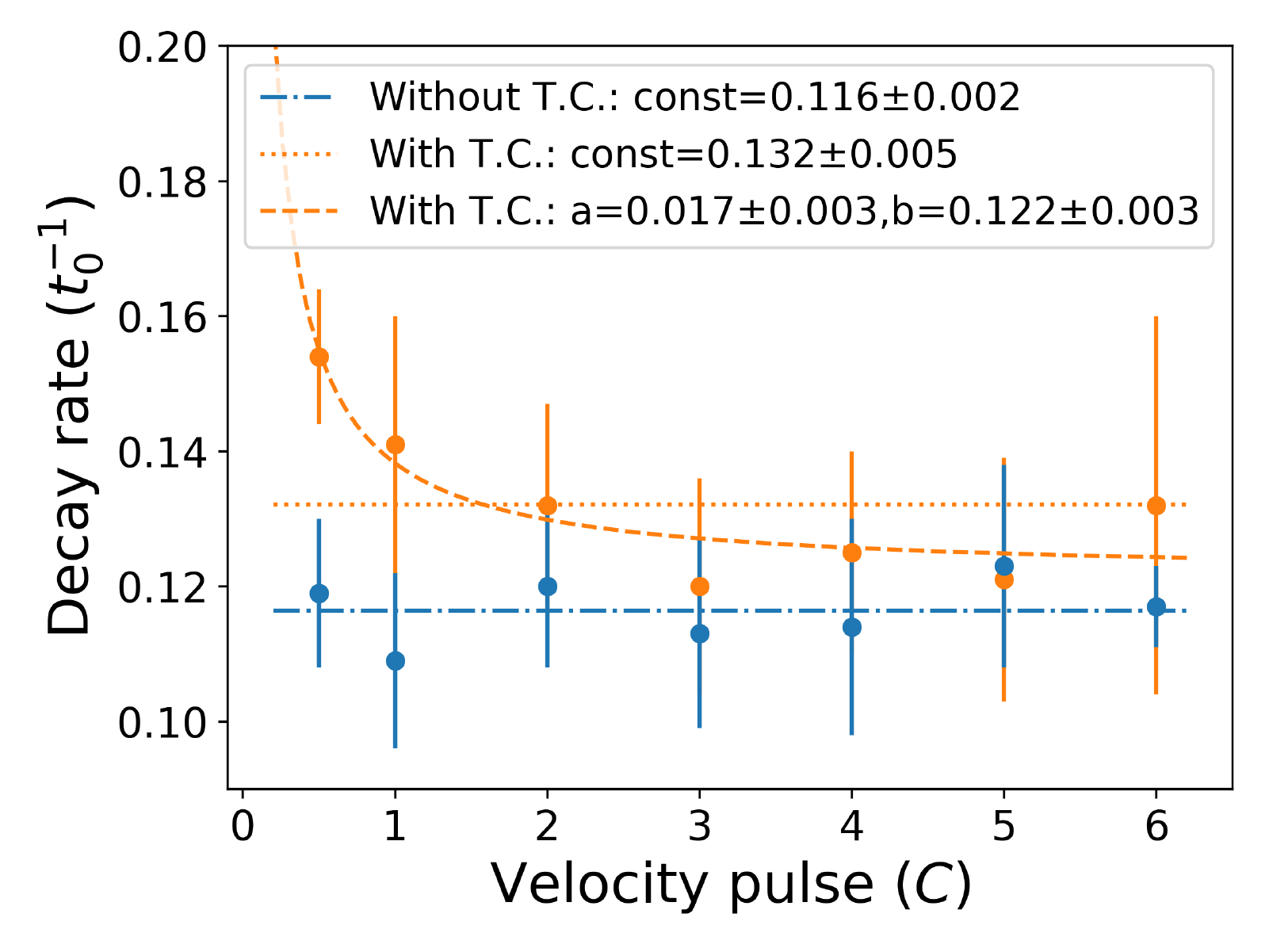}
    \caption{Decay rates of the amplitudes of the $J_z$ current density profiles of Figs \ref{fig:ringwavelet} and \ref{fig:ringwaveletTC}. For the two data sets, we have fitted a constant value for each set of data points (blue dashed line for the set without thermal conduction and orange dotted line for the set with thermal conduction), as well as the function $f(C)=(a/C)+b$ for the set with thermal conduction (orange dashed$-$dotted line). The calculation of the decay rates for each case is shown in the Appendix \ref{sec:appendix}.}\label{fig:decayrate}
\end{figure}

One final comment has to be made about the decay rates of the oscillating $J_z$ profiles. Although our focus in this study has been on the period of oscillatory reconnection, its decay rate is also an important aspect that needs to be addressed. In \citet{Karampelas2022a} it was shown that the decay rates of the oscillation amplitudes were affected by the background magnetic field strength, with stronger magnetic fields leading to a faster decay. As a complementary piece to that analysis, this paper presents the values for the decay rates $a$ with respect to the driver amplitude $C$ in Fig. \ref{fig:decayrate}. A detailed description of the method used in calculating them is included in the Appendix \ref{sec:appendix}. Fig. \ref{fig:decayrate} shows the values of the calculated decay rates for each $J_z$ signal from the  Ring drivers of different amplitude strength. For the set of data without thermal conduction, the results seem  to follow a  trend, with the decay rates fluctuating around a constant value ($0.115 \pm 0.002\,t_0^{-1}$), implying that the decay rates are independent of the driver strength. For the set of data including anisotropic thermal conduction, and in agreement with \citet{Karampelas2022a}, we see that adding thermal conduction leads to higher values of decay rates, by comparison. However, the set of data for anisotropic thermal conduction also has very large error bars resulting from the calculation of the decay rates. Indeed, this set of data is fitted in our study with both a constant value ($0.132\pm0.005\,t_0^{-1}$) and an inverse linear function, the coefficients of which can be seen in Fig. \ref{fig:decayrate}. Thus, the decay rate results presented here, although very important within the context of coronal seismology, need to be treated with caution. Our wavelet analysis has revealed the existence of superimposed periodicities in our signal, which indicates the existence of wave interference at our null, between the oscillatory reconnection process and incoming waves reflected from the boundaries. Through Fourier and wavelet analysis, we have managed to isolate and identify the proper period of oscillatory reconnection. However, the decay rates are calculated through the amplitude of the $J_z$ signals, which are directly affected by the artificial reflections in our system. Therefore, the decay rate analysis needs to be further explored in future studies, better suited to deal with the reflections from the artificial dissipation region and the reflective boundaries before definitive conclusions can be drawn.

To summarize, our study shows for the first time for a magnetic X-point in hot coronal plasma, that the period of the oscillation is independent of the type or strength of the external pulse perturbing the null point. Alongside the connection between the strength of the underlying magnetic field and the frequency of the oscillation \citep{Karampelas2022a}, this new result indicates that the period of oscillatory reconnection depends mainly on the underlying characteristics of the plasma in the vicinity of the null point, further emphasizing the importance of oscillatory reconnection as a diagnostic tool in the context of coronal seismology. 


\begin{acknowledgments}
All authors acknowledge UK Science and Technology Facilities Council (STFC) support from grant ST/T000384/1. K.K. also acknowledges support by an FWO (Fonds voor Wetenschappelijk Onderzoek – Vlaanderen) postdoctoral fellowship (1273221N). This work used the Oswald High Performance Computing facility operated by Northumbria University (UK). 
\end{acknowledgments}


\appendix
\section{Calculating the decay rates of the oscillating $J_z$ profiles} \label{sec:appendix}

\begin{figure*}
    \centering
    \resizebox{\hsize}{!}{
    \includegraphics[trim={0.cm 0.cm 0.cm 0.cm},clip,scale=0.45]{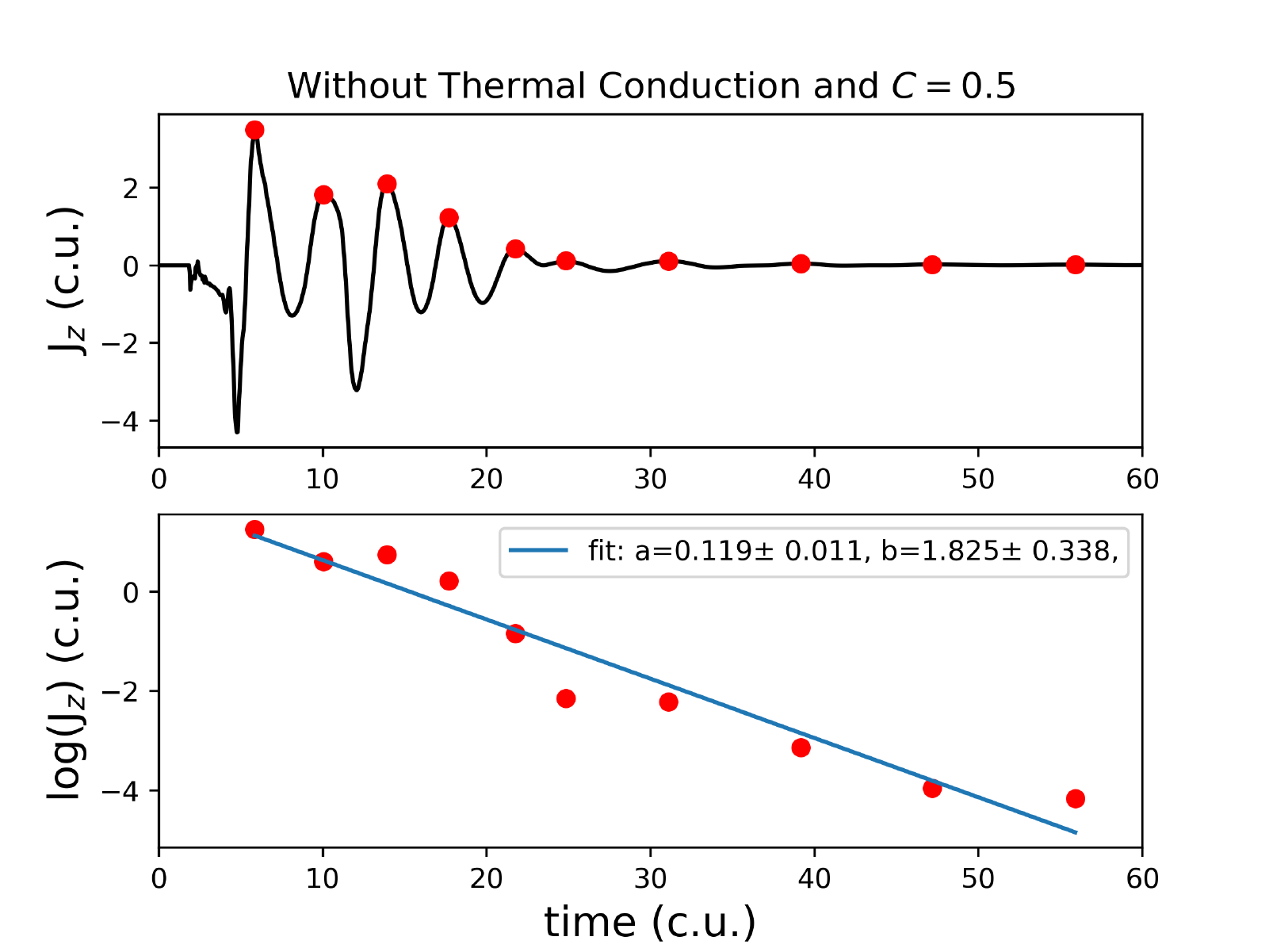}
    \includegraphics[trim={0.cm 0.cm 0.cm 0.cm},clip,scale=0.45]{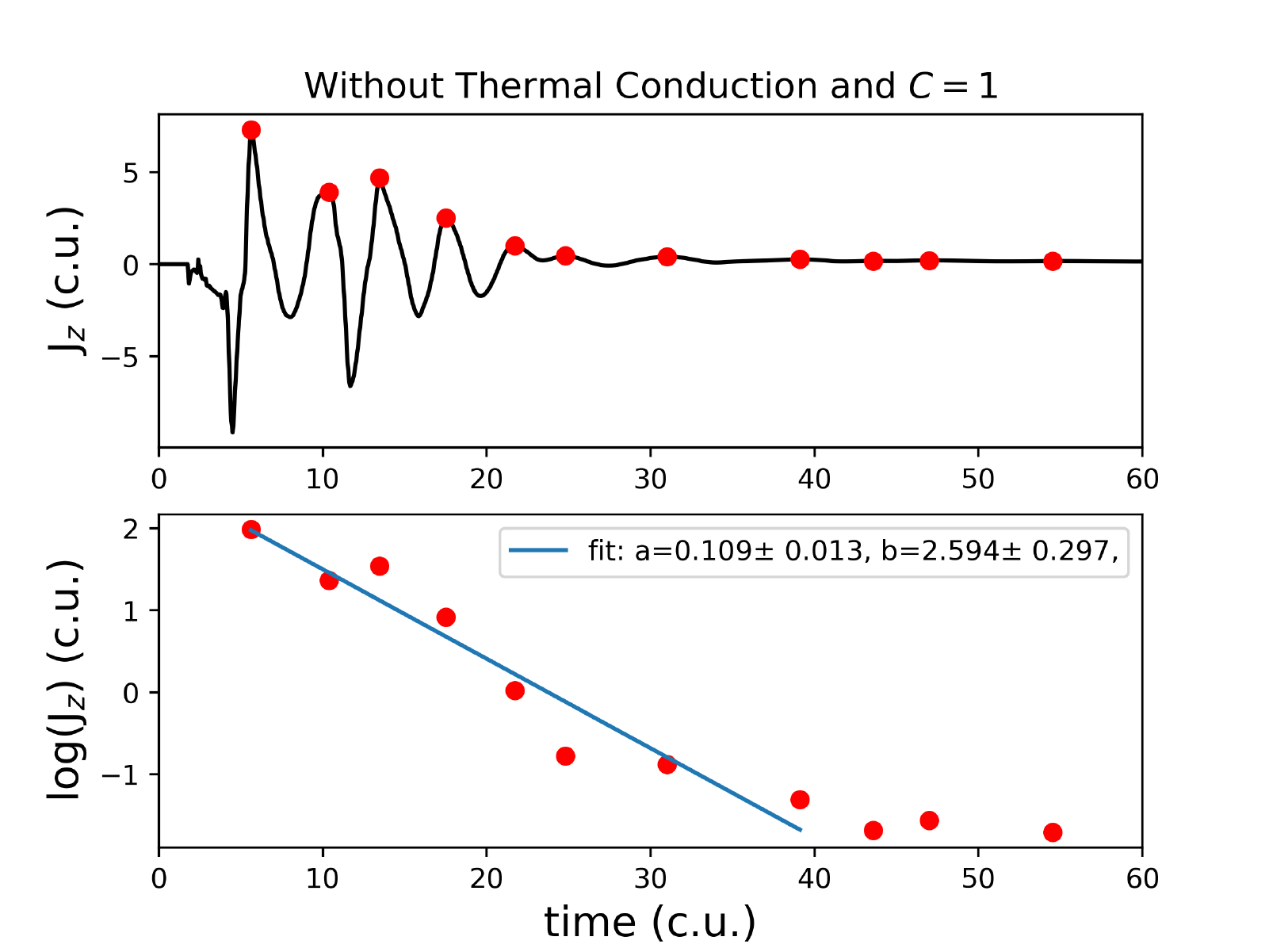}
    \includegraphics[trim={0.cm 0.cm 0.cm 0.cm},clip,scale=0.45]{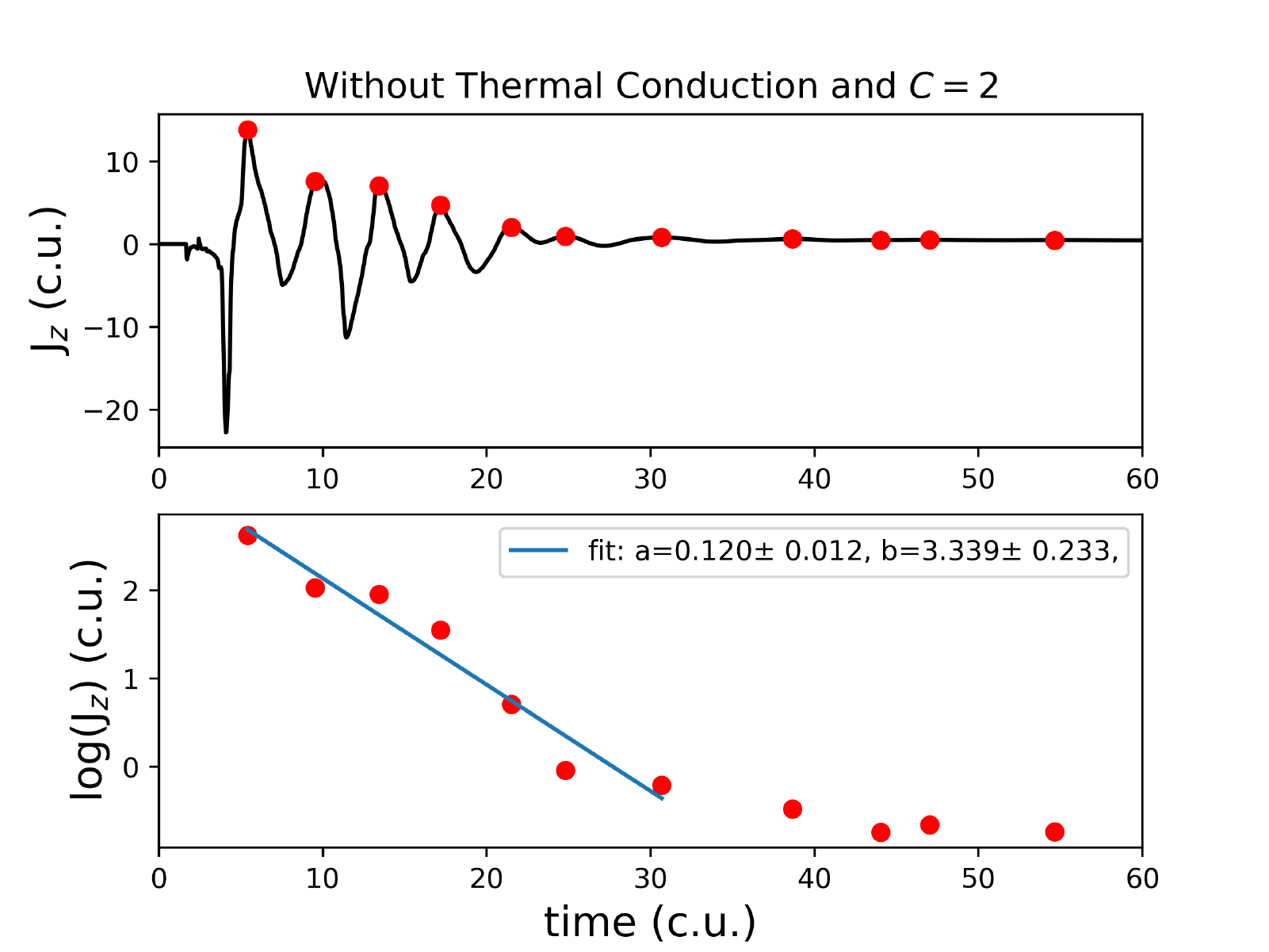}
    }
    \resizebox{\hsize}{!}{
    \includegraphics[trim={0.cm 0.cm 0.cm 0.cm},clip,scale=0.45]{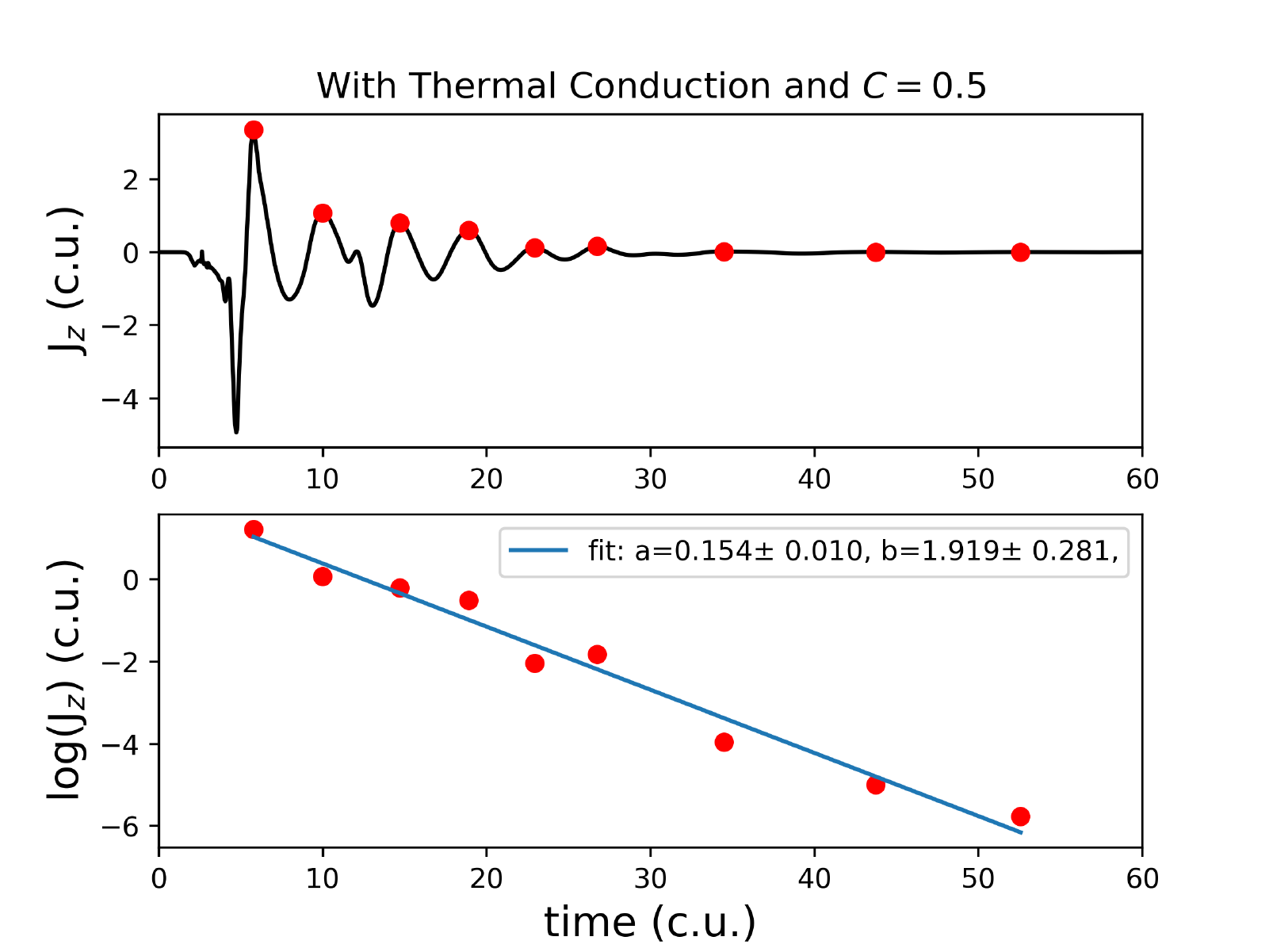}
    \includegraphics[trim={0.cm 0.cm 0.cm 0.cm},clip,scale=0.45]{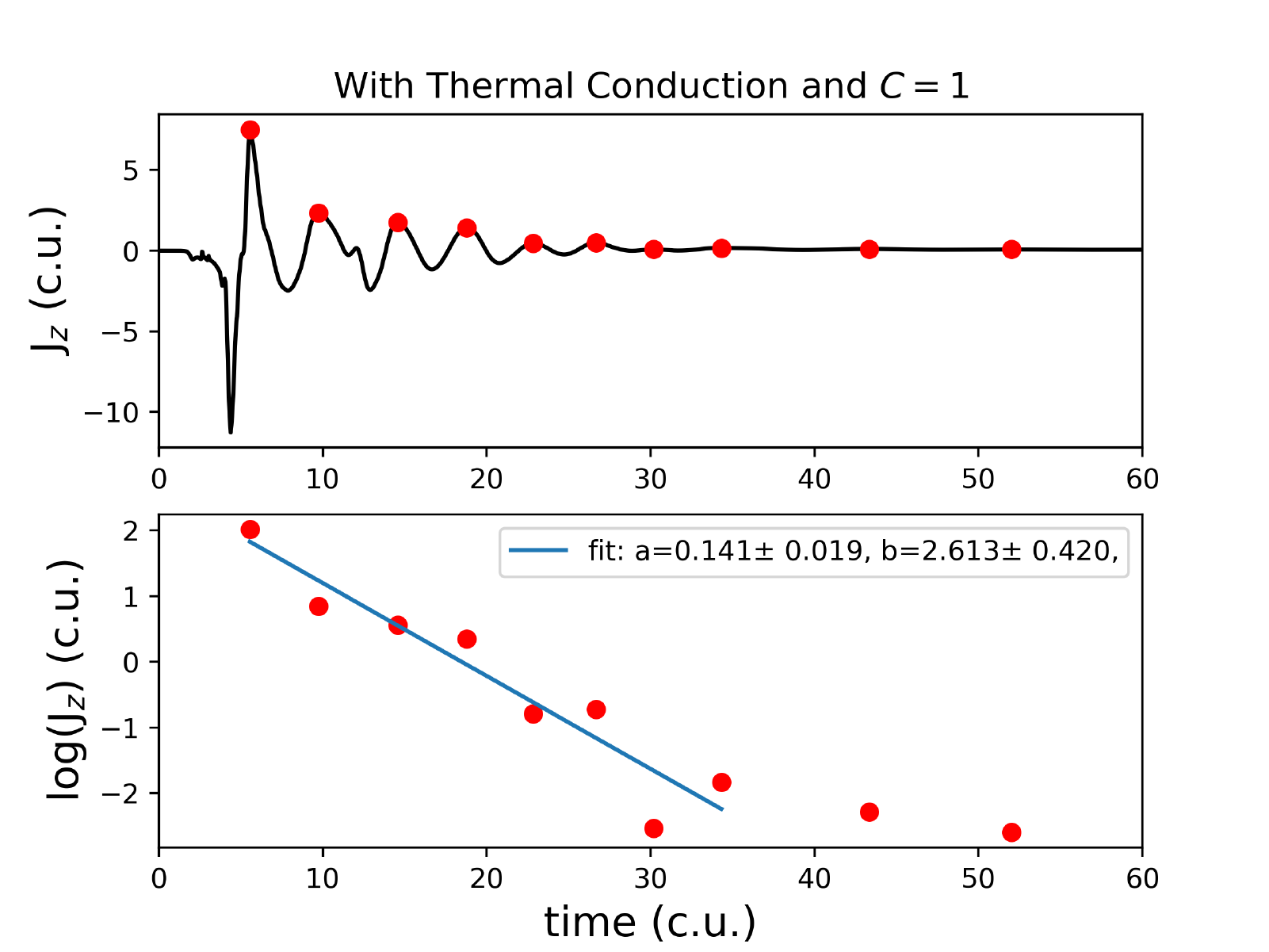}
    \includegraphics[trim={0.cm 0.cm 0.cm 0.cm},clip,scale=0.45]{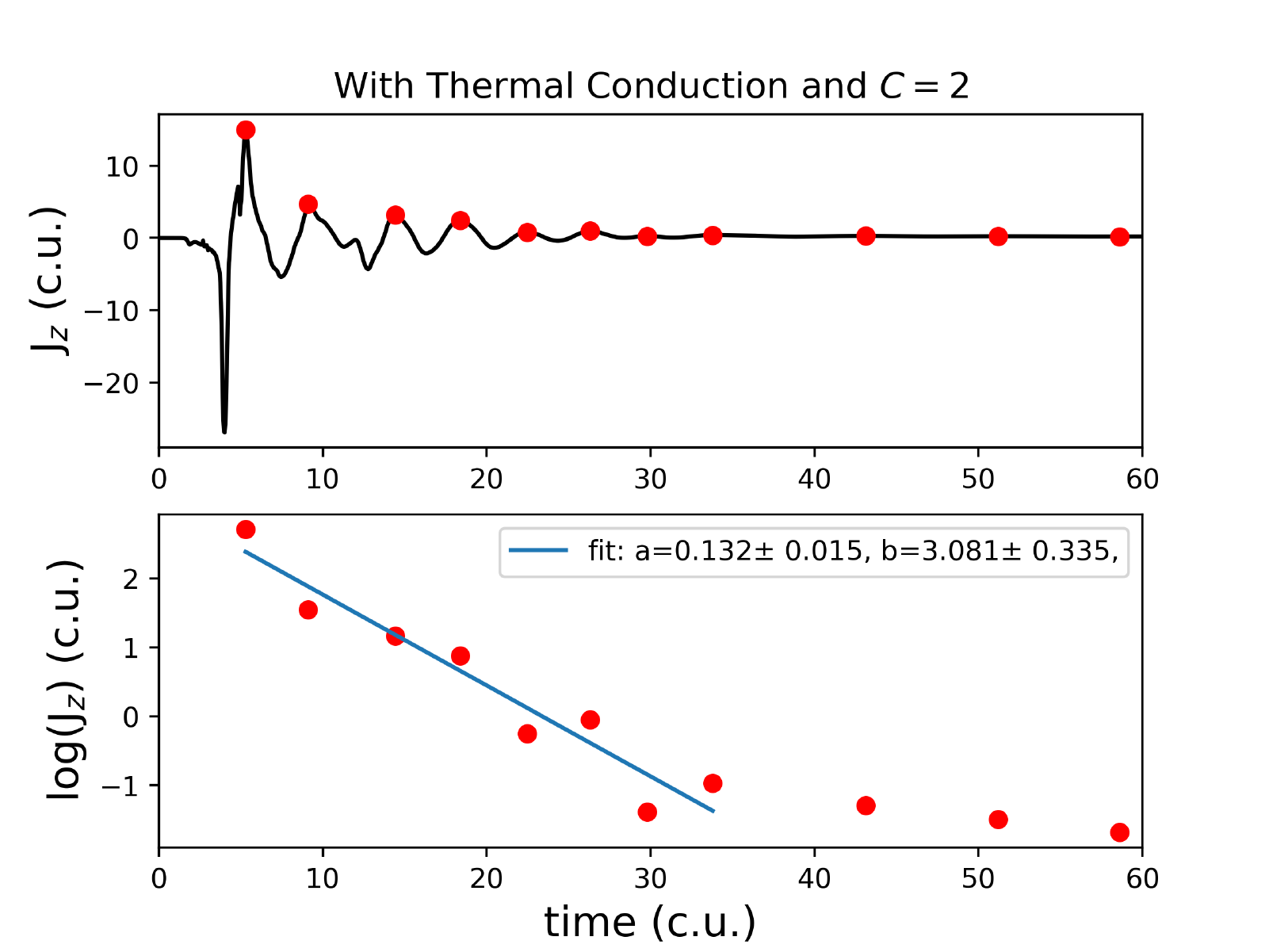}
    }
   \resizebox{\hsize}{!}{
    \includegraphics[trim={0.cm 0.cm 0.cm 0.cm},clip,scale=0.45]{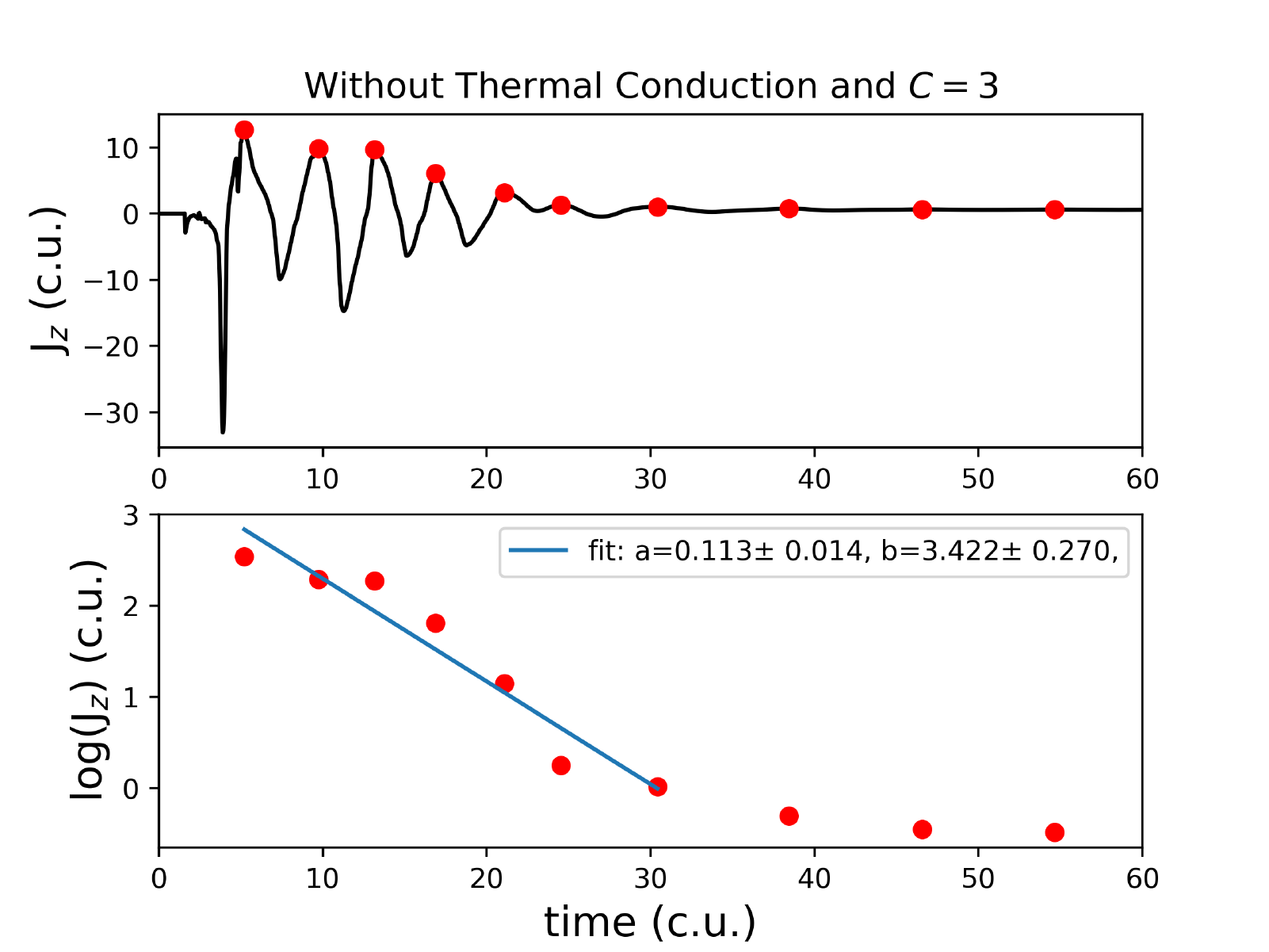}
    \includegraphics[trim={0.cm 0.cm 0.cm 0.cm},clip,scale=0.45]{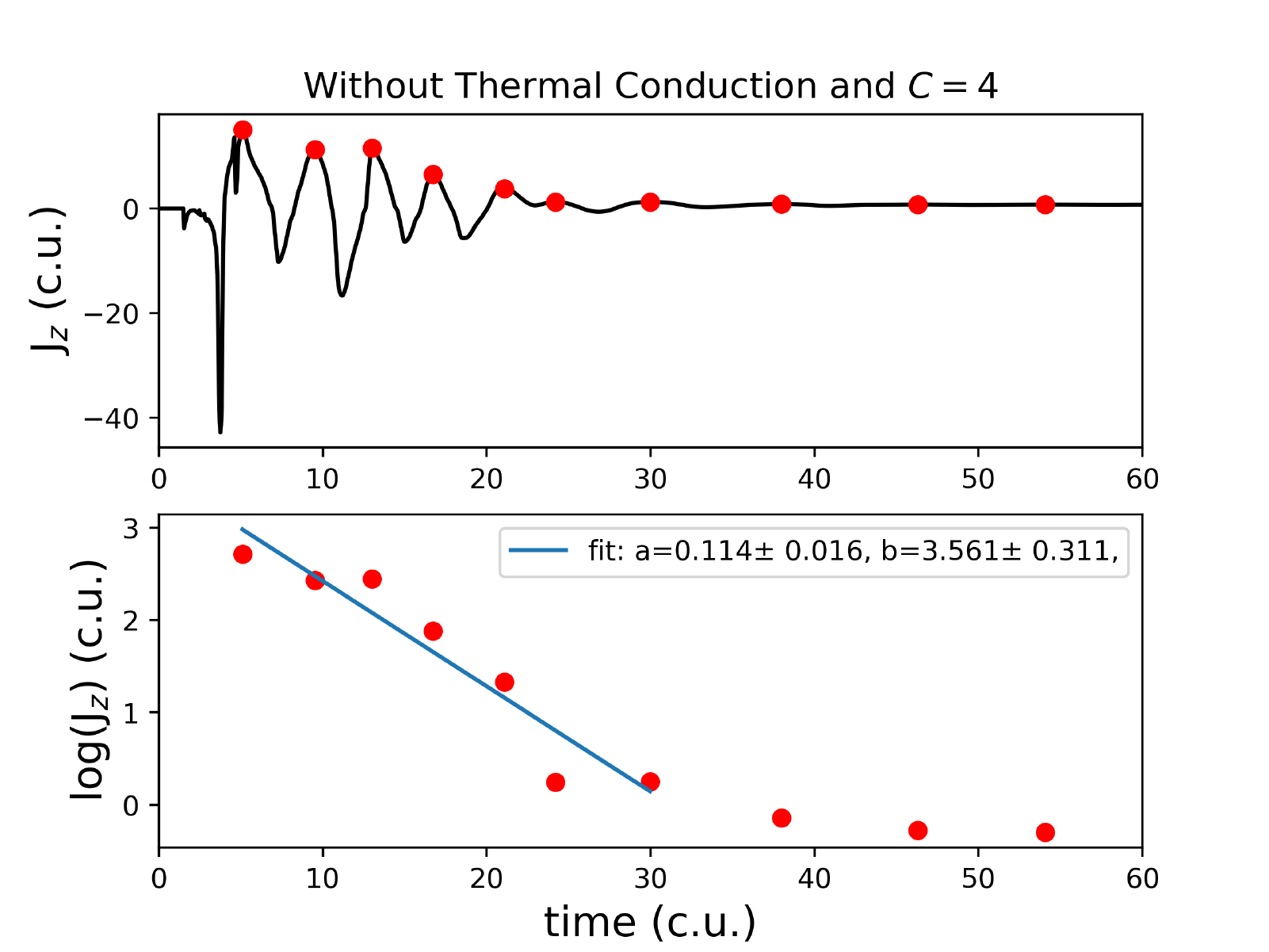}
    \includegraphics[trim={0.cm 0.cm 0.cm 0.cm},clip,scale=0.45]{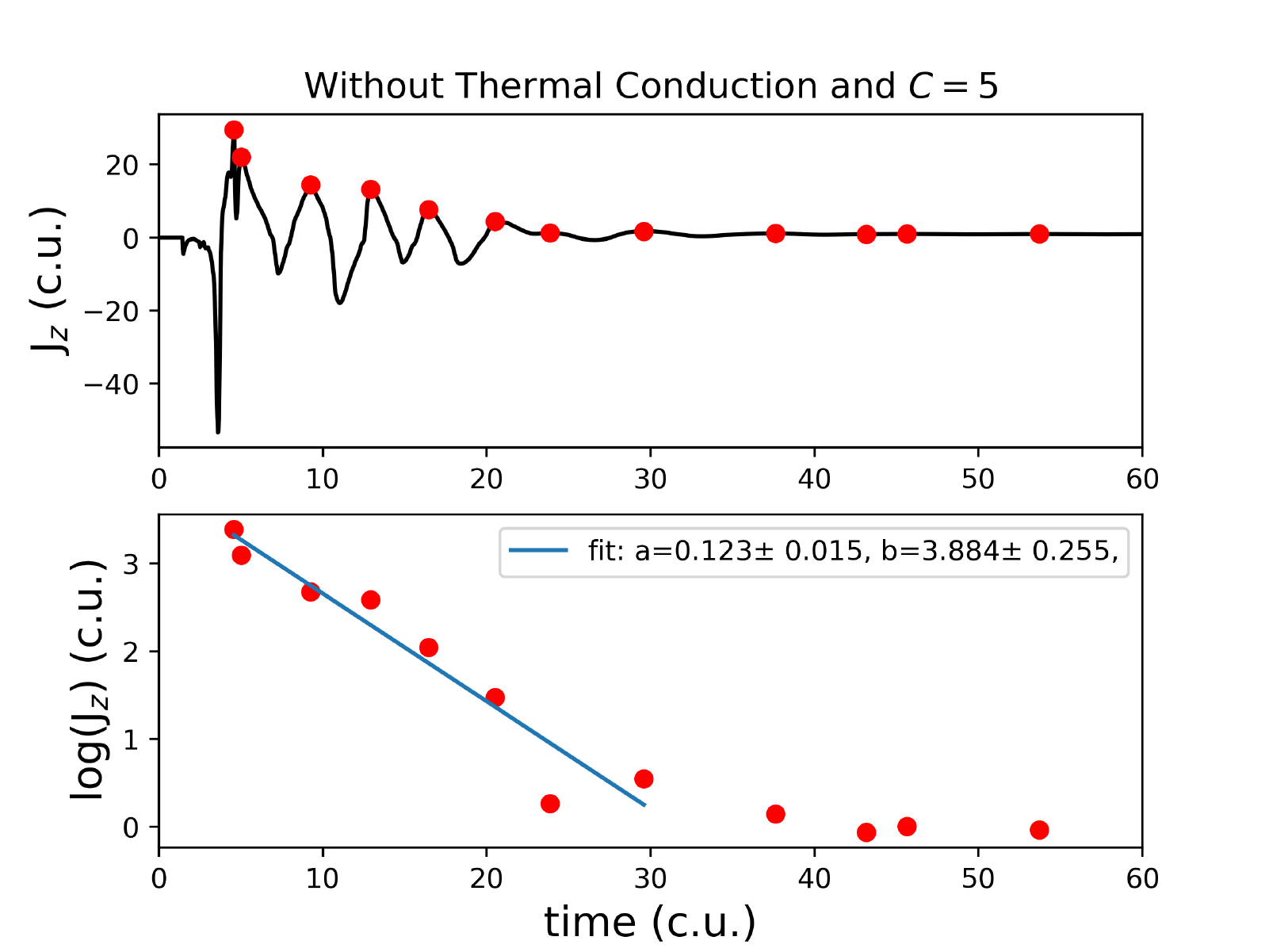}
    }
    \resizebox{\hsize}{!}{
   \includegraphics[trim={0.cm 0.cm 0.cm 0.cm},clip,scale=0.45]{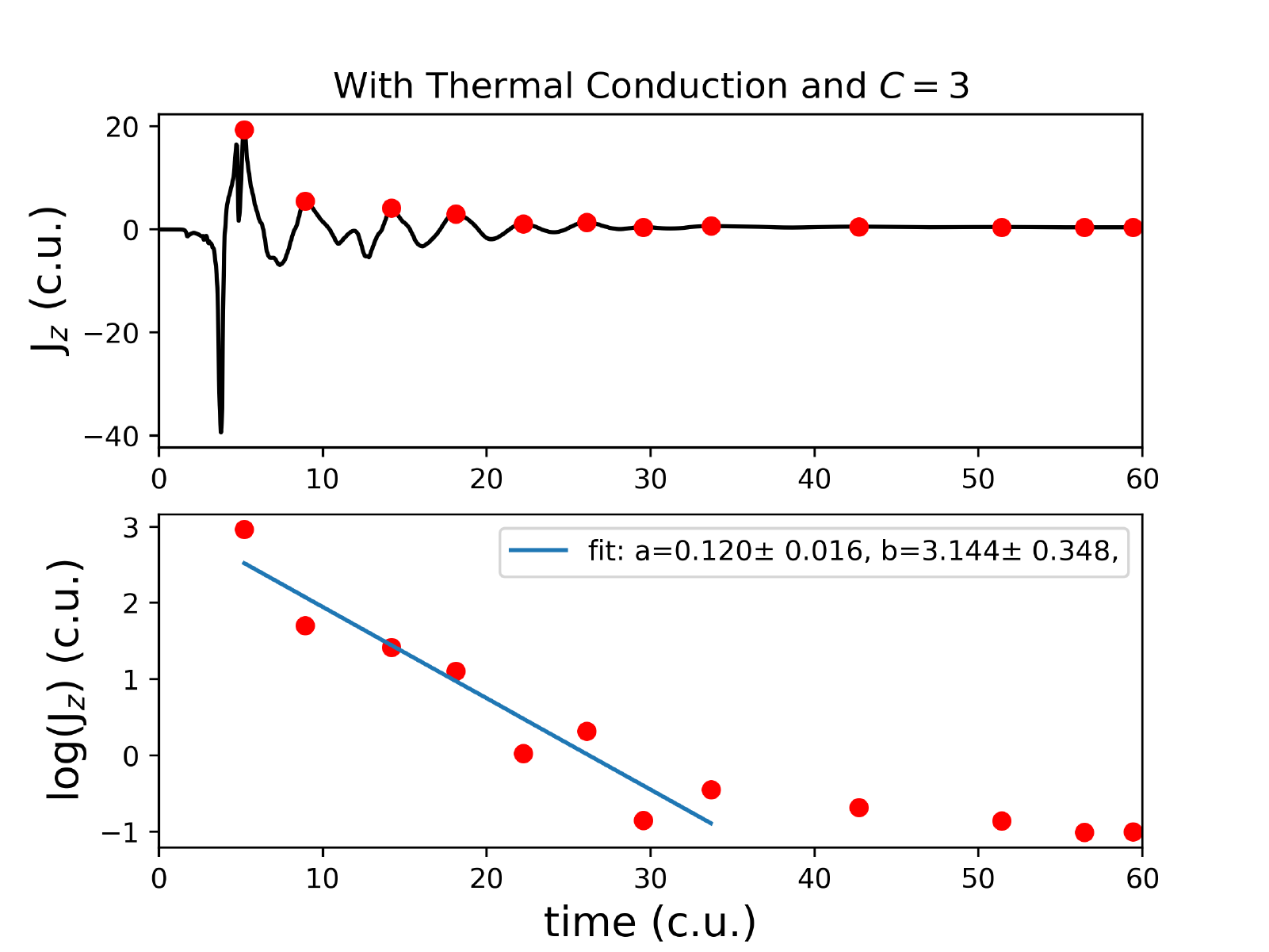}
    \includegraphics[trim={0.cm 0.cm 0.cm 0.cm},clip,scale=0.45]{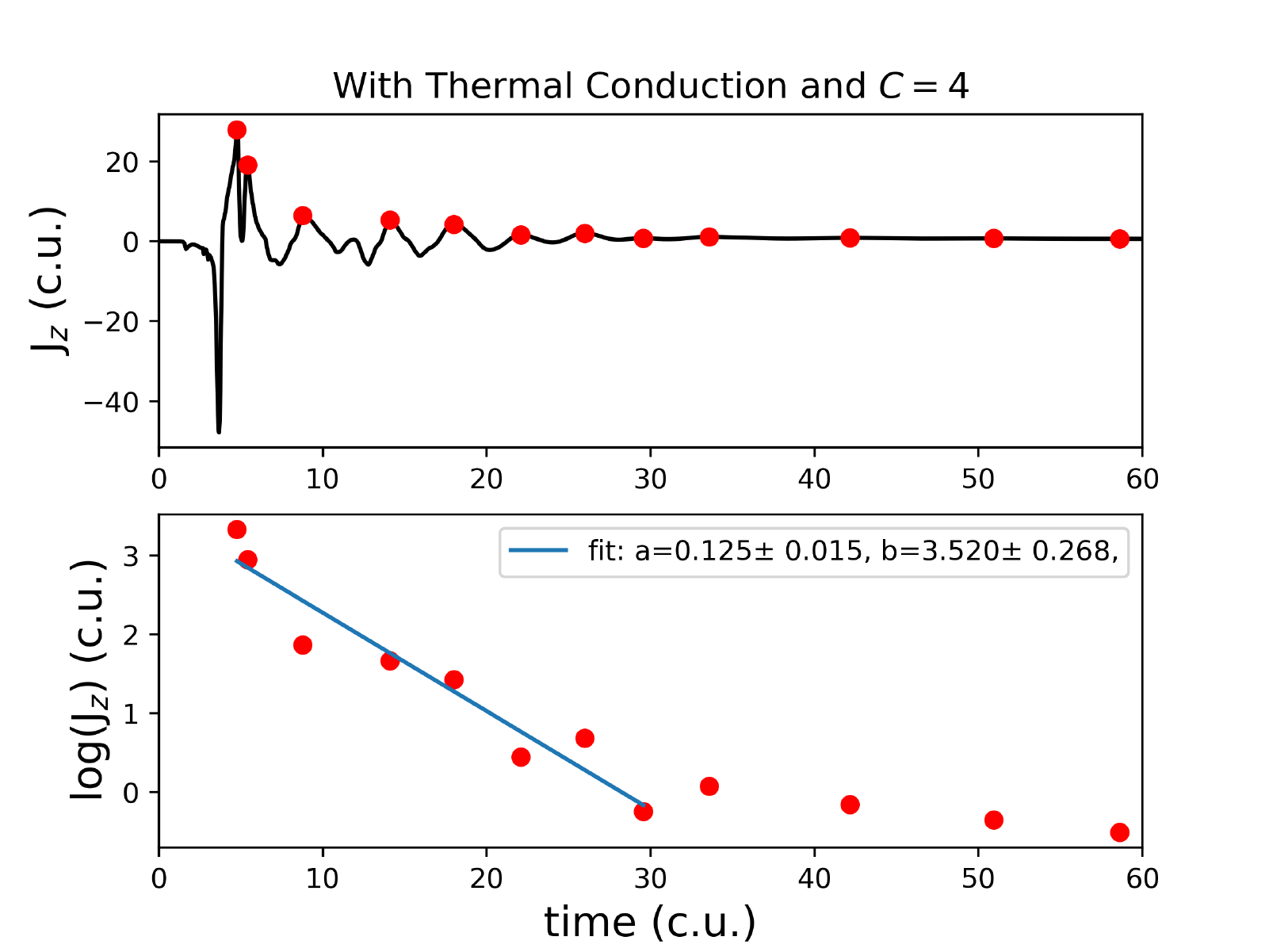}
    \includegraphics[trim={0.cm 0.cm 0.cm 0.cm},clip,scale=0.45]{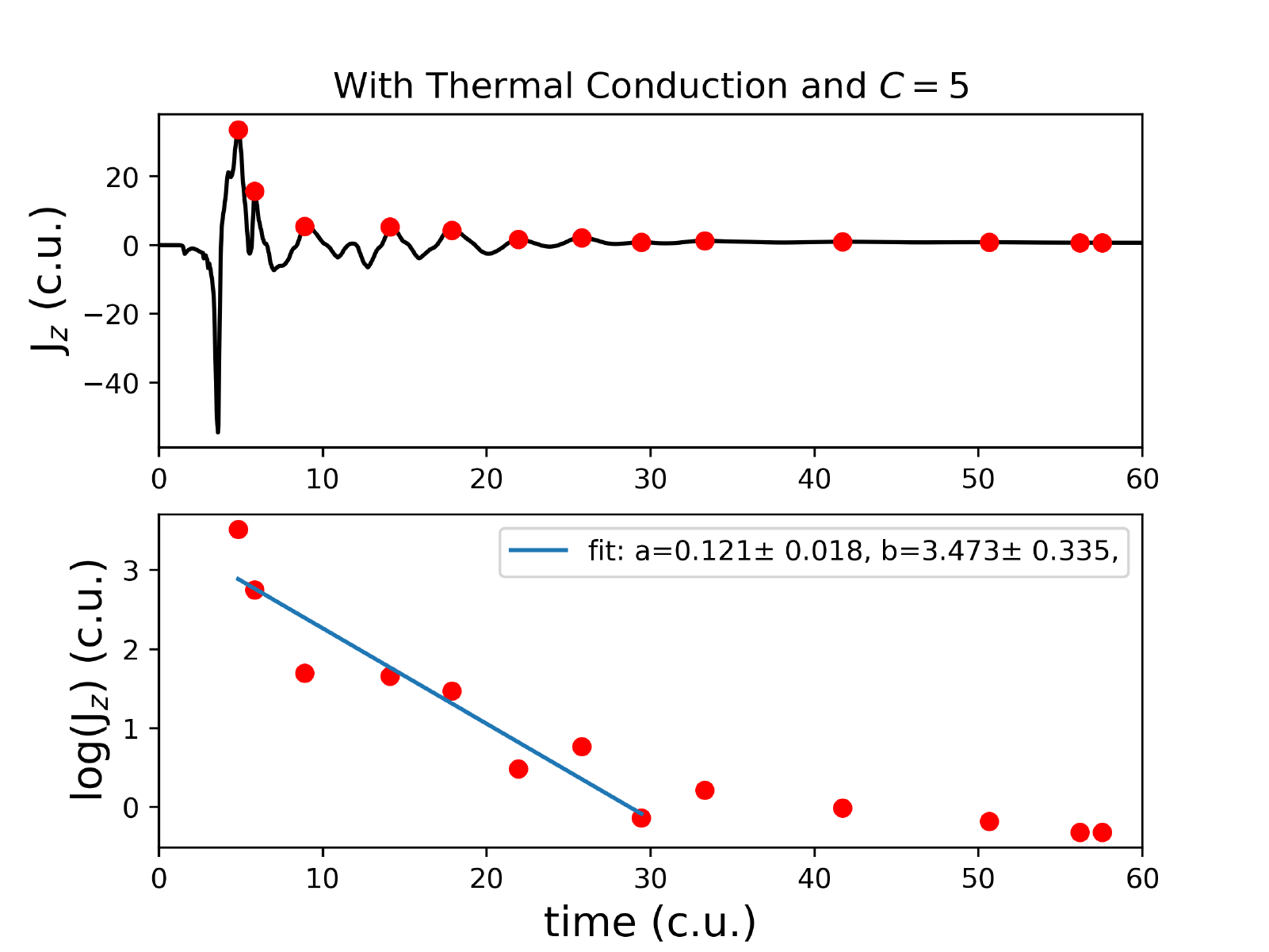}
    }
    \includegraphics[trim={0.cm 0.cm 0.cm 0.cm},clip,scale=0.37]{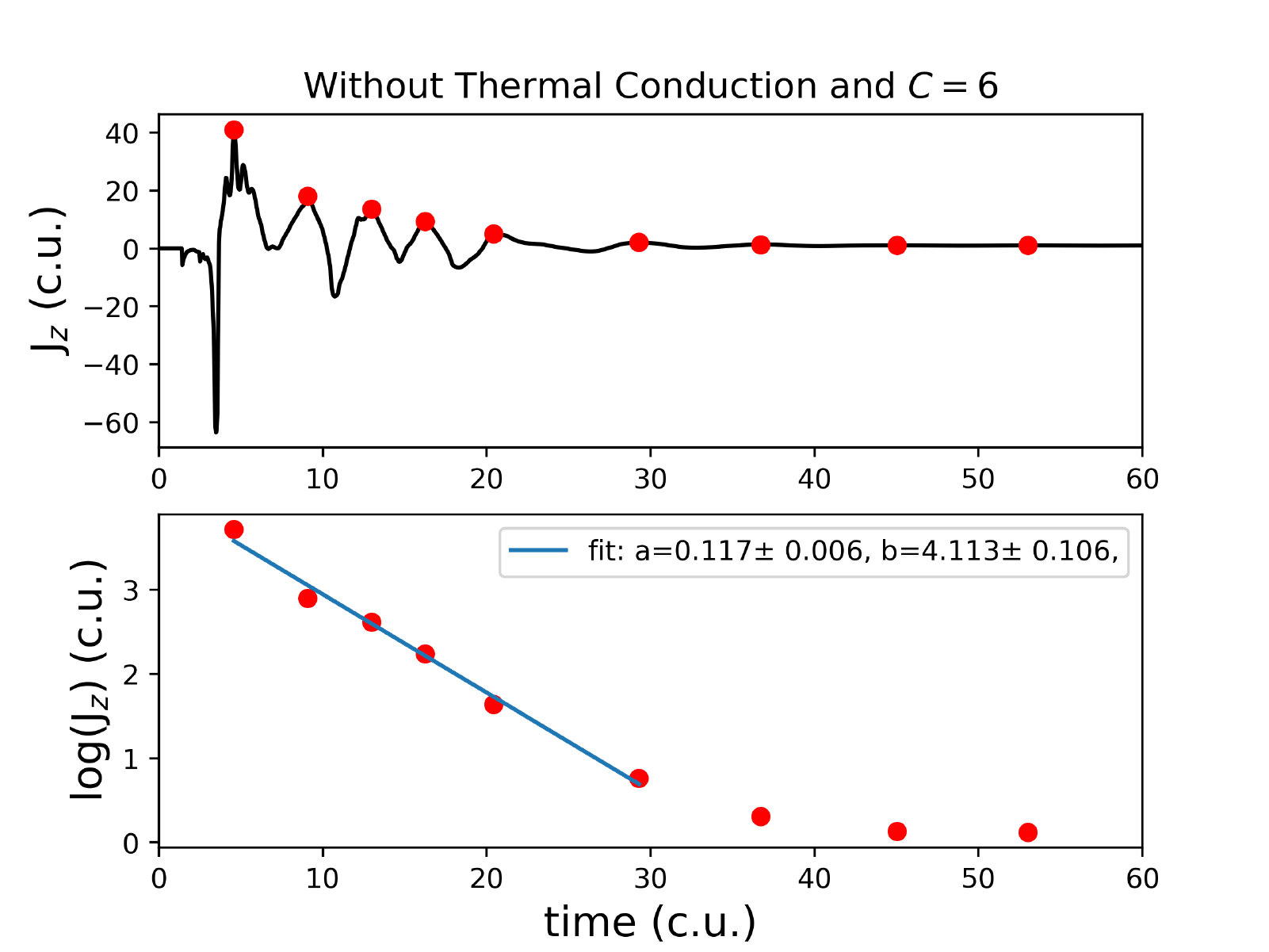}
    \includegraphics[trim={0.cm 0.cm 0.cm 0.cm},clip,scale=0.37] {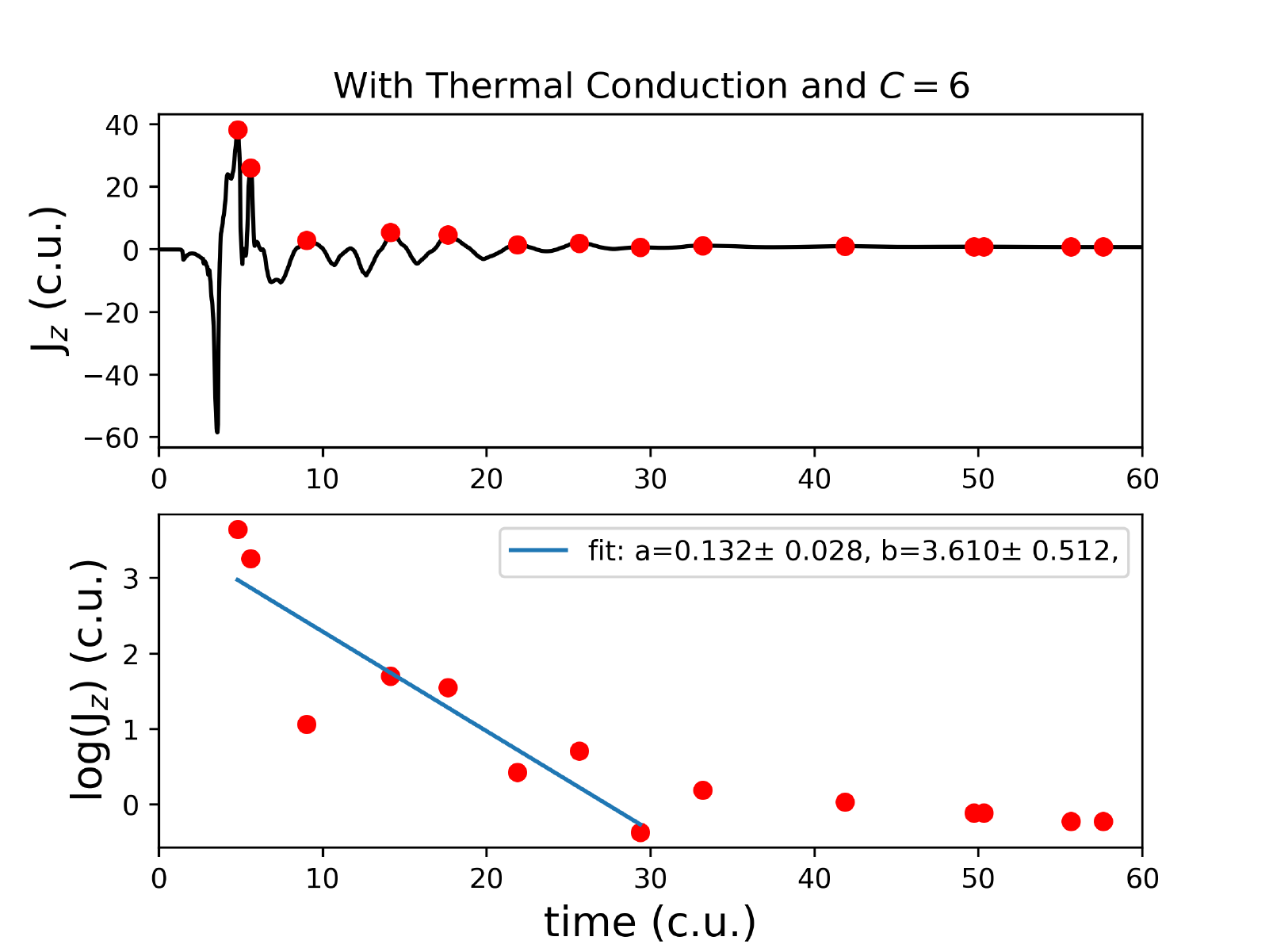}
    \caption{Panels showing the oscillating $J_z$ profiles for all the different cases of the Ring drivers used, together with the panels showing the fits for the decay rates during the decay phase of our signal, plot in logarithmic scale. The panels are plotted in batches, comparing cases with and without thermal conduction.}
    \label{fig:appendixdecay}
\end{figure*}

In order to study the decay rates of the $J_z$ current density profiles for our simulations using the Ring drivers of different initial amplitudes ($C$), we will follow the method used in \citet{Karampelas2022a}. We took the oscillating $J_z$ profiles from each simulation for the different Ring drivers, as shown in Figs. \ref{fig:ringwavelet} and \ref{fig:ringwaveletTC}. All of these profiles are gathered in Fig. \ref{fig:appendixdecay}, where they are grouped into batches consisting of the equivalent setups with and without anisotropic thermal conduction, for an easier comparison and overview.

For each of the oscillating profiles in Fig. \ref{fig:appendixdecay}, we first consider the local maxima that correspond to the oscillation amplitudes. These are plotted over the $J_z$ profiles as red dots, on the top half of each panel of Fig. \ref{fig:appendixdecay}. In the bottom half of each panel, we show the same points plotted on a logarithmic scale. As was shown in \citet{Karampelas2022a}, these values can be divided, as a first-order approximation, in two regions, the first of which being clearly related to the exponential decay of our signal. From now on, we will be referring to this region as the decay phase. 
 
Like in \citet{Karampelas2022a}, we will be focusing on the decay phase of our signal. We assume that our signal exhibits an exponential decay and thus we fit $\exp(- a\,t + b)$ in order to calculate the decay rate $a$. In order to calculate the coefficients $a$ and $b$, we take the natural logarithm of the considered values for the amplitudes, and fit the linear function $f(t)=-a\,t+b$ to them. Through this fit, we can then estimate the decay rates for all the different cases, as well as the standard deviation errors on the parameters. We need to stress here that for the cases with $C=0.5$, with and without thermal conduction, the decay phase was difficult to be distinguished from the following phase of the signal. Therefore, we have decided to use all of the data points for the fit. 

In agreement with \citet{Karampelas2022a}, we see that the cases with thermal conduction exhibit stronger decay rates than the corresponding ones without anisotropic thermal conduction. This was explained in  \citet{Karampelas2022a}, as the result of the faster subtraction of energy from our system that takes place due to the conduction mechanism.

Our data also show that the fit for the set of $J_z$ signals with anisotropic thermal conduction results in errors of increased magnitude with respect to the fit for the $J_z$ signals without thermal conduction. Additionally, the fits for the signals resulting from larger values of the initial velocity pulse amplitude $C$ exhibit larger errors, once the cases with anisotropic thermal conduction are considered.

Gathering all the results of the fits for all simulated $J_z$ signals, we present them in Fig. \ref{fig:decayrate}. As shown in that figure, the decay rates for the set without thermal conduction seem to be fluctuating around a constant value, which can be estimated by fitting a constant value to the data. Considering the decay rates for the set of signals with thermal conduction, this trend is more ambiguous. The large error bars and the value of the decay rate for $C=0.5$ for the case with thermal conduction make it harder to fit a constant value for the decay rate trend. For reference, both a constant value and an inverse linear function $(a/C) + b$ are fitted for these data.

\bibliography{paper}{}
\bibliographystyle{aasjournal}

\end{document}